\documentclass[review]{elsarticle}

\usepackage{lineno,hyperref,amsmath,amsthm,amsfonts}\usepackage{algorithm}% http://ctan.org/pkg/algorithm
\usepackage{algpseudocode}\usepackage{tikz}%---- tikz environment
\usepackage{framed}
\usetikzlibrary{fit,positioning,shapes,arrows,shadows}
\usepackage{bm,times}
\definecolor{candypink}{rgb}{0.89, 0.44, 0.48}
\definecolor{dollarbill}{rgb}{0.52, 0.73, 0.4}%----
\definecolor{fuchsiapink}{rgb}{1.0, 0.47, 1.0}
\modulolinenumbers[5]
\newtheorem*{remark}{Remark}

\journal{arXiv.org}
\usepackage{amssymb, epsfig,amssymb, latexsym}
\usepackage{amsfonts,psfrag,amsmath,bbm,color,url}\usepackage{multirow}
\usepackage{subcaption}

\def\d{{\, \rm d}}

\def\PP{{{\rm l}\kern - .15em {\rm P} }}
\def\PN2{{\PP_{N}-\PP_{N-2}}}

\newcommand{\bx}{\boldsymbol{x}}

\newcommand{\bz}{\boldsymbol{z}}

\newcommand{\deleted}[1]{{}}%\color{red}{#1}}}
\begin{document}

\begin{frontmatter}

\title{An Efficient Data-Driven Multiscale Stochastic Reduced Order Modeling Framework for Complex Systems}
% \title{An Efficient Data-Driven Multiscale Stochastic Reduced Order Modeling Framework for Complex Turbulent Systems}


\author{Changhong Mou}
\address{Department of Mathematics, University of Wisconsin-Madison, 480 Lincoln Drive, Madison, WI 53706, USA.}
\ead{cmou3@wisc.edu}

\author{Nan Chen\fnref{myfootnote}}
\address{Department of Mathematics, University of Wisconsin-Madison, 480 Lincoln Drive, Madison, WI 53706, USA.}
\fntext[myfootnote]{Corresponding author}
\ead{chennan@math.wisc.edu}

\author{Traian Iliescu}
\address{Department of Mathematics, Virginia Tech,  Virginia Tech, Blacksburg, VA 24061, USA.}
\ead{iliescu@vt.edu}




\begin{abstract}
Suitable reduced order models (ROMs) are computationally efficient tools in characterizing key dynamical and statistical features of nature. In this paper, a systematic multiscale stochastic ROM framework is developed for complex systems with strong chaotic or turbulent behavior. The new ROMs are fundamentally different from the traditional Galerkin ROM (G-ROM) or those deterministic ROMs that aim at minimizing the path-wise errors and applying mainly to laminar systems. Here, the new ROM focuses on recovering the large-scale dynamics to the maximum extent while it also exploits cheap but effective conditional linear functions as the closure terms to capture the statistical features of the medium-scale variables and its feedback to the large scales. In addition, physics constraints are incorporated into the new ROM. One unique feature of the resulting ROM is that it facilitates an efficient and accurate scheme for nonlinear data assimilation, the solution of which is provided by closed analytic formulae. Such an analytic solvable data assimilation solution significantly accelerates the computational efficiency and allows the new ROM to avoid many potential numerical and sampling issues in recovering the unobserved states from partial observations. The overall model calibration is efficient and systematic via explicit mathematical formulae. The new ROM framework is applied to complex nonlinear systems, in which the intrinsic turbulent behavior is either triggered by external random forcing or deterministic nonlinearity. It is shown that the new ROM significantly outperforms the G-ROM in both scenarios in terms of reproducing the dynamical and statistical features as well as recovering unobserved states via the associated efficient data assimilation scheme.
\end{abstract}


\end{frontmatter}

%\linenumbers

\section{Introduction}
Modeling complex nonlinear turbulent systems is of practical importance. These complex nonlinear system appear in many areas \cite{vallis2017atmospheric, strogatz2018nonlinear, wilcox1988multiscale, sheard2009principles, ghil2012topics}, such as geophysics, climate science, engineering, atmosphere and ocean science, neural science, and material science. They often involve strong nonlinearities and multiscale structures \cite{wilcox1988multiscale, majda2016introduction, tao2009multiscale, majda2014new}. Energy is transferred across different scales via the intrinsic nonlinearity that triggers regime switching, intermittent instabilities, extreme events and many other chaotic and turbulent features \cite{salmon1998lectures, dijkstra2013nonlinear, palmer1993nonlinear, farazmand2019extreme, trenberth2015attribution, moffatt2021extreme, majda2003introduction, manneville1979intermittency}. As a result, the associated probability density functions (PDFs) are usually non-Gaussian with fat tails. Building appropriate mathematical models for complex systems that succeed in characterizing these key features  not only advances the understanding of nature, but facilitates effective state estimation and skillful forecast as well.
However, the complexity of nature makes it extremely challenging to develop a full order model (FOM) and apply it to forecast many quantities of interest. In fact, due to
the lack of physical understanding or the inadequate resolution in computations, a perfect FOM is never available in practice. Even if a nearly perfect FOM is accessible, the model is often of high dimensionality and contains a complicated model structure \cite{majda2012challenges, majda2018model}. As a consequence, there exist many mathematical and computational difficulties in analyzing and simulating such a FOM.

Suitable reduced order models (ROMs)~\cite{hesthaven2015certified,HLB96,noack2011reduced,quarteroni2015reduced,taira2020modal} are desirable surrogates for the FOMs to reproduce certain key dynamical and statistical features of nature with a much lower computational cost. Linear regression models are possibly the simplest class of %the 
ROMs \cite{yan2009linear, freedman2009statistical} %, which 
that can provide certain skill for short-term forecasts but they usually 
%suffer from characterizing 
{struggle in characterizing} the underlying nonlinear physics. Beyond the regression strategies, one systematic and commonly used approach to developing ROMs is to project the starting complex nonlinear system to the leading a few energetic modes in light of the Galerkin proper orthogonal decomposition methods~\cite{HLB96} or other empirical basis functions such as the principal interaction patterns \cite{hasselmann1988pips,kwasniok1996reduction} and the dynamic mode decomposition \cite{rowley2009spectral,schmid2010dynamic}. With a careful design of the closure terms {(see~\cite{ahmed2021closures} for a survey)} to compensate the truncation error, these ROMs are skillful in resolving certain problems in many areas, including fluids and turbulence \cite{carlberg2013gnat,mou2021data,noack2011reduced, taira2020modal, xie2018data}.
While many traditional ROM strategies focus on seeking for the optimal approximate models with deterministic model structures, stochastic effects have been incorporated into many recent developments of  ROMs for complex turbulent systems. The stochasticity is essential in assessing the uncertainty and model error in ROMs. It also plays a pivotal role in facilitating data assimilation and advancing probabilistic forecast, which is vital for predicting non-Gaussian extreme events. The MTV (named after \underline{M}ajda, \underline{T}imofeyev and \underline{V}anden Eijnden) strategy is one of the most straightforward approaches to incorporate stochastic effects in the ROMs \cite{majda2001mathematical, majda1999models}, the idea of which is to replace the self-interactions among the high frequencies by random noise. %Recently, 
{The recently proposed} nonlinear autoregression (NAR) model emphasizes the random effect in the development of regression-based ROMs and its significance in advancing the statistical forecast \cite{lu2020data, lin2021data}. On the other hand, physics-constrained regression models are a set of nonlinear stochastic ROMs \cite{majda2012physics, harlim2014ensemble,kondrashov2015data}, which take into account the energy conserving nonlinear interactions in the model development that guarantees the well-posednss of long-term behavior of the system. In addition, stochastic parameterizations are powerful ways to reduce the complexity of the unresolved part of the dynamics that nevertheless provides effective statistical feedbacks to the resolved-scale dynamics \cite{palmer2001nonlinear, majda2011improving, crommelin2008subgrid, phillips2004evaluating, branicki2013non}.

The goal of this paper is to build a systematic multiscale stochastic ROM framework for complex systems with strong chaotic or turbulent behavior.  The new ROM framework has several unique features.
First, a new strategy of building effective closure terms is incorporated into the framework that includes both deterministic and stochastic components. Fundamentally different from the traditional closure methods, the development of the closures in this new ROM framework exploits the multiscale characteristics of the underlying system to reach a systematic formulation that is mathematically more concise and tractable than most of the traditional nonlinear closures. Nevertheless, appropriate stochastic closure terms are combined with the deterministic ones in this framework that succeeds 
%\ti{succeeds?} 
in providing effective statistical feedbacks between different scales. These closure terms significantly advance the ROM in reproducing both the dynamics and statistics of nature.
Second, data assimilation is a necessary procedure in understanding and forecasting turbulent systems \cite{evensen2009data, kalnay2003atmospheric, law2015data, majda2012filtering}. It aims at recovering the unobserved state variables using %a 
Bayesian inference, and is the prerequisite of the ensemble forecast for turbulent flows. Different from most of the traditional ROMs %which 
that have to rely on ensemble based data assimilation methods for state estimation and the initialization of the forecast, the ROM with the new concise closures facilitates an efficient and accurate scheme for nonlinear data assimilation, the solution of which is provided by closed analytic formulae. Such an analytic solvable data assimilation solution %accelerates 
{increases} the computational efficiency and allows the new ROM to avoid many potential numerical and sampling issues that usually require careful ad hoc tunings.
Third, physics constraints can be naturally incorporated into this new ROM framework, which is essential for preserving the basic principles in physics and preventing finite-time blow-up solutions in the statistical forecast \cite{majda2012physics, harlim2014ensemble, majda2012fundamental}. Another desirable feature is that the ROM and the associated closures can be systematically determined via a computationally efficient optimization procedure with no ad hoc tuning. Note that the development of the ROM in this framework does not require to know the exact FOM, which is often not accessible in practice. The framework is in fact amenable to data-driven scenarios. Specifically, the partially available information from the FOM can be used to determine the large-scale structure of the ROM, while the remaining part of the ROM including the closures can be inferred from data. Finally, it is worthwhile to highlight that many well-known complex nonlinear systems in geophysics, neural science, and engineering turbulence have the same structure as those belonging to the new ROM framework \cite{chen2018conditional}, which is a further justification of the framework.

The new ROM framework will be applied to complex nonlinear systems that belong to two categories, distinguished by the triggering mechanisms of the turbulent behavior. The first test model is a viscous stochastic Burgers equation \cite{weinan2000invariant, weinan1997probability}. 
% \ti{I think Weinan is his first name and E his last name.  And Burgers should start with a capital letter. Please check references.}
% \cm{They are now fixed.}
In such a complex system, external random forcing interacts with the deterministic nonlinearity to induce intermittent signals in the form of viscous shocks, which will be dissipated via the viscous term before the appearance of discontinuities. Here, the stochastic forcing %is the triggering effects of 
%\ti
{triggers} turbulent features. The second test example is the quasi-geostrophic (QG) equation \cite{vallis2017atmospheric, salmon1998lectures}. The QG equation is a deterministic system and the turbulent features are induced completely by the nonlinear interactions between the state variables at different spatial scales. In both cases, only a small number of the state variables %is 
are involved in the ROM that allows a much lower computational cost to reproduce the key dynamical and statistical features. It is also significant to see the skill of the stochastic closure terms in effectively characterizing the turbulent features triggered completely by the deterministic nonlinearity.
%\ti{Is this true only for the QGE or Burgers as well?}

The rest of the paper is organized as follows. Section \ref{Sec:ROM_Framework} includes the new ROM framework, the associated data assimilation scheme, and the model calibration. Section \ref{Sec:Test_Examples} demonstrates the skill of the new ROMs in reproducing the key dynamical and statistical features of both the viscous stochastic Burgers equation and the QG system. The comparison with the state-of-the-art ROMs in terms of both the model simulation and the data assimilation skill is presented in this section as well. The paper is concluded in Section \ref{Sec:Conclusion}, together with a discussion of potential extensions of the current framework. All the technical details are included in Appendix.

\section{The New Reduced Order Modeling (ROM) Framework}\label{Sec:ROM_Framework}
\subsection{The mathematical framework of the ROM with a conditional linear closure}\label{Subsec:CGROM}
Many complex turbulent nonlinear systems in nature have the following abstract mathematical structure \cite{majda2016introduction,vallis2017atmospheric, salmon1998lectures, kalnay2003atmospheric},
\begin{equation}\label{eq:abs_formu}
\frac{\mathrm{d}\mathbf{u}}{\mathrm{d}t}=\left(L+D\right)\mathbf{u}+B\left(\mathbf{u},\mathbf{u}\right) +\mathbf{F}\left(t\right)+\boldsymbol{\sigma}\left(t\right)\dot{\mathbf{W}}\left(t\right),
\end{equation}
where  $\mathbf{u}\in\mathbb{C}^{N}$ is the multidimensional state variable and $t$ is time. In \eqref{eq:abs_formu}, $L\mathbf{u}$ and $D\mathbf{u}$ represent linear dispersion and dissipation effects, respectively, where $L$ is skew-symmetric and $D$ is negative-definite.
The nonlinear effect is introduced through an energy-conserving quadratic form with $\mathbf{u}\cdot B\left(\mathbf{u},\mathbf{u}\right)=0$. The system is also subject to external forcing effects that contain a deterministic component, $\mathbf{F}\left(t\right)$, and possibly a stochastic disturbance as well, which is assumed to be a Gaussian random noise, $\boldsymbol{\sigma}\left(t\right)\dot{\mathbf{W}}\left(t\right)$, for simplicity.
The model in \eqref{eq:abs_formu} can be understood as discretizing a (stochastic) partial differential equation by projecting it onto a given set of basis functions $\{\boldsymbol\varphi_1,\boldsymbol\varphi_2,\ldots,\boldsymbol\varphi_N\}$, resulting in a high-dimensional system, where the stochastic forcing accounts for the truncation error.
Due to the intrinsic turbulent behavior, the system in \eqref{eq:abs_formu} is highly nonlinear and the state variable $\mathbf{u}$ can possess strongly non-Gaussian statistics in both the marginal and joint PDFs.

Decompose the state variable $\mathbf{u}$ into $\mathbf{u} = (\mathbf{v}^\mathtt{T},\mathbf{w}^\mathtt{T}, \mathbf{z}^\mathtt{T})^\mathtt{T}$ with $\cdot^\mathtt{T}$ being the vector transpose, where $\mathbf{v}\in\mathbb{C}^{N_1}$, $\mathbf{w}\in\mathbb{C}^{N_2}$, $\mathbf{z}\in\mathbb{C}^{N_3}$, and $N_1+N_2+N_3=N$. A typical criterion for such a decomposition is based on the rank of the scales for the state variables. In other words, $\mathbf{v}$ includes the leading %a 
few modes corresponding to the large-scale components of the state variables, $\mathbf{w}$ contains the next a few modes describing the medium-scale features, while $\mathbf{z}$ involves the remaining modes that are treated as the small-scale variables. Note that different from many traditional ROM development, where the state variables are  decomposed into only the large- and small-scale components, a three-scale decomposition is utilized here. Such a new decomposition facilitates a %more smoothed 
smoother transition from large- to small-scale variables by means of a set of intermediate-scale variables. It also allows the feedback from medium- to large-scale variables to be explicitly incorporated into the ROM, which is often crucial in enhancing the model accuracy in characterizing energy transfer and intermittency. Finally, the choice of $N_1, N_2$, and $N_3$ is  case dependent but in a typical scenario $\mathbf{z}$ accounts for the majority number of the modes.

A natural way to build ROMs with a reduced dimension of the coupled system in \eqref{eq:abs_formu} is to project $\mathbf{u}$ onto its large- and medium-scale components $\mathbf{v}$ and $\mathbf{w}$,
% \ti{Below (1), we said that it is a discretized set of (S)PDE.  Below, do we project vectors?}
\begin{subequations}\label{ROM_Res}
\begin{align}
 \frac{\d\mathbf{v}}{\d t} &= \mathbf{A}^{(v)}_{v}\mathbf{v} + \mathbf{v}^*\tilde{\mathbf{B}}^{(v)}_{vv}\mathbf{v} + \mathbf{v}^*\tilde{\mathbf{B}}^{(v)}_{vw}\mathbf{w} + \mathbf{w}^*\tilde{\mathbf{B}}^{(v)}_{ww}\mathbf{w} + \mbox{residual}_1, \\
 \frac{\d\mathbf{w}}{\d t} &= \mathbf{A}^{(w)}_{w}\mathbf{w} + \mathbf{v}^*\tilde{\mathbf{B}}^{(w)}_{vv}\mathbf{v} + \mathbf{v}^*\tilde{\mathbf{B}}^{(w)}_{vw}\mathbf{w} + \mathbf{w}^*\tilde{\mathbf{B}}^{(w)}_{ww}\mathbf{w} + \mbox{residual}_2,
\end{align}
\end{subequations}
where $\cdot^*$ is the conjugate transpose. 
% \ti{Should we comment on the stochastic disturbance in (1)? And should we explicitly define residual1 and residual2?}
It is seen that the linear and the quadratic nonlinear terms in \eqref{eq:abs_formu} have been explicitly expressed in \eqref{ROM_Res}. Therefore, $\mathbf{A}^{(\cdot)}_{\cdot}$ and $\tilde{\mathbf{B}}^{(\cdot)}_{\cdot}$ in \eqref{ROM_Res} are all constant matrices and tensors while the two residual terms include the contribution from the variable $\mathbf{z}$ that has been truncated here. The notation in \eqref{ROM_Res} is slightly abused in the sense that the quadratic nonlinear terms between $\mathbf{v}$ and $\mathbf{w}$ are simply written as $\mathbf{v}^*\tilde{\mathbf{B}}^{(v)}_{vw}\mathbf{w}$ and $\mathbf{v}^*\tilde{\mathbf{B}}^{(w)}_{vw}\mathbf{w}$ while the other parts of the associated nonlinear terms, say $\mathbf{w}^*\tilde{\tilde{\mathbf{B}}}^{(v)}_{vw}\mathbf{v}$ and $\mathbf{w}^*\tilde{\tilde{\mathbf{B}}}^{(w)}_{vw}\mathbf{v}$, are omitted in \eqref{ROM_Res} for notation simplicity. Note that if the state variables are all real-valued, then $\mathbf{v}^*\tilde{\mathbf{B}}^{(v)}_{vw}\mathbf{w}$ and $\mathbf{v}^*\tilde{\mathbf{B}}^{(w)}_{vw}\mathbf{w}$ are sufficient to characterize  the quadratic nonlinear interactions between $\mathbf{v}$ and $\mathbf{w}$. Since each complex-valued variable can always be rewritten as two real-valued variables, representing its real and imaginary parts, all the state variables in the following will be regarded as real-valued. Similarly, the notation $\cdot^*$ hereafter simply denotes the transpose of a real-valued vector.
In the development of ROMs, various methods have been proposed to deal with the residual terms. If these residual terms are simply ignored, then the resulting ROM becomes the so-called Garlerkin ROM (G-ROM) \cite{hasselmann1988pips,kwasniok1996reduction}. However, it has been shown that the G-ROM often introduces inaccuracy in reproducing even the large-scale dynamics of the FOM. Such a finding implies the importance of building approximate functions to compensate the contribution from the residuals. If these functions depend only on the state variables of the ROM, then they are known as the closures since they allow the ROM to be closed \cite{ahmed2021closures,carlberg2013gnat, noack2011reduced, taira2020modal, xie2018data}.
In most of the existing methods, the closure terms are assumed to be a linear or quadratic function of the entire ROM state variables, which are $(\mathbf{v},\mathbf{w})$ in the setup of \eqref{ROM_Res}.

The new ROM to be developed here has several fundamental differences compared with the existing %ones
ROMs. First, only the first three terms on the right hand side of \eqref{ROM_Res} are retained in the new ROM framework, which represent the basic dynamics. In other words, the self nonlinear interactions of $\mathbf{w}$ %is 
are omitted. One motivation %of 
for such a simplification is that $\mathbf{w}$ is a smaller scale variable compared with $\mathbf{v}$ and therefore it is often a faster variable and contains less energy. As a consequence, %the role of 
$\mathbf{w}^*\tilde{\mathbf{B}}^{(v)}_{vw}\mathbf{w}$ and $\mathbf{w}^*\tilde{\mathbf{B}}^{(w)}_{vw}\mathbf{w}$ %is 
are expected to %be 
have a weaker role than the other nonlinear terms in \eqref{ROM_Res}. Another important reason is related to the development of an efficient data assimilation scheme for the ROM, which will be discussed in Section \ref{Subsec:DA}. Next, a hybrid closure configuration is introduced to approximate the residual part as well as the %ignorance
%\ti
{contribution} of the self nonlinear interactions of $\mathbf{w}$. Here, the closure includes both a set of deterministic functions and a stochastic component. The building block of the deterministic part of the closure in this new ROM framework is very distinct from%\ti
{that in} the existing methods, which is either linear or quadratic nonlinear in terms of both $\mathbf{v}$ and $\mathbf{w}$. Here, a special type of %the 
nonlinear functions are adopted, which are fully nonlinear with respect to $\mathbf{v}$ but are conditional linear %in regard 
%\ti
{with respect} to $\mathbf{w}$. The motivation is again that, due to the relatively faster nature of $\mathbf{w}$, a conditional linear function and stochastic terms can effectively approximate the statistical behavior of $\mathbf{w}$ and its feedback to the large-scale component, $\mathbf{v}$. With these new ingredients, the resulting ROM with the closure reads:
\begin{subequations}\label{ROM_CG_1}
\begin{align}
 \frac{\d\mathbf{v}}{\d t} &= \mathbf{A}^{(v)}_{v}\mathbf{v} + \mathbf{v}^*\mathbf{B}^{(v)}_{vv}\mathbf{v} + \mathbf{v}^*\mathbf{B}^{(v)}_{vw}\mathbf{w} + (\boldsymbol\tau^{(v)}_{w}(\mathbf{v})\mathbf{w} + \boldsymbol\tau^{(v)}_{v}(\mathbf{v})) + \boldsymbol\sigma_v \dot{\mathbf{W}}_v,\label{ROM_CG_1a}\\
 \frac{\d\mathbf{w}}{\d t} &= \mathbf{A}^{(w)}_{w}\mathbf{w} + \mathbf{v}^*\mathbf{B}^{(w)}_{vv}\mathbf{v} + \mathbf{v}^*\mathbf{B}^{(w)}_{vw}\mathbf{w} + (\boldsymbol\tau^{(w)}_{w}(\mathbf{v})\mathbf{w} + \boldsymbol\tau^{(w)}_{v}(\mathbf{v})) + \boldsymbol\sigma_w \dot{\mathbf{W}}_w,\label{ROM_CG_1b}
\end{align}
\end{subequations}
where $\dot{\mathbf{W}}_v$ and $\dot{\mathbf{W}}_w$ are the standard Gaussian white noises. The noise coefficients $\boldsymbol\sigma_v$ and $\boldsymbol\sigma_w$ are constant matrices and for simplicity they are diagonal. The four matrix or vector functions in the closure terms $\boldsymbol\tau^{(v)}_{w}(\mathbf{v})$, $\boldsymbol\tau^{(v)}_{v}(\mathbf{v})$, $\boldsymbol\tau^{(w)}_{w}(\mathbf{v})$, and $\boldsymbol\tau^{(w)}_{v}(\mathbf{v})$ can be arbitrarily nonlinear functions of $\mathbf{v}$. It is worthwhile to highlight that the conditional linear functions of $\mathbf{w}$ are not linear functions of $\mathbf{w}$. The conditional %linear 
linearity means that $\mathbf{w}$ is linear with $\mathbf{v}$ being fixed. Therefore, the closure in \eqref{ROM_CG_1} includes the effects from the nonlinear interactions between $\mathbf{v}$ and $\mathbf{w}$. Here, for the consistency with the quadratic nonlinearity in the starting model \eqref{ROM_Res}, the closure terms $\boldsymbol\tau^{(v)}_{w}(\mathbf{v})\mathbf{w} + \boldsymbol\tau^{(v)}_{v}(\mathbf{v})$ and $\boldsymbol\tau^{(w)}_{w}(\mathbf{v})\mathbf{w} + \boldsymbol\tau^{(w)}_{v}(\mathbf{v})$ are assumed to be quadratic, and therefore
\begin{equation}\label{Closure_Terms}
\begin{split}
  \boldsymbol\tau^{(v)}_{w}(\mathbf{v}) = \mathbf{v}^*\mathbf{C}^{(v)}_{w\tau} + \mathbf{D}^{(v)}_{w\tau},&\qquad
  \boldsymbol\tau^{(v)}_{v}(\mathbf{v}) = \mathbf{v}^*\mathbf{B}^{(v)}_{v\tau}\mathbf{v} + \mathbf{C}^{(v)}_{v\tau}\mathbf{v} + \mathbf{D}^{(v)}_{v\tau}\\
  \boldsymbol\tau^{(w)}_{w}(\mathbf{v}) = \mathbf{v}^*\mathbf{C}^{(w)}_{w\tau} + \mathbf{D}^{(w)}_{w\tau},&\qquad
  \boldsymbol\tau^{(w)}_{v}(\mathbf{v}) = \mathbf{v}^*\mathbf{B}^{(w)}_{v\tau}\mathbf{v} + \mathbf{C}^{(w)}_{v\tau}\mathbf{v} + \mathbf{D}^{(w)}_{v\tau}
\end{split}
\end{equation}
where $\mathbf{D}^{(v)}_{v\tau}$ and $\mathbf{D}^{(w)}_{v\tau}$ are constant vectors, while %the other 
$\mathbf{B}^{(\cdot)}_{\cdot}$, $\mathbf{C}^{(\cdot)}_{\cdot}$, and $\mathbf{D}^{(\cdot)}_{\cdot}$ are constant matrices.

It is worthwhile to highlight that many complex nonlinear systems have already fitted into the framework of \eqref{ROM_CG_1}.
Some well-known classes of the models are \cite{chen2018conditional, chen2016filtering}:
\begin{itemize}
  \item Physics-constrained nonlinear stochastic models. Examples include the noisy versions of Lorenz models, low-order models of Charney-DeVore flows, and a paradigm model for topographic mean flow interaction.
  \item Stochastically coupled reaction-diffusion models in neuroscience and ecology. Examples include stochastically coupled FitzHugh-Nagumo models and stochastically coupled SIR epidemic models.
  \item Multi-scale models in turbulence, fluids and geophysical flows. Example include the Boussinesq equations with noise and stochastically forced rotating shallow water equation.
\end{itemize}
Other applications that share similar motivations or ideas as \eqref{ROM_CG_1} include predicting the intermittent time-series of the monsoon intraseasonal variabilities \cite{chen2014predicting, chen2015predicting}, recovering the turbulent ocean flows with noisy observations from Lagrangian tracers \cite{chen2014information, chen2015noisy}, cheap  solvable forecast models in dynamic stochastic superresolution  \cite{branicki2013dynamic, keating2012new}, and stochastic superparameterization for geophysical turbulence \cite{majda2014new}. These examples further justify that the new ROM framework \eqref{ROM_CG_1} is appropriate to characterize or approximate many nonlinear and non-Gaussian systems in various disciplines.

\subsection{Efficient data assimilation utilizing the new ROM}\label{Subsec:DA}
One of the major goals of building a ROM is to use it as a cheap forecast model. For turbulent systems, the forecast skill %replies 
%\ti
{relies} heavily on the accuracy of the initial values. However, it is often the case that only partial observations are available for complex systems. These partial observations usually correspond to the large-scale modes as they are the easiest to identify, while the variables that are below certain spatial scales are often hard to be%\ti
{approximated.} %unresolved. 
Therefore, it is important to combine the ROM with the partial observations for recovering the unobserved state variables in an accurate fashion,%\ti
{which then can be used as} %using which as the 
initializations for the forecast~\cite{kaercher2018reduced,maday2015parameterized,pagani2017efficient,popov2021multifidelity,stefanescu2015pod,xiao2018parameterised,zerfas2019continuous}. This is known as data assimilation (or filtering) \cite{kalnay2003atmospheric, lahoz2010data, majda2012filtering, evensen2009data, law2015data}.

Ensemble based data assimilation schemes \cite{evensen2009data, whitaker2002ensemble}, such as the ensemble Kalman filter, have been widely utilized in practice%\ti
{eliminate the rest of this sentence?} and lead to many successes. However, there are a few intrinsic difficulties in applying these classical data assimilation methods. The main challenge comes from the sampling error as the state estimation is completely determined by the ensembles. To prevent the filter divergence, localization, noise inflation, and many other ad hoc tuning procedures are usually inevitable as effective remedies when using the ensemble based methods \cite{hunt2007efficient, houtekamer2016review}. In addition, if a large number of ensembles are used in the system to guarantee the accuracy of the data assimilation results, then the computational cost can %be 
significantly increase.

Different from the traditional data assimilation methods, the new ROM in \eqref{ROM_CG_1} allows an efficient and accurate data assimilation scheme. The solution of the data assimilation result is given by closed analytic formulae and therefore it avoids all the ad hoc tuning procedures. In fact, the ROM in \eqref{ROM_CG_1} can be written into a more abstract form,
\begin{subequations}\label{CGNS}
\begin{align}
  \frac{\d\mathbf{v}(t)}{\d t} &= \Big[\mathbf{A}_\mathbf{0}(\mathbf{v},t) + \mathbf{A}_\mathbf{1}(\mathbf{v},t) \mathbf{w}(t)\Big]  + \mathbf{B}_\mathbf{1}(\mathbf{v},t)\dot{\mathbf{W}}_\mathbf{1}(t),\label{CGNS_X}\\
  \frac{\d\mathbf{w}(t)}{\d t} &= \Big[\mathbf{a}_\mathbf{0}(\mathbf{v},t) + \mathbf{a}_\mathbf{1}(\mathbf{v},t) \mathbf{w}(t)\Big]  + \mathbf{b}_\mathbf{2}(\mathbf{v},t)\dot{\mathbf{W}}_\mathbf{2}(t),\label{CGNS_Y}
\end{align}
\end{subequations}
where $\mathbf{A}_0, \mathbf{a}_0, \mathbf{A}_1, \mathbf{a}_1, \mathbf{B}_1$, and $\mathbf{b}_2$ are matrices and vectors that can depend nonlinearly on the state variables $\mathbf{v}$. Despite being highly nonlinear and non-Gaussian, the process of $\mathbf{w}$ in \eqref{CGNS} is linear by design if a realization of the time series $\mathbf{v}(s)$ for $s\in[0,t]$ is given. Due to such a conditional linear property, the conditional distribution of $\mathbf{w}(t)$, given a realization of $\mathbf{v}(s)$ for $s\in[0,t]$, is Gaussian,
\begin{equation}\label{CGNS_PDF}
    p(\mathbf{w}(t)|\mathbf{v}(s),s\leq t) = \mathcal{N}(\boldsymbol\mu(t),\mathbf{R}(t)).
\end{equation}
One desirable feature of the ROM framework \eqref{CGNS} is that the conditional mean $\boldsymbol\mu$ and the conditional covariance $\mathbf{R}$ can be solved via the following \emph{closed analytic formulae} \cite{liptser2013statistics, chen2018conditional, chen2016filtering}
\begin{subequations}\label{CGNS_Stat}
\begin{align}
  \frac{\d \boldsymbol{\mu}}{\d t} &= (\mathbf{a}_\mathbf{0} + \mathbf{a}_\mathbf{1} \boldsymbol{\mu}) + (\mathbf{R}\mathbf{A}_\mathbf{1}^* ) (\mathbf{B}_1\mathbf{B}_1^*)^{-1} \left[\frac{\d\mathbf{v}}{\d t} - (\mathbf{A}_\mathbf{0} + \mathbf{A}_\mathbf{1}\boldsymbol{\mu})\right],\label{CGNS_Stat_Mean}\\
  \frac{\d\mathbf{R}}{\d t} &= \mathbf{a}_\mathbf{1} \mathbf{R} + \mathbf{R}\mathbf{a}_\mathbf{1}^* + \mathbf{b}_2\mathbf{b}_2^* - ( \mathbf{R}\mathbf{A}_\mathbf{1}^*)(\mathbf{B}_1\mathbf{B}_1^*)^{-1}(\mathbf{A}_\mathbf{1}\mathbf{R}).\label{CGNS_Stat_Cov}
\end{align}
\end{subequations}
The explicit formulae in \eqref{CGNS_PDF}--\eqref{CGNS_Stat} correspond to the optimal nonlinear filter (namely the online data assimilation) solution of the state variable $\mathbf{w}(t)$ given a realization of the observed time series $\mathbf{v}(s)$ for $s\in[0,t]$.
Thus,  $\boldsymbol\mu$ and  $\mathbf{R}$ are also known as the filter posterior mean and the filter posterior covariance, respectively. The closed analytic formulae for the filtering solution allow an efficient and accurate online state estimation of the unobserved state $\mathbf{w}$ from the observed time series of $\mathbf{v}$ and the model structure of the ROM. The real-time recovery of the unobserved state provides the initial condition of $\mathbf{w}$, which together with the observed initial value of $\mathbf{v}$, advances the forecast utilizing the ROM. Note that the classical Kalman-Bucy filter \cite{kalman1961new} is the simplest special example of \eqref{CGNS_Stat}.

Due to the solvable conditional Gaussian statistics in \eqref{CGNS_PDF}--\eqref{CGNS_Stat}, the new ROM is named as the conditional Gaussian ROM (CG-ROM).
%Figure \ref{fig:schematic-cg} includes a schematic illustration of this new CG-ROM framework and its roles in studying complex systems.

% \begin{figure}[htb]%[H]
%     \centering
%      \input{figures/tikz/tikz_cg_frame}
%     \caption{Schematic illustration of the new CG-ROM framework: Model development and applications to data assimilation and forecast.  
%     % \ti{I think we should use capital letters for all the words in the schematic, for consistency.}
%     % \cm{They are fixed now.}
%     }
%     \label{fig:schematic-cg}
% \end{figure}

\subsection{Data-driven model calibration \label{section:ss-data-driven model calibration}}
What remains is to determine the matrices and vectors in the ROM \eqref{ROM_CG_1}, which correspond to the model parameters.
As was discussed above, partial observation means only a time series of the observed large-scale variable $\mathbf{v}$ is available in practice. Yet, it is assumed that a time series of both $\mathbf{v}$ and $\mathbf{w}$ is accessible in the model calibration stage to simplify the procedure of determining the model parameters. This can be explained as using the reanalysis data in many practical applications, where the historical trajectories of the unobserved data has been inferred via certain Bayesian approaches \cite{saha2010ncep}. Otherwise, an iteration process involving an alternation between state estimation and parameter estimation can be adopted for the model calibration. The basic algorithm of parameter estimation remains the same but the entire process is more complicated, which will diverge from the ROM development here, and is therefore treated as %a 
future work%. See 
(see the discussion in Section \ref{Sec:Conclusion}). Nevertheless, only the value of the observed variable $\mathbf{v}$ is available in the forecast stage such that data assimilation becomes essential in recovering the initial value of $\mathbf{w}$.

The parameters in the ROM \eqref{ROM_CG_1} can be divided into three sets:
\begin{enumerate}
  \item [(A).] The parameters in the basic dynamics: $\mathbf{A}^{(v)}_{v}$, $\mathbf{B}^{(v)}_{vv}$, $\mathbf{B}^{(v)}_{vw}$, $\mathbf{A}^{(w)}_{w}$, $\mathbf{B}^{(w)}_{vv}$, $\mathbf{B}^{(w)}_{vw}$ in \eqref{ROM_CG_1}
  \item [(B).] The parameters in the deterministic part of the closure: $\mathbf{B}^{(\cdot)}_{\cdot}$, $\mathbf{C}^{(\cdot)}_{\cdot}$ and $\mathbf{D}^{(\cdot)}_{\cdot}$ in \eqref{Closure_Terms}
  \item [(C).] The parameters in the stochastic part of the closure: $\boldsymbol\sigma_v$ and $\boldsymbol\sigma_w$ in \eqref{ROM_CG_1}
\end{enumerate}
The parameters in Set (A) are inherited from the FOM if the FOM is available. Otherwise,  they can be handled in the same way as those in Sets (B) and (C), which are computed using the following optimization algorithm. For the simplicity of discussion, consider the discrete approximation of the original continuous version of the ROM \eqref{ROM_CG_1} by applying the Euler-Maruyama scheme:
\begin{subequations}\label{ROM_CG_1_update}
\begin{align}
 \mathbf{v}^{j+1} &= \underbrace{\mathbf{v}^{j}+\biggl[ \mathbf{A}^{(v)}_{v}\mathbf{v}^{j} + (\mathbf{v}^j)^*\mathbf{B}^{(v)}_{vv}\mathbf{v}^{j} + (\mathbf{v}^j)^*\mathbf{B}^{(v)}_{vw}\mathbf{w}^j
 \biggr]\Delta t}_{\mathbf{C}_L^j}
 + \nonumber
 \\
 &\underbrace{
 \biggl[\boldsymbol\tau^{(v)}_{w}(\mathbf{v}^j)\mathbf{w}^j + \boldsymbol\tau^{(v)}_{v}(\mathbf{v}^j)\biggr]\Delta t}_{\mathbf{M}^j_L(\theta_L)}
 + (\boldsymbol\sigma_v^j \sqrt{\Delta t})\varepsilon^j_{\mathbf{v}},\label{ROM_CG_1a_update}
 \\
 %%%%
  \mathbf{w}^{j+1} &=
  \underbrace{
  \mathbf{w}^{j}+\biggl[ \mathbf{A}^{(w)}_{w}\mathbf{v}^{j} + (\mathbf{v}^j)^*\mathbf{B}^{(w)}_{vv}\mathbf{v}^{j} + (\mathbf{v}^j)^*\mathbf{B}^{(w)}_{vw}\mathbf{w}^j
 \biggr]\Delta t
 }_{\mathbf{C}_S^j}
 + \nonumber
 \\
 &\underbrace{\biggl[\boldsymbol\tau^{(w)}_{w}(\mathbf{v}^j)\mathbf{w}^j + \boldsymbol\tau^{(w)}_{v}(\mathbf{v}^j)\biggr]\Delta t}_{\mathbf{M}^j_S(\theta_S)}
 + (\boldsymbol\sigma_w^j \sqrt{\Delta t})\varepsilon^j_{\mathbf{w}},\label{ROM_CG_1b_update}
\end{align}
\end{subequations}
where $\varepsilon^j_{\mathbf{v}}$ and $\varepsilon^j_{\mathbf{w}}$ are standard independent and identically distributed Gaussian random variables at time $t_j$. Assume the noise coefficients are constants, namely $\boldsymbol\sigma_v^j:=\boldsymbol\sigma_v$ and $\boldsymbol\sigma_w^j:=\boldsymbol\sigma_w$ for all $j$. Denote $\boldsymbol{\Sigma}$ a block diagonal matrix with entries $(\boldsymbol\sigma_v)^2\Delta t$ and $(\boldsymbol\sigma_w)^2\Delta t$. Note that in many practical situations $\boldsymbol\Sigma$ is a diagonal matrix, from which $\boldsymbol\sigma_v$ and $\boldsymbol\sigma_w$ can be inferred. Further denote
\begin{equation}\label{defined_matrices}
    \boldsymbol\theta =\begin{bmatrix}
     \boldsymbol\theta_L\\
     \boldsymbol\theta_S
    \end{bmatrix}
    \qquad
    \mathbf{C} =\begin{bmatrix}
     \mathbf{C}_L^j\\
     \mathbf{C}_S^j
    \end{bmatrix}
    \qquad
    \mathbf{M}^j\boldsymbol\theta =\begin{bmatrix}
     \mathbf{M}_L^j\boldsymbol\theta_L&0\\
     0&\mathbf{M}_S^j\boldsymbol\theta_S
    \end{bmatrix},
\end{equation}
where $\boldsymbol\theta_L$ contains all the parameters in $\boldsymbol\tau^{(v)}_{w},\boldsymbol\tau^{(v)}_{v}$ and $\boldsymbol\theta_S$ contains all those in $\boldsymbol\tau^{(w)}_{w},\boldsymbol\tau^{(w)}_{v}$. Here, the parameters are assumed to appear as multiplicative pre-factors of functions consisting of state variables $\mathbf{M}^j$ such that the closure terms can be written in an abstract form $\mathbf{M}^j\boldsymbol\theta$. The vector $\mathbf{C}^j$ in \eqref{defined_matrices} includes all those terms on the right hand side of \eqref{ROM_CG_1} that do not contain parameters.  Then the optimization problem based on a linear regression can be formulated as{~\cite{chen2020learning}}: 
%\ti{Should we add a reference here?}
\begin{align}
    & \widetilde{\mathcal{L}}
    =
     \min_{\boldsymbol\theta,\boldsymbol{\Sigma}}
     \biggl(
     \sum_{j=1}^J
     \biggl(
      J \log|\boldsymbol{\Sigma}|
     +\frac{1}{2}
     \sum_{j}
     \,
     \bigl(
     \bz^{j+1}-\mathbf{M}^j\boldsymbol\theta-\mathbf{C}^j
     \bigr)^\ast
     \mathbf{R}^{-1}
       \bigl(
     \bz^{j+1}-\mathbf{M}^j\boldsymbol\theta-\mathbf{C}^j
     \bigr)
     \,
     \biggr)
     \biggr),
     \label{eqn:unconstrained-optimization-1}
     \end{align}
where $\bz^{j+1} = [\mathbf{v}^{j+1};\mathbf{w}^{j+1}]$ and $ J $ is the number of training snapshots. To solve the optimization problem~\eqref{eqn:unconstrained-optimization-1}, we use the iterative method to compute the following:
\begin{align}
    &\boldsymbol{\Sigma}
    =
    \frac{1}{J}
    \sum_{j=1}^J
    \left(
    \mathbf{z}^{j+1}-\mathbf{M}^j\boldsymbol\theta-\mathbf{C}^j
    \right)
        \left(
    \mathbf{z}^{j+1}-\mathbf{M}^j\boldsymbol\theta-\mathbf{C}^j
    \right)^\ast,\\
    & \boldsymbol\theta = \biggl(\sum_{j}^J\, (\mathbf{M}^j)^\ast \boldsymbol{\Sigma}^{-1}\mathbf{M}^j \biggr)^{-1}
    \, \biggl(\sum_{j}^J\, (\mathbf{M}^j)^\ast \boldsymbol{\Sigma}^{-1}\bigl( \mathbf{z}^{j+1}-\mathbf{C}^j\bigr)\biggr).
\end{align}

\subsection{Incorporating physics constraints into the ROM \label{sectoin:ss-physics-constraints}}
Physics constraints correspond to the conservation of the energy in the quadratic nonlinear terms in \eqref{eq:abs_formu}, namely, $\mathbf{u}\cdot B\left(\mathbf{u},\mathbf{u}\right)=0$. Incorporating the physics constraint into the ROM is central in preventing finite-time blow-up of the solutions and facilitates a skillful medium- to long-range forecast \cite{majda2012physics, harlim2014ensemble, majda2012fundamental}.

The physics constraints (or in general any constraint) for the model can be written as the following matrix equality:
\begin{align}
\mathbf{H}\boldsymbol\theta
    =\mathbf{0},
\end{align}
where $\mathbf{H}$ contains those functions that are involved in the constraints. Then the optimization problem in \eqref{eqn:unconstrained-optimization-1} is reformulated as a constrained optimization problem,
\begin{align}
&\begin{aligned}
     \widetilde{\mathcal{L}}
    =
     \min_{\boldsymbol\theta,\boldsymbol{\Sigma}}
     \biggl(
     -\frac{ J }{2}\log|&\boldsymbol{\Sigma}^{-1}|
     +
     \\
     &\frac{1}{2}
     \sum_{j}
     \,
     \bigl(
     \bz^{j+1}-\mathbf{M}^j\boldsymbol\theta-\mathbf{C}^j
     \bigr)^\ast
     \boldsymbol{\Sigma}^{-1}
       \bigl(
     \bz^{j+1}-\mathbf{M}^j\boldsymbol\theta-\mathbf{C}^j
     \bigr)
     \,
     \biggr) 
\end{aligned}
\\
     &s.t. \hspace{2cm} \mathbf{H}\boldsymbol\theta =\mathbf{0}.
\end{align}
The constrained optimization problem can be solved by the Lagrangian multiplier method \cite{boyd2004convex}, where the Lagrangian function yields
\begin{align}
\begin{aligned}
     f(\boldsymbol\theta,\boldsymbol{\Sigma}^,\lambda)
    =
 \biggl(
     -\frac{ J }{2}\log|\boldsymbol{\Sigma}^{-1}|
     +\frac{1}{2}
     \sum_{j}
          \,
     \bigl(
     \bz^{j+1}-\mathbf{M}^j\boldsymbol\theta-\mathbf{C}^j
     \bigr)^\ast
     \boldsymbol{\Sigma}^{-1}
     \\
       \cdot\bigl(
     \bz^{j+1}-\mathbf{M}^j\boldsymbol\theta-\mathbf{C}^j
     \bigr)
         \,
     \biggr)
     +
    \lambda^\ast \mathbf{H}\boldsymbol\theta.
\end{aligned}
\end{align}
The solution can be obtained using the iterative method:
\begin{align}
    \boldsymbol{\Sigma}
    &=
    \frac{1}{J}
    \sum_{j=1}^J
    \left(
    \mathbf{z}^{j+1}-\mathbf{M}^j\boldsymbol\theta-\mathbf{C}^j
    \right)
        \left(
    \mathbf{z}^{j+1}-\mathbf{M}^j\boldsymbol\theta-\mathbf{C}^j
    \right)^\ast,
    \\
    \lambda &= \left( \mathbf{H}\mathcal{K}^{-1} \mathbf{H}^\ast\right)^{-1}\left(\mathbf{H}\mathcal{K}^{-1}\mathbf{b}\right),
    \\
    \boldsymbol\theta &= \mathcal{K}^{-1}(\mathbf{b}-\mathbf{H}^\ast\lambda),
\end{align}
where
\begin{align}
    \mathcal{K}&= \sum_{j}\, (\mathbf{M}^j)^\ast \boldsymbol{\Sigma}^{-1}\mathbf{M}^j\,,
    \\
    \mathbf{b}&=\sum_{j}\, (\mathbf{M}^j)^\ast \boldsymbol{\Sigma}^{-1}{(\mathbf{z}^{j+1}-\mathbf{C}^j)}\,.
\end{align}
% \ti{I think we have a lot of material in this section.  Maybe we can summarize the models and the corresponding equations in a table?  Or include a flow chart?}
%\red{Yes}
\begin{figure}[H]
\centering
\begin{framed}
\includegraphics[width=.95\textwidth]{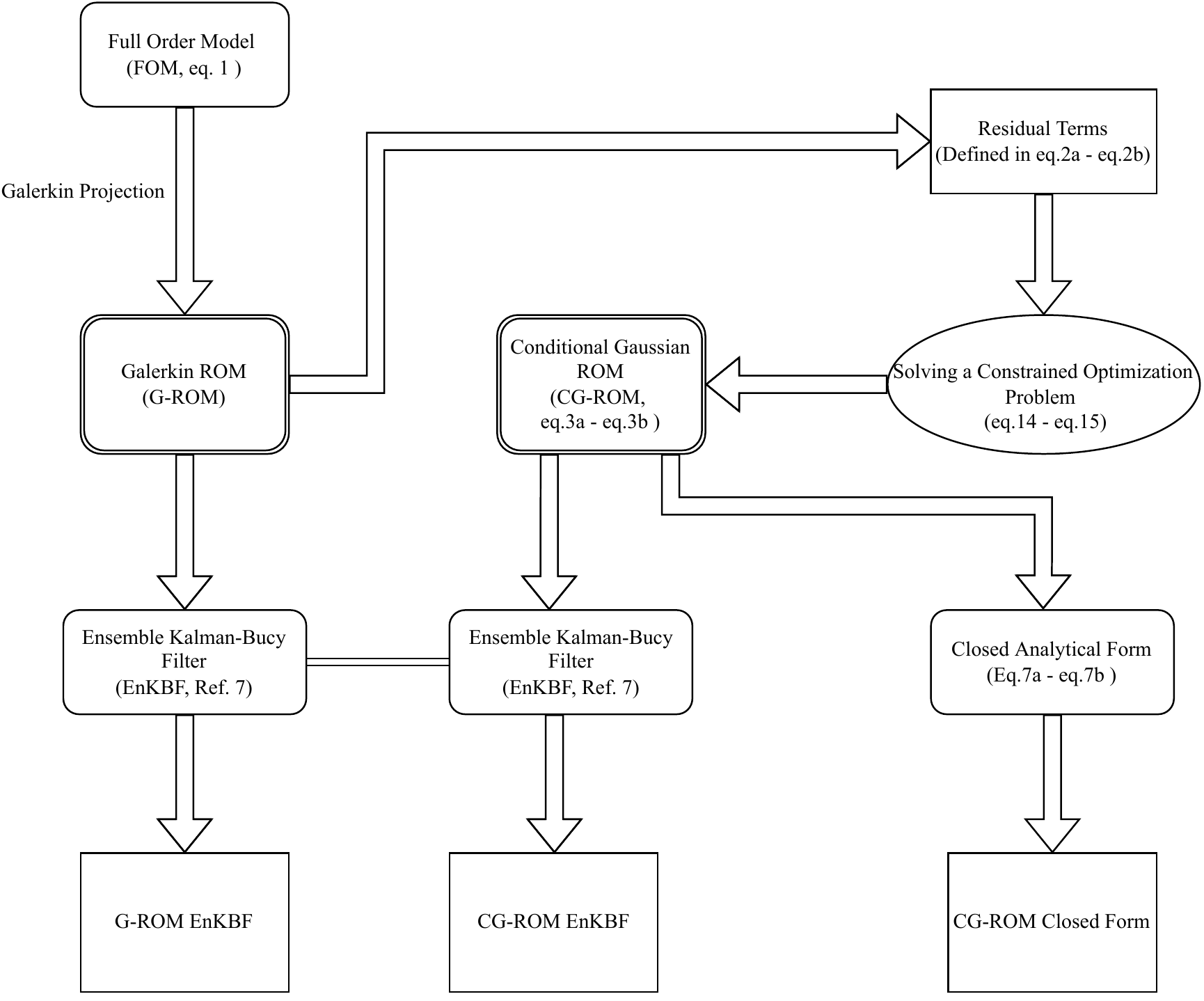}
\end{framed}
    \caption{Flowchart of conditional Gaussian framework: a summary of models and the corresponding equations.
 \label{fig:flow-chart}
    }
\end{figure}

In Figure~\ref{fig:flow-chart}, we summarize all the models and corresponding equations used in this paper.

\section{Applications to Complex Turbulent Flows}\label{Sec:Test_Examples}
% In Section~\ref{ss-burgers}, the numerical results of 
% \ti{Maybe summarize in a paragraph the plan of this section?}
\subsection{The viscous stochastic Burgers equation~\label{ss-burgers}}

The first application of the CG-ROM framework is the following viscous stochastic Burgers equation \cite{weinan2000invariant, weinan1997probability},
\begin{equation}\label{eqn:sbe}
    \frac{\partial u}{\partial t} = \nu u_{xx} + \lambda u - \gamma u u_x + \sigma \dot{W}_t,
\end{equation}
which is posed on the spatial domain $(0,L)$ supplemented with the Dirichlet boundary conditions $u(0,t)  = u(L,t) = 0$ for $t\geq0$ and a suitable initial condition $u(x,0)  = u_0(x)$ for $x\in(0,L)$.
In \eqref{eqn:sbe}, $\nu>0$ is the viscosity constant, $\lambda$ is the linear drag coefficient, $\gamma$ is the advection strength, $\sigma$ stands for the level of the stochastic noise, and $\dot{W}_t$ is a spatiotemporal white noise source.

The viscous stochastic Burgers equation \eqref{eqn:sbe} is very different from the standard deterministic inviscid Burgers equation, which is well-known for the development of discontinuities, namely shock waves. In \eqref{eqn:sbe}, the external random forcing interacts with the deterministic nonlinearity to induce intermittent bursts of the signals in the form of viscous shocks, which will be dissipated via the viscous term before the appearance of discontinuities. Therefore, the viscous stochastic Burgers equation is a suitable test model that mimics turbulent behavior in nature with intermittency. It is also noticeable that the stochastic forcing is %the triggering effects of 
%\ti
{triggering} turbulent features.

\subsubsection{Numerical solver of the FOM}

%More details of the numerical solver of the FOM \eqref{eqn:sbe} can be found in section 3.1 of reference \cite{iliescu2018regularized}.

The viscous stochastic Burgers equation \eqref{eqn:sbe} is solved by a semi-implicit Euler-Maruyama scheme. At each time step the nonlinearity $uu_x = (u^2)_x/2$ and the noise term $\sigma \, \dot{W}_t$ are treated explicitly, where the random noises are drawn independently from a normal distribution $\mathcal{N}(0,1)$, while the other terms are treated implicitly.
The Laplacian operator is discretized using the standard second-order central difference approximation.

The spatial domain considered in the following tests is $x\in(0,2\pi)$. In the numerical discretization, the total number of grid points is $N_x = 512$, 
%{\red{(Should this be 512? Similar issue in the QG model.)}}{\color{blue}( I think here it should be $512$ and in QG, it should be $256\times 512$ in spectral method)}
and therefore $\Delta x \approx 1.2\times 10^{-2}$. The simulations of the viscous stochastic Burgers equation \eqref{eqn:sbe} are performed with a numerical integration time step $\Delta t=10^{-3}$ and %is 
are integrated up to $T=10^4$. 
The FOM solutions are recorded every \(5\times 10^{-2}\) simulation time units, %where 
and the initial condition is given by
\begin{align}
    u(x,0) = 0.1 \sqrt{\frac{1}{\pi}} \sin(\frac{x}{2} )+0.1 \sqrt{\frac{1}{\pi}} \sin(2x).
\end{align}
%{\red{(How the stochastic forcing is added?)}}
{%\color{blue}
In particular, the stochastic noise is added
as 
\begin{align}
\sigma
=
\sum_{k_s =1}^{N_k} \hat{\sigma}_i
\sqrt{\frac{2}{L}}\sin \left(\frac{k_s\pi x}{L}\right),    
\label{eqn:def-sigma-burgers}
\end{align}
where $N_k$ is chosen to be $4$ and $\hat{\sigma}_i$ is a constant for $i=1,2,3,4$.
%\ti{Is $i$ the same index as that in \eqref{eqn:def-sigma-burgers}?  If so, why does it take these values?}

}
\subsubsection{ROM formulation}

To build the standard Galerkin ROM, the simplest way is to adopt the Fourier sine modes as the basis functions due to the Dirichlet boundary conditions:
\begin{align}
   \varphi_k = \sqrt{\frac{2}{L}}\sin\left(\frac{k\pi x}{L}\right), k=1,2,\cdots,R.
   \label{eqn:rom-basis}
\end{align}
Then, the G-ROM yields the following:
\begin{align}
    \left(du,\varphi_k\right)  +\bigl[\left(\nu \partial_x u,\partial_x\varphi_k\right) - \left(\lambda u,\varphi_k\right) +\left( \gamma u u_x,\varphi_k\right) \bigr] dt &= \left(\sigma  dW_t,\varphi_k\right),
\label{eqn:G-ROM-weak-form}\\  &k=1,2,\cdots,R.  \notag
\end{align}
With the sought solution, $u_r = \sum_{l=1}^r a_l(t)\varphi_l(\bx) $, where $r$ is the dimension of reduced space and $\{a_l\}_{l=1}^r$ are the sought ROM coefficients varying with time, one can replace $u$ with $u_r$ in the equation~\eqref{eqn:G-ROM-weak-form} and hence obtain the Galerkin ROM (G-ROM):
\begin{equation}\label{GROM_Projection}
    da_k = \sum_{l=1}^r A_{kl} a_l\, dt+\sum_{l=1}^{r}\sum_{m=1}^{r} B_{lmk}a_la_m\, dt +\sigma_k dW_t,
    \qquad k=1,\cdots,r,
\end{equation}
where $A$ is a matrix and $B$ is a tensor which are defined as
\begin{align}
   &A_{kl} = -\nu\bigl(\frac{\partial\varphi_l}{\partial x},\frac{\partial\varphi_k}{\partial x}\bigr)+\lambda\bigl(\varphi_l,\varphi_k\bigr)
   \label{eqn:G-ROM-A}
   \\
   &B_{lmk} =-\gamma \bigl(\varphi_l\frac{\partial\varphi_m}{\partial x},\varphi_k\bigr).
      \label{eqn:G-ROM-B}
\end{align}

Thanks to the explicit formulation of ROM basis functions  (i.e., they are sine functions~\eqref{eqn:rom-basis}),
we can explicitly %write down the formulations 
%\ti
{calculate the formulas} in~\eqref{eqn:G-ROM-A} and~\eqref{eqn:G-ROM-B}:
\begin{align}
    &\bigl(\varphi_l,\varphi_k\bigr) = \delta_{lk},\quad  \biggl(\frac{\partial\varphi_l}{\partial x},\frac{\partial\varphi_k}{\partial x}\biggr) =\frac{k^2\pi^2}{L^2} \delta_{lk},\quad l,k=1,\cdots, r,\label{eqn:rom-linear}
    \\
    &\biggl(\varphi_l\frac{\partial\varphi_m}{\partial x},\varphi_k\biggr) = \begin{cases}
       \frac{m\pi}{4}\biggl(\frac{2}{L}\biggr)^{3/2}&\text{if $l+m-k=0$ or $l-m-k=0$}, \\
       -\frac{m\pi}{4}\biggl(\frac{2}{L}\biggr)^{3/2} &\text{if $l-m+k=0$,} \\
      0 & \text{otherwise.}
   \end{cases}
   \label{eqn:rom-nonlinear}
\end{align}

The CG-ROM starts from \eqref{GROM_Projection}. It drops the quadratic nonlinear interaction between the medium-scale variables but adds closure terms as was described in Section \ref{Subsec:CGROM}. Physics constraints are imposed in the development of the CG-ROM,  while they are automatically included in the G-ROM.
\subsubsection{Dynamical regimes}

Two different regimes are considered in the following tests, with the associated parameters %are 
listed in Table~\ref{tab:regime-sbe}. The two dynamical regimes %are differed by 
%\ti
{use a different} %the 
linear drag coefficient, $\lambda$. A larger (i.e., more positive value of) $\lambda$ is more prone to induce linear instability and, %is correspondingly more turbulent. 
%\ti
{thus, yields more turbulent features.}
Linear analysis shows that only one eigenvalue is positive in Regime I, while three positive eigenvalues are observed in Regime II. Therefore, Regime I is a weakly chaotic regime, while Regime II is a much more challenging regime with strong instabilities.
%%
% {\red{(Changhong, I interchange the two Regimes. The Regime I here is your original Regime II. In the figures, do not list the parameters in the title, but call that Regime I/II. Also, if your $\sigma$ is a constant, then that means you have white noise to ALL modes. There must be something wrong in the notation/presentation here, which need to be fixed.)}}
% {\color{blue}( $\sigma$ is replaced with $\hat\sigma$ which is defined in equation~\eqref{eqn:def-sigma-burgers}.)}

\begin{table}[H]
    \centering
    \begin{tabular}{c|c|c|c|c}
    \hline
        Parameters
        & $\nu$ &$\lambda$
        & $\gamma$ & %$\sigma$ 
        {
        %\color{blue} 
        $\hat{\sigma}$}
        \\
    \hline\hline
        Regime I &  $0.005$ & $0.00375$ & $1$ & $0.003$ \\
        Regime II & $0.005$ & $0.01125$ & $1$ & $0.003$ \\
        \hline
    \end{tabular}
    \caption{Parameters of the two dynamical regimes of the viscous stochastic Burgers equation \eqref{eqn:sbe}. }
    \label{tab:regime-sbe}
\end{table}

\subsubsection{Results: Comparison of the FOM and %the 
ROMs in model simulations and statistics}

To present numerical results of ROMs, we mainly consider the ROMs with $r=5$ modes, which%\ti
{account for 96.05\% of the FOM energy for Regime I,  and 93.81\% of the FOM energy for Regime II, where} %of the energy in 
the FOM system %that 
has in total $512$ modes. In the CG-ROM, the %number 
%\ti
{dimension} of the observed variable $\mathbf{v}$ is $r_1=2$ and therefore the %number 
%\ti
{dimension} of the unobserved medium-scale variable $\mathbf{w}$ is $r_2=3$, such that the total number of the modes in the CG-ROM is the same as that in the G-ROM. For the simplicity of the notation, these leading five modes are denoted by $a_1,\ldots, a_5$.

\begin{figure}[htb]%[H]
    \centering
            \includegraphics[width=\textwidth]{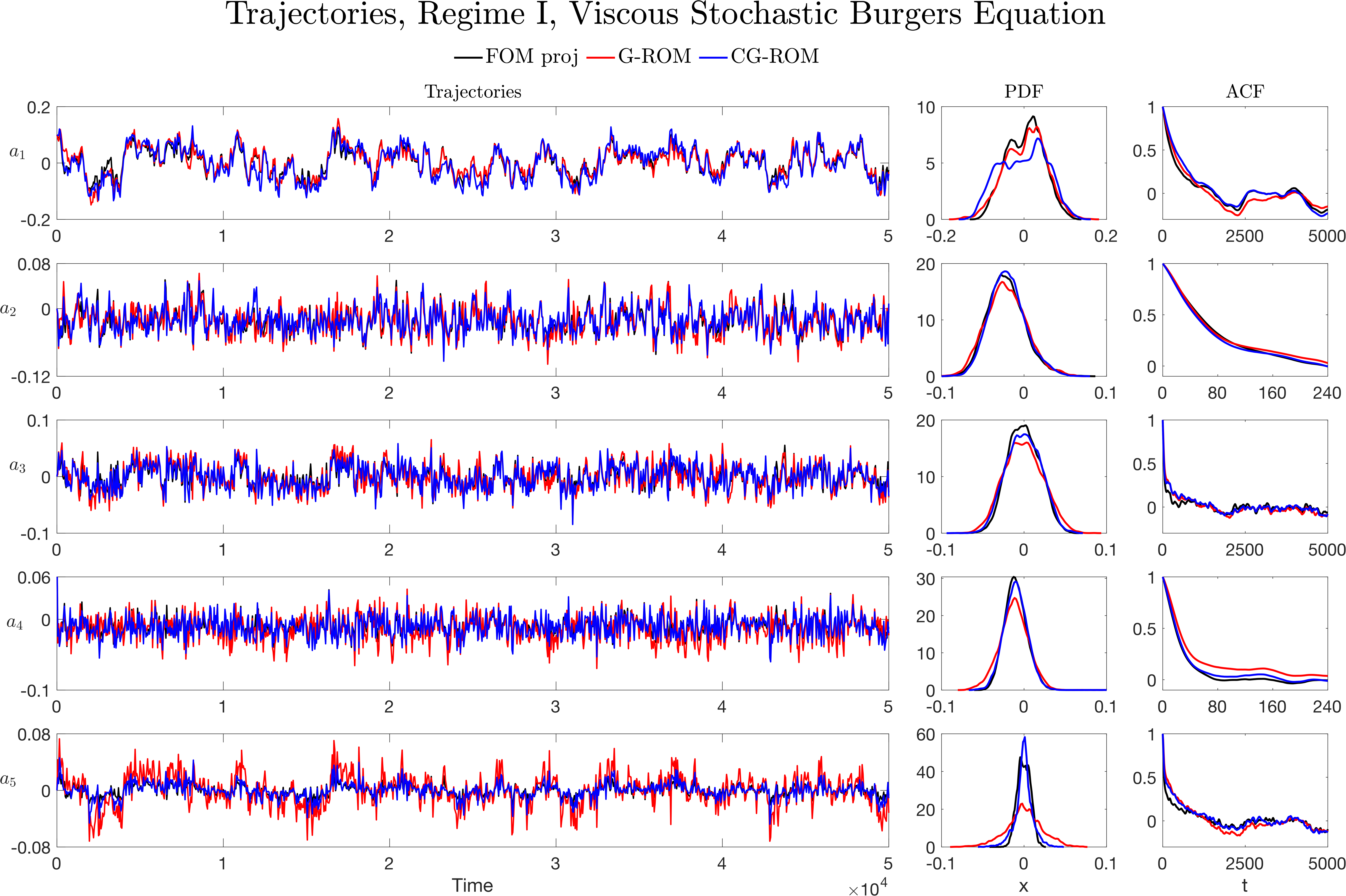}
    \caption{Regime I of the viscous stochastic Burgers equation. Comparison of the trajectories, PDFs, and ACFs of the FOM (black), G-ROM (red), and %the 
    CG-ROM (blue) for $r=5$ and $r_1=2$. The same random noise source is utilized in the three models. The PDFs and the ACFs are computed based on a finite %length of the trajectories with 
    %\ti
    {finite trajectory length of} $5\times 10^4$ units.  }
        \label{fig:sbe-rom-trajectories-2}
\end{figure}

Figure \ref{fig:sbe-rom-trajectories-2} compares the trajectories, PDFs, and autocorrelation functions (ACFs) of the FOM (black), G-ROM (red), and %the 
CG-ROM (blue) for Regime I. In order to contrast the path-wise behavior, the same random noise source is utilized in the three models. Yet, it is important to notice that in practice the noise is unknown and therefore the statistics, namely the PDFs and the ACFs, are more suitable metrics for model comparison. The G-ROM is nearly as accurate as the CG-ROM for the first two modes in terms of both the simulated trajectories and the statistics. However, due to the lack of %the 
closure terms, the G-ROM loses its skill in capturing the statistics of modes $a_3, a_4$, and $a_5$. The amplitudes of the corresponding trajectories are also overestimated. Table \ref{table-cg-rom-l2-errors} includes the $L^2$ errors for different $r_1$ and $r$ values in the ROMs. The results in %the 
Table \ref{table-cg-rom-l2-errors} consistently show that the CG-ROM outperforms the G-ROM for both the large-scale modes $\mathbf{v}$ and the medium-scale modes $\mathbf{w}$, indicating the robustness of the CG-ROM performance.

\begin{figure}[htb]%[H]
    \centering
            \includegraphics[width=\textwidth]{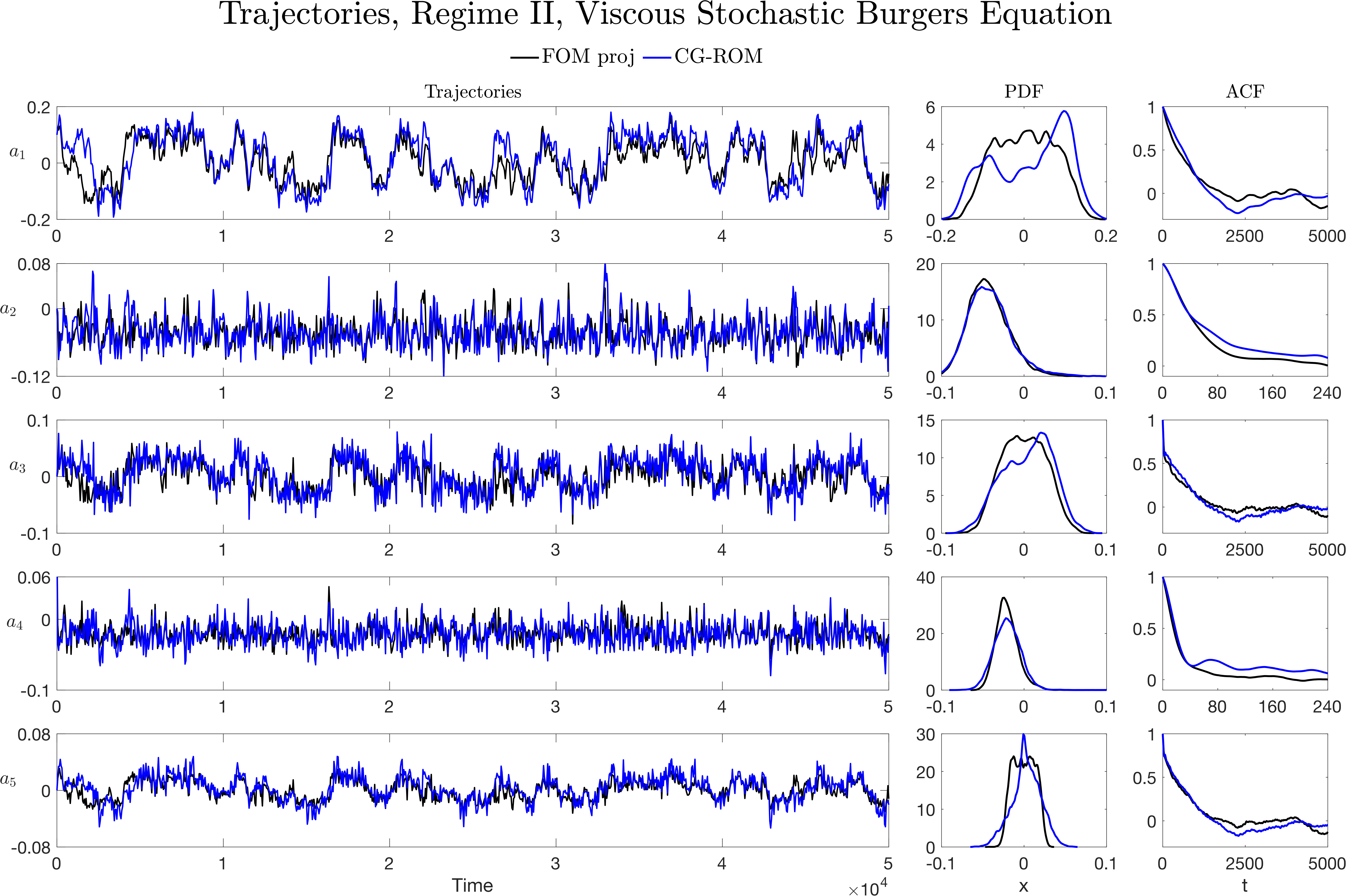}
    \caption{Regime II of the viscous stochastic Burgers equation. Comparison of the trajectories, PDFs, and ACFs of the FOM (black) and the CG-ROM (blue) for $r=5$ and $r_1=2$. The same random noise source is utilized in the two models. The PDFs and the ACFs are computed based on a finite %length of the trajectories with 
    %\ti
    {finite trajectory length of} $5\times 10^4$ units. The path-wise solution of the G-ROM blows up %\ti
    {around time unit $100$} and is therefore omitted here.}
        \label{fig:sbe-rom-trajectories-1}
\end{figure}

\begin{figure}[htb]%[H]
    \centering
            \includegraphics[width=\textwidth]{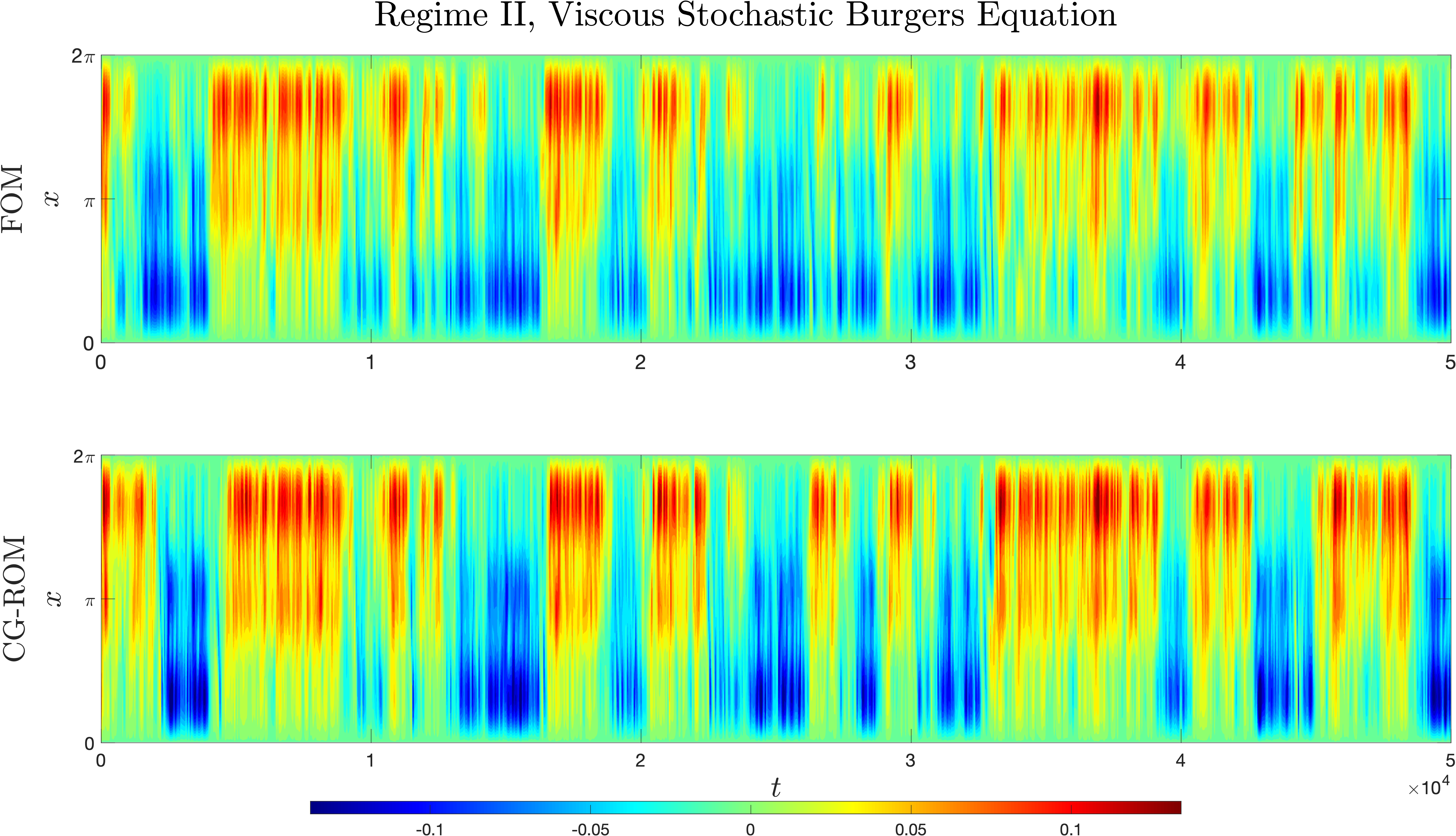}
    \caption{Regime II of the viscous stochastic Burgers equation. Reconstructed spatiotemporal patterns of the FOM and %the 
    CG-ROM.}
        \label{fig:sbe-rom-st-field-1}
\end{figure}

Figure \ref{fig:sbe-rom-trajectories-1} compares the trajectories and the statistics %between 
of the FOM (black) and the CG-ROM (blue) in Regime II, which is a much tougher regime for the test. Despite being strongly turbulent with significant non-Gaussian statistics in all the leading $r=5$ modes, the CG-ROM succeeds in reproducing both the trajectories (by assigning the same random forcing as the FOM) and the overall non-Gaussian PDFs. Note that the solution of the G-ROM suffers from strong instability and blows up in a finite time in such a test case. This further indicates the necessity of incorporating the closure terms in the development of the ROMs. Figure \ref{fig:sbe-rom-st-field-1} shows the reconstructed spatiotemporal patterns. The CG-ROM is able to reproduce the intermittent spatiotemporal patterns of the viscous shocks, which highly resemble those from the FOM with a pattern correlation that is above $0.95$.

\begin{table}[H]
    \centering
    \begin{tabular}{c|c|c|c|c}
    \hline
        $r_1,r$ & ROMs & $\mathcal{E}\left(\boldsymbol{v};L^2\right)$& $\mathcal{E}\left(\boldsymbol{w};L^2\right)$&  $\mathcal{E}\left(\boldsymbol{u};L^2\right)$
        \\
         \hline\hline
         \multirow{2}{*}{$r_1=2,r=4$}
         & G-ROM ($r$ dim)&5.818e-2& 6.476e-2 &8.705e-2
          \\
         &CG-ROM&2.051e-2 &1.924e-2 & 2.812e-2
         \\
         \hline
         \multirow{2}{*}{$r_1=2,r=5$}
         & G-ROM ($r$ dim)& 5.109e-2 &  5.919e-2  & 7.819e-2
         \\
         &CG-ROM& 2.251e-2&   2.086e-2 & 3.069e-2
         \\
         \hline
         \multirow{2}{*}{$r_1=3,r=6$}
         & G-ROM ($r$ dim) &2.193e-2&3.302e-2& 3.964e-2
         \\
         & CG-ROM& 2.957e-2& 1.668e-2&3.395e-2
         \\
         \hline
    \end{tabular}
    \caption{Regime I of the viscous stochastic Burgers equation. The  $L^2$ errors for different $r_1$ and $r$ values in the ROMs. The same random noise source is utilized in the FOM and the two ROMs.}
    \label{table-cg-rom-l2-errors}
\end{table}

\subsubsection{Results: Comparison of the FOM and %the 
ROMs in data assimilation}\label{Subsubsec:SBE_DA}
The focus of this subsection is on the data assimilation (i.e., filtering) of the unobserved variable $\mathbf{w}=(a_3,a_4,a_5)^{\mathtt{T}}$ using ROMs given one realization of the observed trajectories of the leading two modes $\mathbf{v}=(a_1,a_2)^\mathtt{T}$ generated from the FOM. As in the previous subsection, the two ROMs utilized here are the G-ROM and the CG-ROM. Recall that the closed analytic formulae \eqref{CGNS_Stat} associated with the CG-ROM provide an efficient and accurate solution of the filter posterior distribution. However, no such a closed formula can be applied to the G-ROM. Therefore, the ensemble Kalman-Bucy filter (EnKBF) \cite{bergemann2012ensemble}, which is a widely used ensemble based data assimilation scheme for continuous-in-time observations, is utilized for the G-ROM. As will be seen below, the solutions using G-ROM with the EnKBF are much less accurate %solution 
than those using the CG-ROM with closed analytic filter formulae. To understand if the data assimilation scheme or the ROM itself is the main source accounting for the poor behavior of the former, the EnKBF is also applied to the CG-ROM. In other words, three filters will be tested in %the following
%\ti
{our numerical investigation}:
\begin{itemize}
  \item CG-ROM Closed Form: using the closed analytic formulae \eqref{CGNS_Stat} as the data assimilation scheme for the CG-ROM,
  \item G-ROM EnKBF: using the EnKBF as the data assimilation scheme for the G-ROM, and
  \item CG-ROM EnKBF: using the EnKBF as the data assimilation scheme for the CG-ROM.
\end{itemize}
For the EnKBF, the size of ensemble is taken to be $100$, which has been tested to be sufficient to reach  reasonably accurate and robust results. Note that although the G-ROM or CG-ROM are often relatively low dimensional, running such a moderate ensemble size forward through the system via EnKBF is still much more computationally expensive than the closed form in the CG-ROM.
% \ti{Can we add a table with the computational cost?  The reviewers may ask about the computational cost.}

\begin{figure}[htb]%[H]
    \centering
            \includegraphics[width=\textwidth]{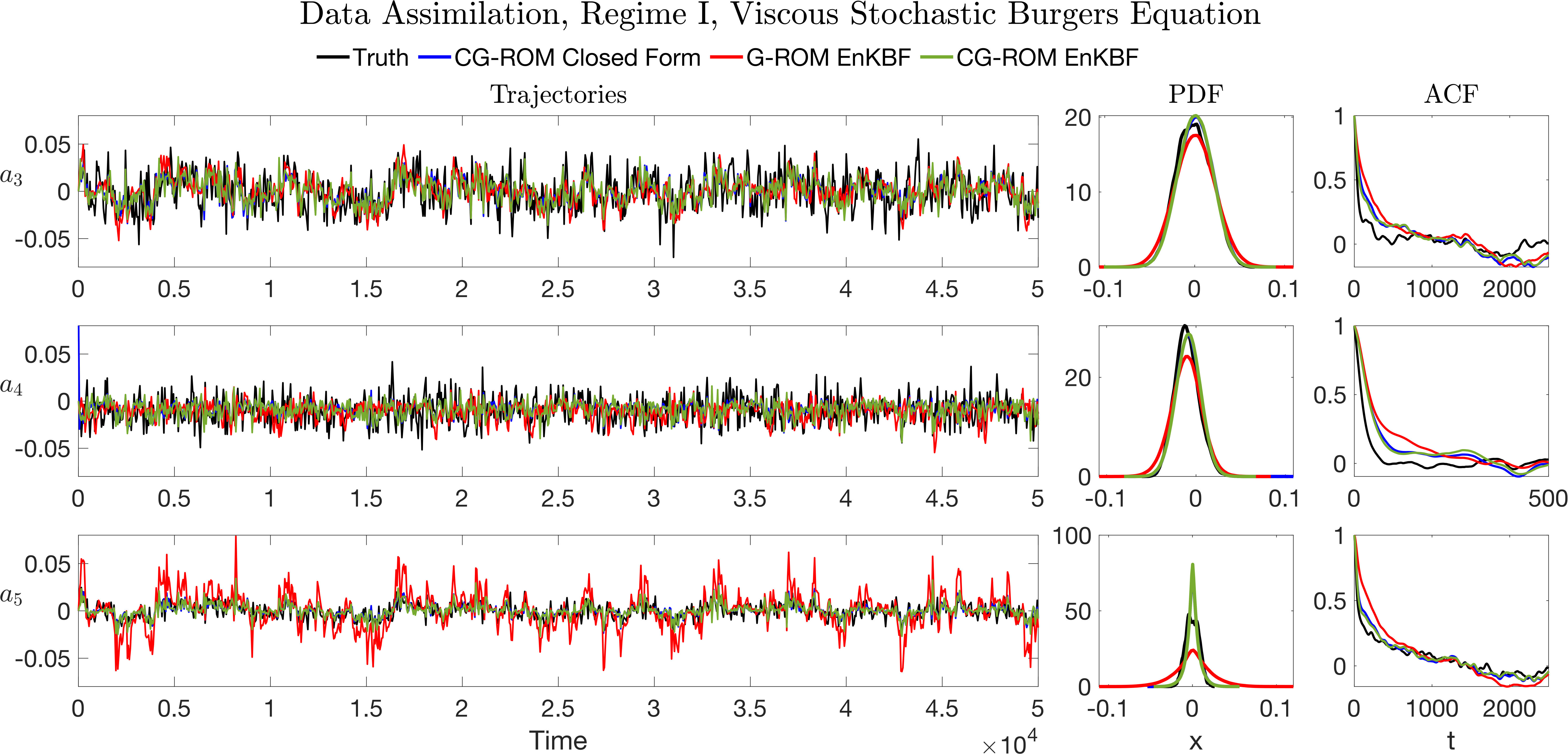}
    \caption{Regime I of the viscous stochastic Burgers equation. Truth and %the 
    posterior mean time series of $\mathbf{w}=(a_3,a_4,a_5)^\mathtt{T}$ from data assimilation using the three different filters.}
        \label{fig:sbe-da-trajectories-2}
\end{figure}

\begin{figure}[htb]%[H]
    \centering
            \includegraphics[width=\textwidth]{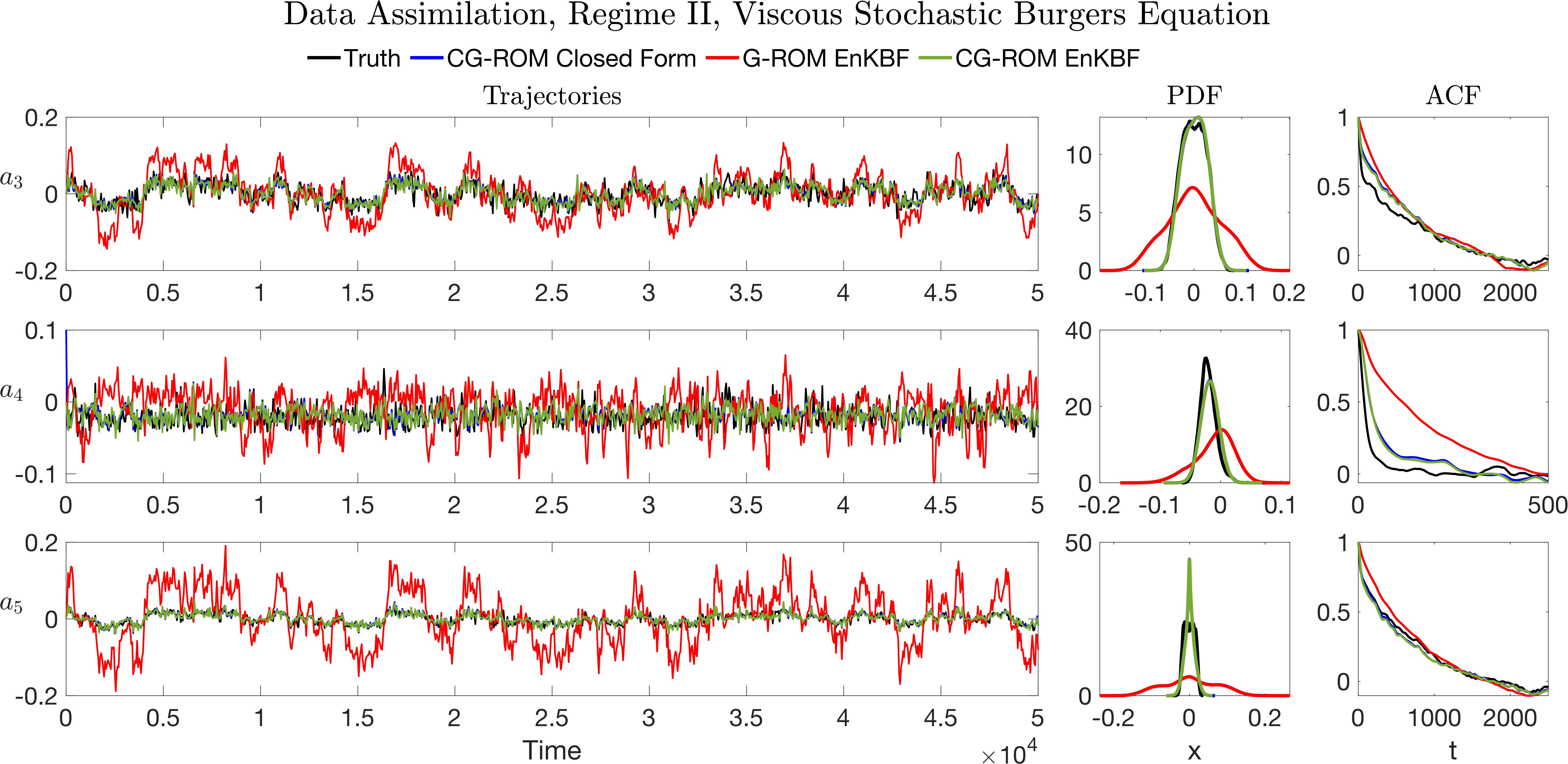}
    \caption{Regime II of the viscous stochastic Burgers equation. Truth and %the 
    posterior mean time series of $\mathbf{w}=(a_3,a_4,a_5)^\mathtt{T}$ from data assimilation using the three different filters.}
        \label{fig:sbe-da-trajectories-1}
\end{figure}

\begin{figure}[htb]%[H]
    \centering
            \includegraphics[width=\textwidth]{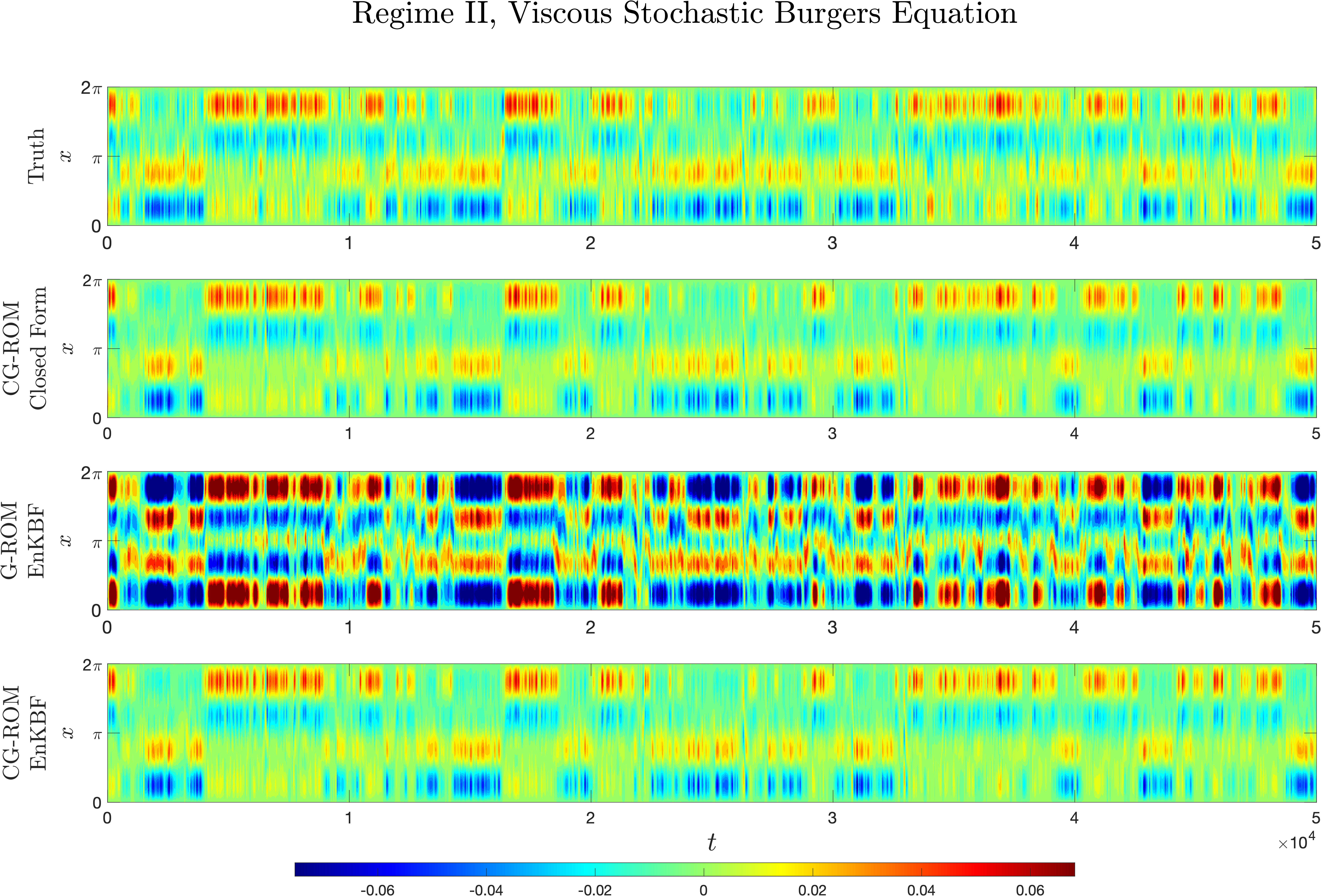}
    \caption{Regime II of the viscous stochastic Burgers equation. Reconstructed spatiotemporal field based on the three modes $\mathbf{w}=(a_3,a_4,a_5)^\mathtt{T}$ from data assimilation and the truth, where the truth is computed by projecting the true signal %to 
    on the basis functions associated with these three modes.}
        \label{fig:sbe-da-st-field-1}
\end{figure}

Figure \ref{fig:sbe-da-trajectories-2} shows the posterior mean time series of $\mathbf{w}=(a_3,a_4,a_5)^\mathtt{T}$ from data assimilation using the three different filters in Regime I. Consistent with the model simulation results presented in Figure \ref{fig:sbe-rom-trajectories-2}, the G-ROM EnKBF is significantly worse than the CG-ROM Closed Form, especially for $a_5$. As the results of CG-ROM EnKBF are only slightly worse than those of CG-ROM Closed Form in such a case, it is conclusive that the main error source for the unskillful behavior in the G-ROM EnKBF is the model error in the G-ROM, not the filtering method EnKBF. In fact, the bare truncation in the G-ROM %\ti
{effectively} breaks the coupling between $\mathbf{v}$ and $\mathbf{w}$ to a large extent. Therefore, it is more difficult to infer the unobserved states from the observations via a model with a poorly described coupling relationship.
Similar qualitative conclusions are reached in Figure \ref{fig:sbe-da-trajectories-1} that illustrates the results of Regime II. However, the G-ROM EnKBF is quantitatively much worse than its counterparts due to the stronger turbulent behavior in this regime. Note that, despite the big error, the data assimilation results using the G-ROM EnKBF do not suffer from the finite-time blow-up issue as in the forward run of the G-ROM. The reason is that in data assimilation observations are %the 
an additional input, which constantly helps mitigate the pathological behavior of the model.
Finally, Figure \ref{fig:sbe-da-st-field-1} displays a spatiotemporal reconstruction of the three unobserved modes $\mathbf{w}=(a_3,a_4,a_5)^\mathtt{T}$, using the recovered results from data assimilation, which is compared with the truth. It further validates that recovered fields from the CG-ROM Closed Form and from the CG-ROM EnKBF are comparably accurate, while that from the G-ROM is completely biased.

%{\red{(For figures, move the legend to the right to zoom in the figures.)}}

\clearpage

\subsection{The QG model}
The second application is the quasi-geostrophic (QG) equation:
\begin{equation}\label{eq:qge}
\begin{split}
\frac{\partial \omega}{\partial t}+J(\omega,\psi)-Ro^{-1}\frac{\partial \psi}{\partial x} &= Re^{-1} \Delta \omega+Ro^{-1}F,\\
\omega &=-\Delta \psi,
\end{split}
\end{equation}
which is a widely used model %to describe 
%\ti
{that describes} the large scale ocean circulation \cite{vallis2017atmospheric, salmon1998lectures}.
In the QG equation \eqref{eq:qge}, $\omega$ is the vorticity, $\psi$ is the streamfunction, $F$ is the external forcing, and the%\ti
{Jacobian} is defined as $J(A,B)=\partial A/\partial x \, \partial B/\partial y-\partial A/\partial y \, \partial B/\partial x$. There are two non-dimensional numbers in \eqref{eq:qge}: %, where 
$Re$ is the Reynolds number and $Ro$ is the Rossby number. The former is the ratio of inertial forces to viscous forces, representing the strength of turbulence, while the latter stands for the ratio of inertial force to Coriolis force.
Note that, different from the viscous stochastic Burgers equation, the QG equation is a deterministic system and its turbulent features are induced completely by the nonlinear interactions between the state variables at different spatial scales.

In the following, a symmetric double-gyre wind forcing is imposed to the QG equation \eqref{eq:qge}  \cite{greatbatch2000four,mou2020data,san2015stabilized,san2011approximate} with $F = \sin(\pi(y-1))$. A %rectangle 
rectangular domain is considered here with $\Omega = [0,1]\times [0,2]$.
Homogeneous Dirichlet boundary conditions are applied for both $\psi$ and $\omega$:
\begin{align} \label{eq:qge:bdry_cond}
\psi(t, x,y) =0,\qquad \omega(t, x,y)=0 \qquad \text{for}\quad (x,y)\in\partial \Omega \; \text{ and } \;  t \ge 0.
\end{align}
The two non-dimensional numbers are set to be $Re=450$ and $Ro = 0.0036$, such that the system has moderate turbulent behavior and a fast rotation.

\subsubsection{Numerical solver of the FOM}
\begin{figure}[htb]%[H]
    \centering
        \begin{subfigure}[b]{\textwidth}
        \centering
        \includegraphics[width=.8\textwidth]{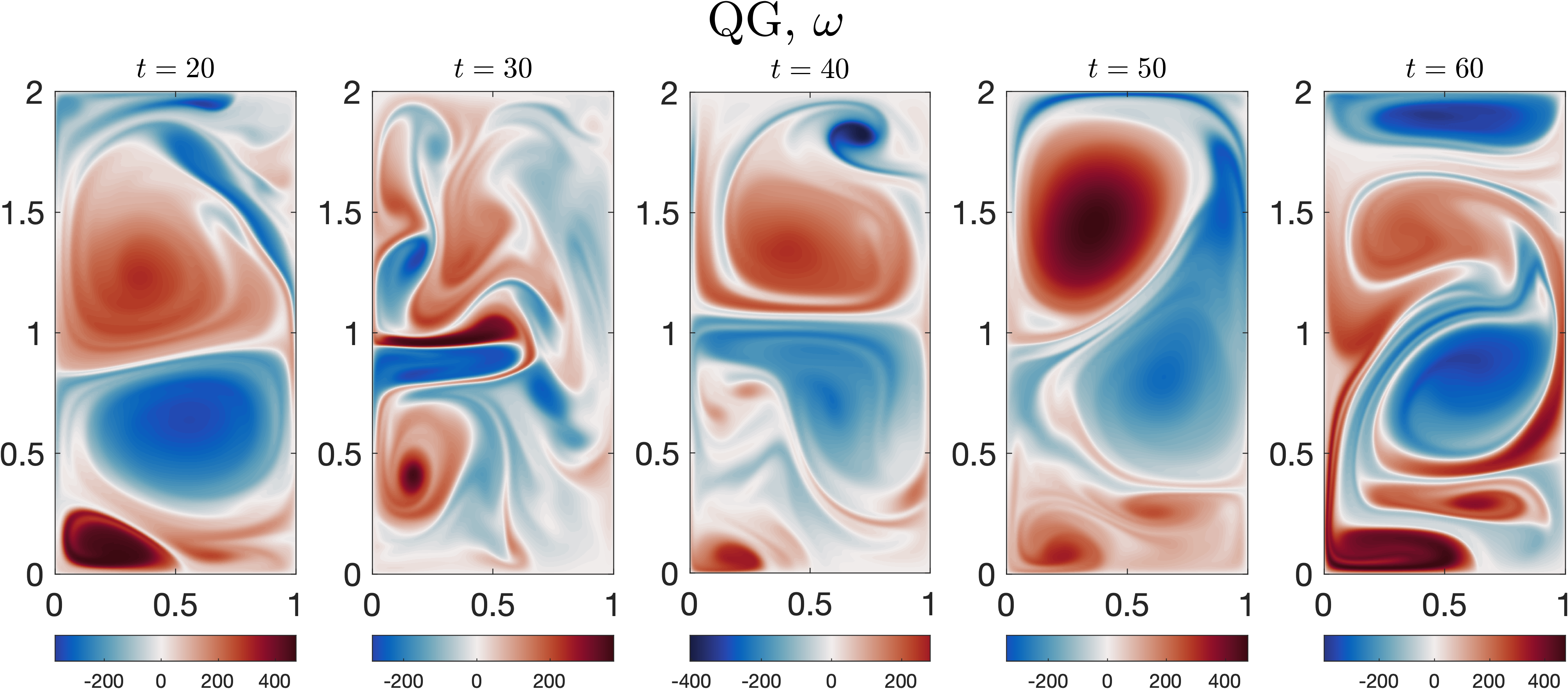}
        \caption{Vorticity}
        \end{subfigure}
        \begin{subfigure}[b]{\textwidth}
        \centering
        \includegraphics[width=.8\textwidth]{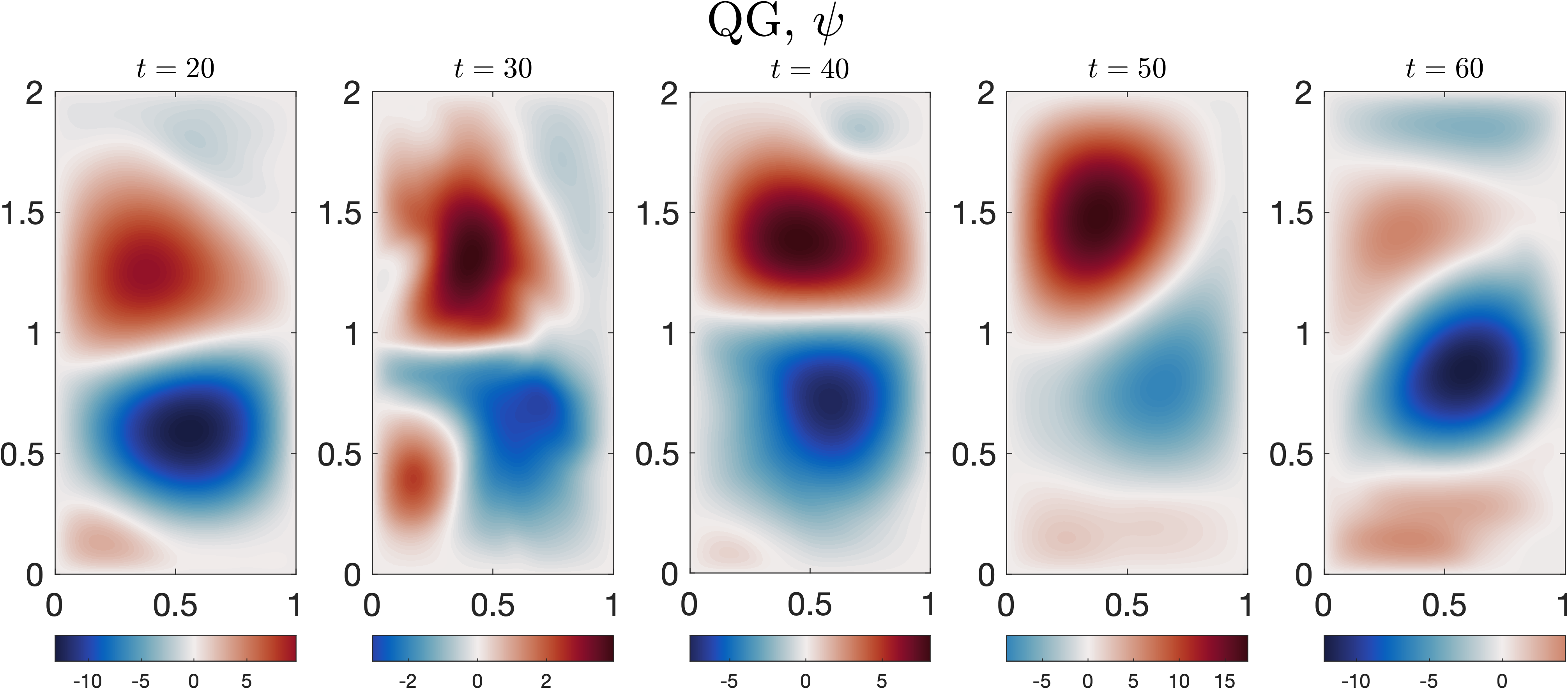}
        \caption{Streamfunction}
        \end{subfigure}
    \caption{The QG equation. Snapshots of vorticity, $\omega$ (first row) and streamfunction $\psi$, (second row) at %three 
    %\ti
    {five} different time instances}
    \label{fig:qg-fom-snapshots}
\end{figure}

A spectral method with a {$256 \times 512$} spatial resolution %{\red{(Again check if this should be 256x512)}}
and an explicit Runge-Kutta method are utilized to solve the QG model.
The FOM is simulated for $80$ %\ti
{time} units. %, where the 
The solution displays a transient behavior on the time interval $[0,10]$, and then converges to a statistically steady state on the time interval $[10,80]$.
The FOM solutions are recorded on the time interval $[10,80]$ every \(10^{-2}\) simulation time units.
which ensures that the snapshots used in the construction of the ROM basis are equally spaced. In Figure~\ref{fig:qg-fom-snapshots}, the vorticity and streamfunction snapshots at $t=20,30,40,50,60$ are plotted.

\subsubsection{ROM formulation}
The ROMs start from projecting the vorticity $\omega$ to its POD bases and keeping the leading $r$ modes. Correspondingly, a reduced set of $r$ basis functions are also introduced for $\psi$, which are  subordinate to the above POD basis functions for $\omega$ via $\phi_i(x,y)  = - \Delta^{-1} \varphi_i(x,y)$, i.e., they solve the following Poisson equation:
\begin{align}
\begin{aligned}
 - \Delta \phi_i(x,y)  = \varphi_i(x,y), \quad \text{ subject to } &\quad  \phi_i(x,y) =0,    \\
 &\text{for}\quad (x,y)\in\partial \Omega, \qquad i=1,2,\cdots,r.
\end{aligned}
\label{eqn:basis-streamfunction}
\end{align}
Note that while the POD basis $\{\varphi_i\}$ for the vorticity $\omega$ is an orthonormal basis under the $L^2$ inner product, the basis $\{\phi_i\}$ for the streamfunction $\psi$ is not orthogonal.
Given the G-ROM approximation $\omega_r = \sum_{i=1}^r a_i(t) \varphi_i(x,y)$ of $\omega$, the corresponding $\psi$ is approximated by $\psi_r = \sum_{i=1}^r a_i(t) \phi_i(x,y)$, which results from the ansatz $\psi_r = -\Delta^{-1}\omega_r$ and  {definition~\eqref{eqn:basis-streamfunction}} of the basis function $\phi_i$. With the above notations, the $r$-dimensional G-ROM for the problem \eqref{eq:qge}--\eqref{eq:qge:bdry_cond} is given by:
\begin{align}
\left(\frac{\partial \omega_r}{\partial t},\varphi_i \right)+\left( J(\omega_r,\psi_r),\varphi_i\right)-Ro^{-1}\left(\frac{\partial \psi_r}{\partial x} ,\varphi_i\right)+Re^{-1} \left( \nabla\omega_r,\nabla\varphi_i\right)=Ro^{-1}\left(F,\varphi_i\right) ,
\label{eq:rom:qge}
\end{align}
where $(\cdot,\cdot)$ denotes the $L^2$ inner product over the spatial domain, and $i = 1,\cdots, r$.
The component-wise form of the G-ROM is as follows:
for $i=1,2,\cdots,r$
\begin{equation}\label{GROM_Projection2}
\dot{a}_i(t) = b_i +\sum_{m=1}^r A_{im}a_m(t) +\sum_{m=1}^r\sum_{n=1}^rB_{imn} a_m(t)a_n(t) ,
\end{equation}
where
\begin{equation*}
\begin{split}
    b_i &= Ro^{-1} (F,\varphi_i), \, \\
A_{im } &= Ro^{-1} \left(\frac{\partial \phi_m}{\partial x},\varphi_i\right)-Re^{-1}\left(\nabla\varphi_m,\nabla\varphi_i\right), \, \\
B_{imn}&= -\left(J(\varphi_m, \phi_n),\varphi_i\right) .
\end{split}
\end{equation*}
In the above procedure, $3500$ equally spaced FOM vorticity snapshots in the time interval $[10,45]$ are utilized to construct the ROM bases \cite{mou2020data}. For computational efficiency, the FOM vorticity is interpolated onto a uniform mesh with the resolution $257\times 513$ over the spatial domain $\Omega = [0,1]\times [0,2]$, i.e., with a mesh size $\Delta x =\Delta y = 1/256$. Based on the interpolated snapshots,  the corresponding eigenvalue problem%\ti
{is solved} to generate the ROM basis.
The fourth-order Runge-Kutta scheme (RK4) is utilized for the temporal discretization. To ensure the numerical stability of the time discretization, a time step size $\Delta t = 0.001$ is utilized. The ROM data is stored every ten time steps to match the FOM sampling rate.

On the other hand, the CG-ROM follows the procedure described in Section \ref{Subsec:CGROM}. It starts from \eqref{GROM_Projection2} and then drops the quadratic nonlinear interaction between the medium-scale variables but adds a closure term. Physics constraints are also included in the CG-ROM.
%\ti{As I said before, I really think we should have all the CG-ROM equations summarized in a table/flowchart at the end of Section 2.  That would make it easier for the reader to follow.}

\begin{figure}[htb]%[H]
    \centering
        \includegraphics[width=\textwidth]{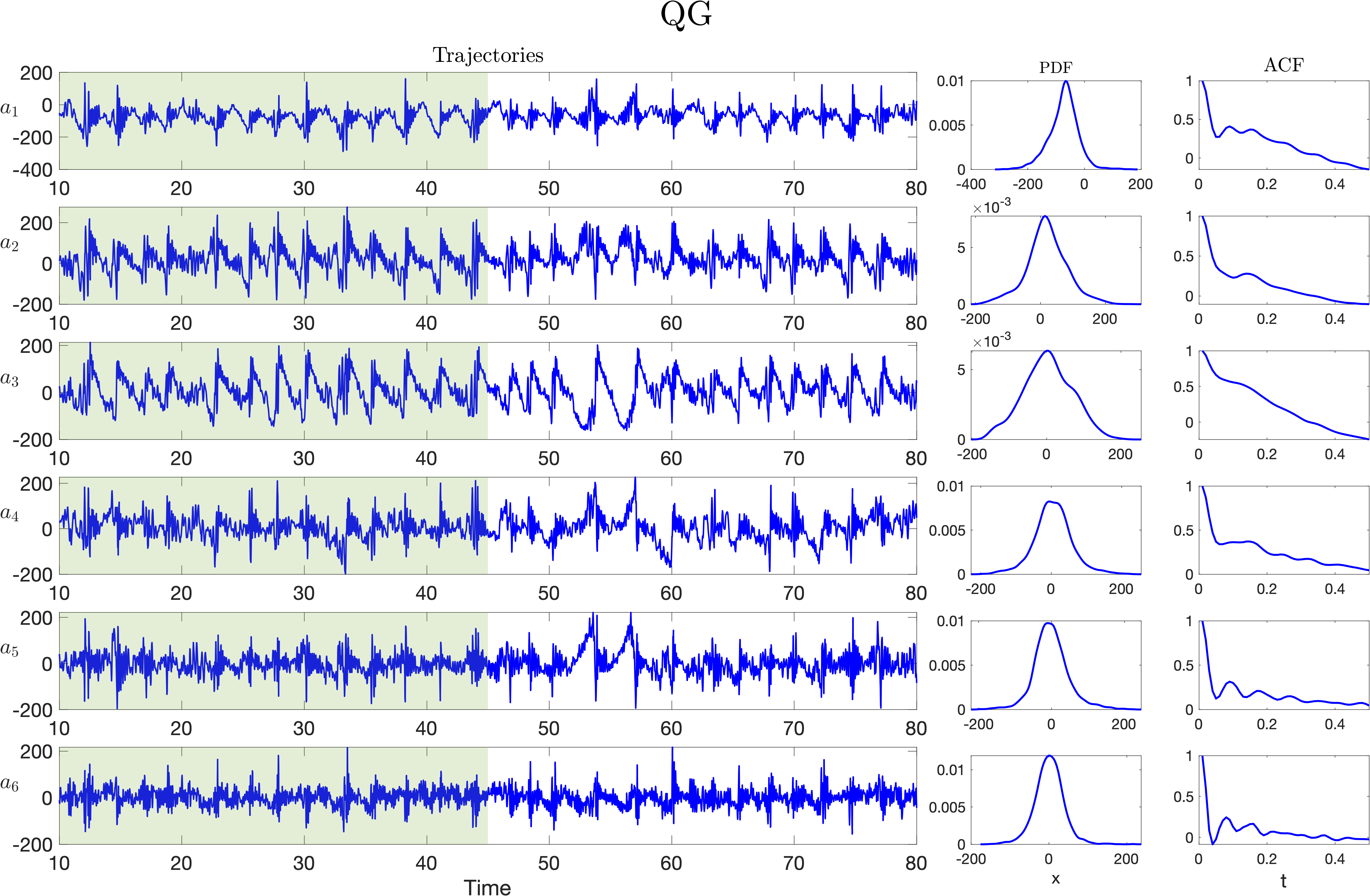}
    \caption{The time series and the associated PDFs and ACFs of the leading $6$ POD modes of the QG equation. The shading period up to $t=45$ is utilized to calibrate the ROMs, while the remaining time series is utilized to test for the data assimilation skill.}
    \label{fig:qg-fom-trajectories}
\end{figure}

Figure \ref{fig:qg-fom-trajectories} shows the time series and the associated statistics of the leading $6$ POD modes of the FOM. Fast oscillations and turbulent behavior can be observed in these time series. In addition, the intermittency introduces extreme events, which lead to non-Gaussian PDFs with fat tails.  The shading period up to $t=45$ is utilized to calibrate the ROMs, while the remaining time series is utilized to test for the data assimilation skill.

In the following, $r=20$ modes are kept in both the G-ROM and the CG-ROM. For notation simplicity, these $20$ modes are denoted by%\ti
{identified with their time-dependent coefficients,} $a_1,\ldots, a_{20}$. Assume that the first $r_1=10$ modes are accurately resolved in the observations.
Therefore, $\mathbf{v}=(a_1,\ldots,a_{10})$ and $\mathbf{w}=(a_{11},\ldots, a_{20})$.

\subsubsection{Results: Comparison of the FOM and the ROMs in model statistics}
%Unlike 
For the viscous stochastic Burgers equation, %where 
the random noise source can be artificially assumed to be known, which %that 
allows a path-wise comparison between the ROMs and the FOM. %, 
In contrast, the turbulent nature of the QG equation makes the trajectories of different models diverge very quickly. Therefore,%\ti
{in this case} it is more justified to compare the statistics of these models.

\begin{figure}[htb]%[H]
    \centering
            \includegraphics[width=.96\textwidth]{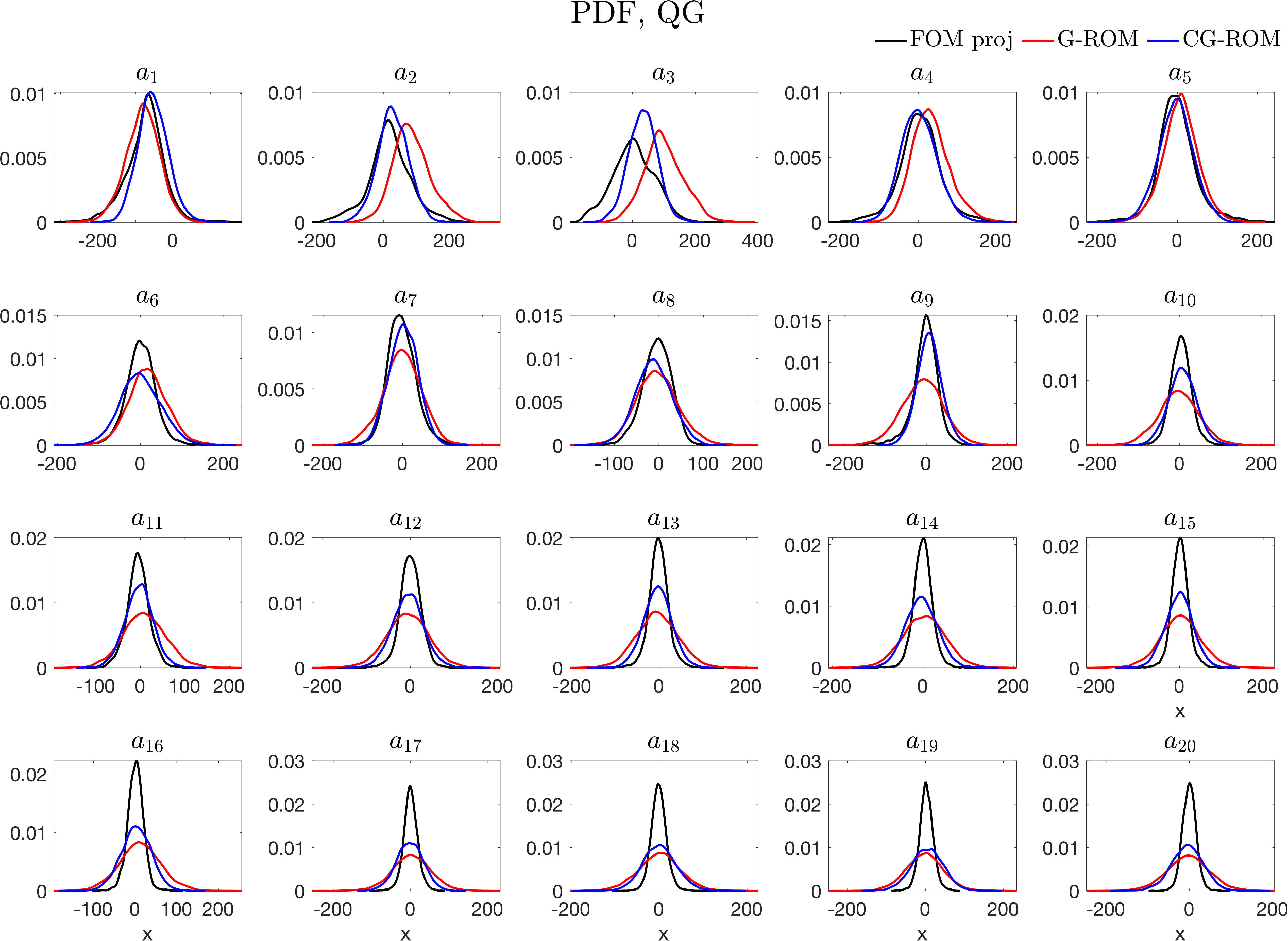}
    \caption{Comparison of the PDFs using the FOM and the two ROMs for different POD modes.}
        \label{fig:qg-rom-trajectories-pdf}
\end{figure}

Figure \ref{fig:qg-rom-trajectories-pdf} shows the comparison of the PDFs using the FOM and the two ROMs for different POD modes. For the first $r_1=10$ modes, the CG-ROM captures the true statistics quite accurately. Particularly, the non-Gaussian behavior in the PDFs is well characterized by the CG-ROM. In contrast, the resulting PDFs from the G-ROM are more biased. For $a_2$ and $a_3$, there is a severe mean shift in the result from G-ROM. For $a_7, a_8, a_9$ and $a_{10}$, the variance of the G-ROM is significantly overestimated. For the remaining $10$ medium-scale modes, neither the G-ROM nor the CG-ROM is very accurate. Nevertheless, the CG-ROM is still more skillul than the G-ROM in recovering the statistics.
\subsubsection{Results: Comparison of the FOM and the ROMs in data assimilation}
\begin{figure}[htb]%[H]
    \centering
            \includegraphics[width=\textwidth]{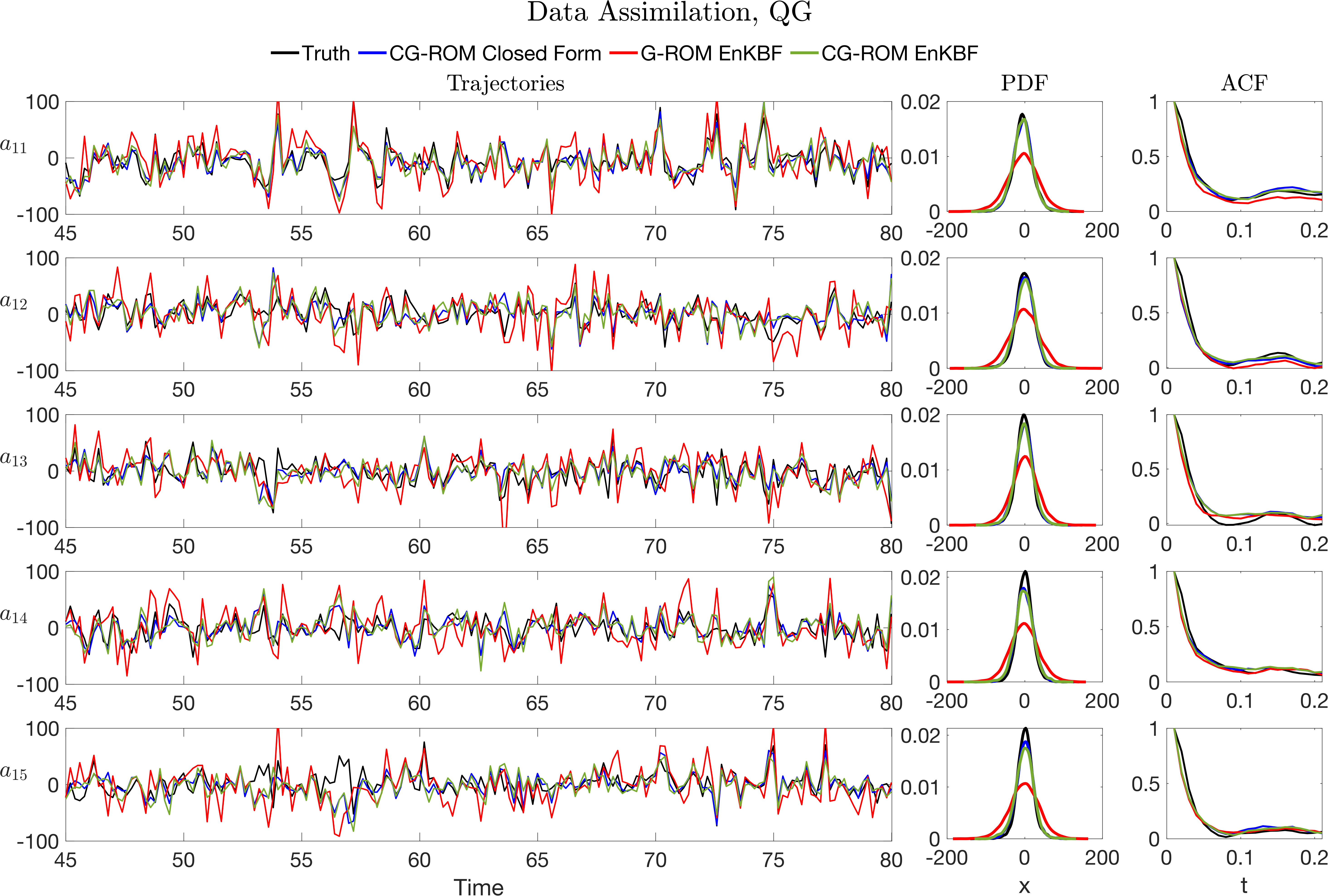}
    \caption{The data assimilation results of the POD modes $a_{11}$ to $a_{15}$ in the QG equation. Comparison of the posterior mean time series and the associated statistics using different filters with the truth. }
        \label{fig:qg-da-trajectories-1}
\end{figure}

\begin{figure}[htb]%[H]
    \centering
            \includegraphics[width=\textwidth]{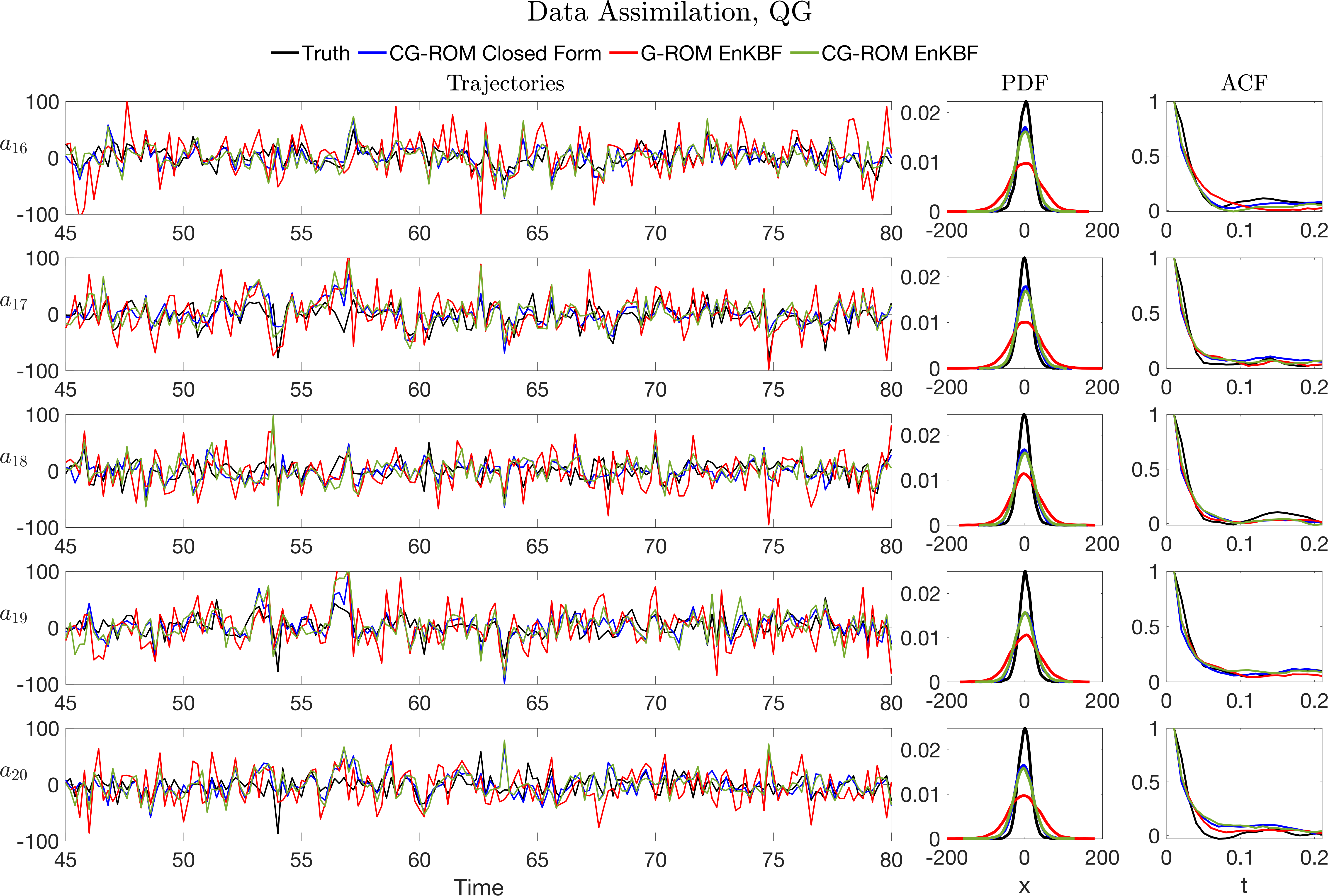}
    \caption{Similar to Figure \ref{fig:qg-da-trajectories-1} but for the POD modes $a_{16}$ to $a_{20}$.}
        \label{fig:qg-da-trajectories-2}
\end{figure}

As in Section \ref{Subsubsec:SBE_DA}, the three filters --- CG-ROM Closed Form, G-ROM EnKBF and CG-ROM EnKBF --- are utilized for assimilating the trajectories of the $10$ unobserved state variables $a_{11},\ldots, a_{20}$.
Figures \ref{fig:qg-da-trajectories-1} and \ref{fig:qg-da-trajectories-2} include the comparison of the assimilated  posterior mean time series and the associated statistics using different filters. Despite that the statistics related to the model free run using the CG-ROM contain certain bias for all these variables, the data assimilation solutions using the CG-ROM remain reasonably accurate. This is not surprising since the observations help correct the model error in the ROM simulations. In contrast to the filters based on the CG-ROM, the G-ROM EnKBF is significantly worse.

\begin{figure}[htb]%[H]
    \centering
            \includegraphics[width=.8\textwidth]{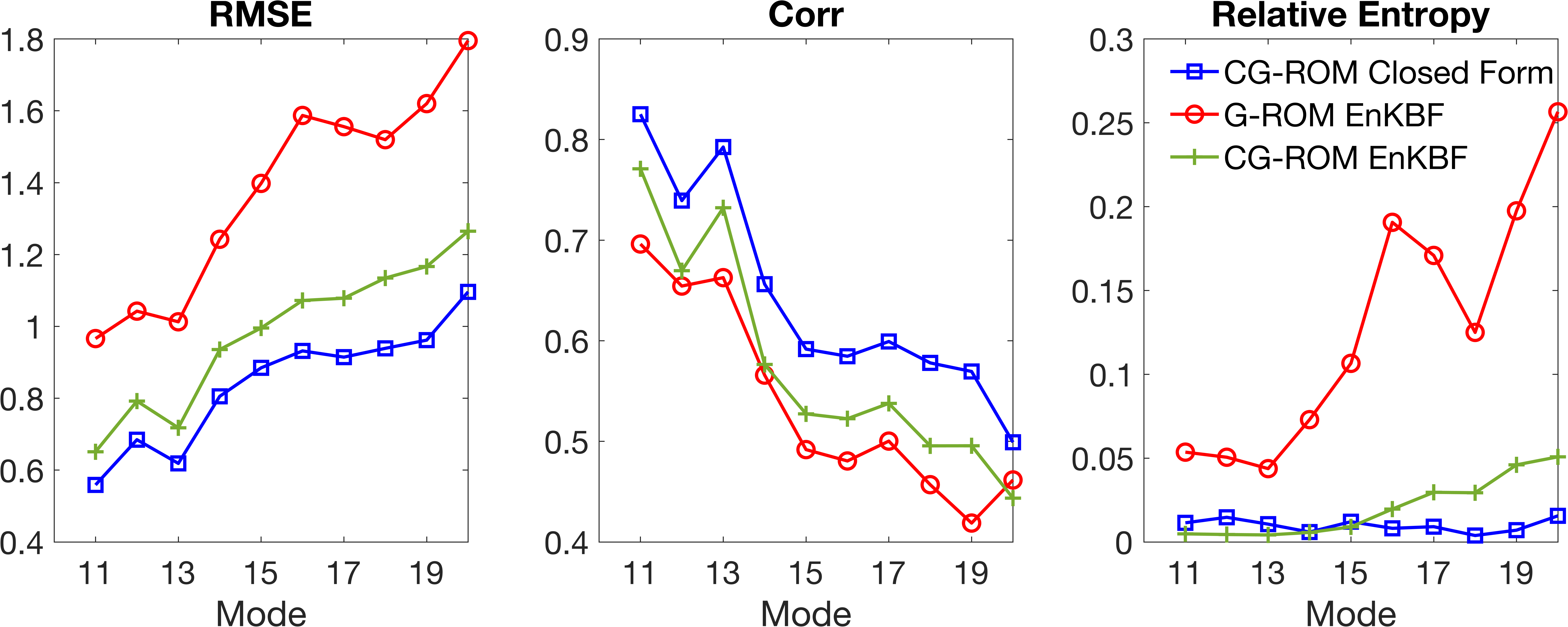}
    \caption{The skill scores of the posterior mean time series. The three panels show the root-mean-square error (RMSE), the pattern correlation (Corr), and the relative entropy. }
        \label{fig:qg-da-rmse-corr-1}
\end{figure}

For the sake of a qualitative comparison between the filtered posterior mean time series and the truth using different filters, three skill scores are utilized here: the root-mean-square error (RMSE), the pattern correlation (Corr), and the relative entropy. They are defined as follows:
\begin{equation}\label{SkillScores}
\begin{split}
  \mbox{Corr} &= \frac{\sum_{i=1}^n(u^M_i-\bar{u}^M)(u_i^{ref}-\bar{u}^{ref})}{\sqrt{\sum_{i=1}^n(u^M_i-\bar{u}^M)^2}\sqrt{\sum_{i=1}^n(u^{ref}_i-\bar{u}^{ref})^2}},\\
  \mbox{RMSE} &=  \sqrt{\frac{\sum_{i=1}^n(u^M_i-u^{ref}_i)^2}{n}},\\
  \mbox{Relative Entropy} &=\mathcal{P}(p^{ref}(u),p^M(u)) = \int p^{ref}(u)\ln\frac{p^{ref}(u)}{p^M(u)}du,
\end{split}
\end{equation}
where $u^M_i$ and $u^{ref}_i$ are the assimilated solution and the truth, respectively, at time $t=t_i$.  The time averages of the assimilated and the true time series are denoted by $\bar{u}^M$ and $\bar{u}^{ref}$, respectively. The RMSE and the Corr are widely used metrics in practice to quantify the path-wise error. A smaller RMSE and a larger Corr correspond to a skillful assimilated time series. On the other hand, the relative entropy is an information criterion \cite{kullback1951information, kullback1987letter, kleeman2011information, majda2005information}, which is adopted to quantify the statistics error between the two distributions formulated by the assimilated time series and the true signal, respectively.
The relative entropy has many attractive features. First, $\mathcal{P}(p,p^M)\geq0$  with equality if and only if $p=p^M$. Second, $\mathcal{P}(p,p^M)$ is invariant under general nonlinear changes of variables. A smaller relative entropy value corresponds to a smaller statistical error.
Figure \ref{fig:qg-da-rmse-corr-1} shows the skill scores using different filters.
First, the larger RMSE and the smaller Corr in the G-ROM EnKBF confirm the intuition from Figures \ref{fig:qg-da-trajectories-1} and \ref{fig:qg-da-trajectories-2} that the path-wise error using the G-ROM EnKBF is much %bigger 
larger than the filters using the CG-ROMs. Second, the relative entropy, which quantifies the error in the PDFs formed by the posterior mean time series and the true signal, indicates that the G-ROM has a bigger statistical error for all the modes as well. It is also worthwhile to mention that the difference between CG-ROM Closed Form and CG-ROM EnKBF implies that closed analytic formulae in the CG-ROM is advantageous in not only %saving 
decreasing the computational time but reducing the sampling error as well.

\begin{figure}[htb]%[H]
    \centering
    \begin{subfigure}[b]{0.48\textwidth}
        \includegraphics[width=\textwidth]{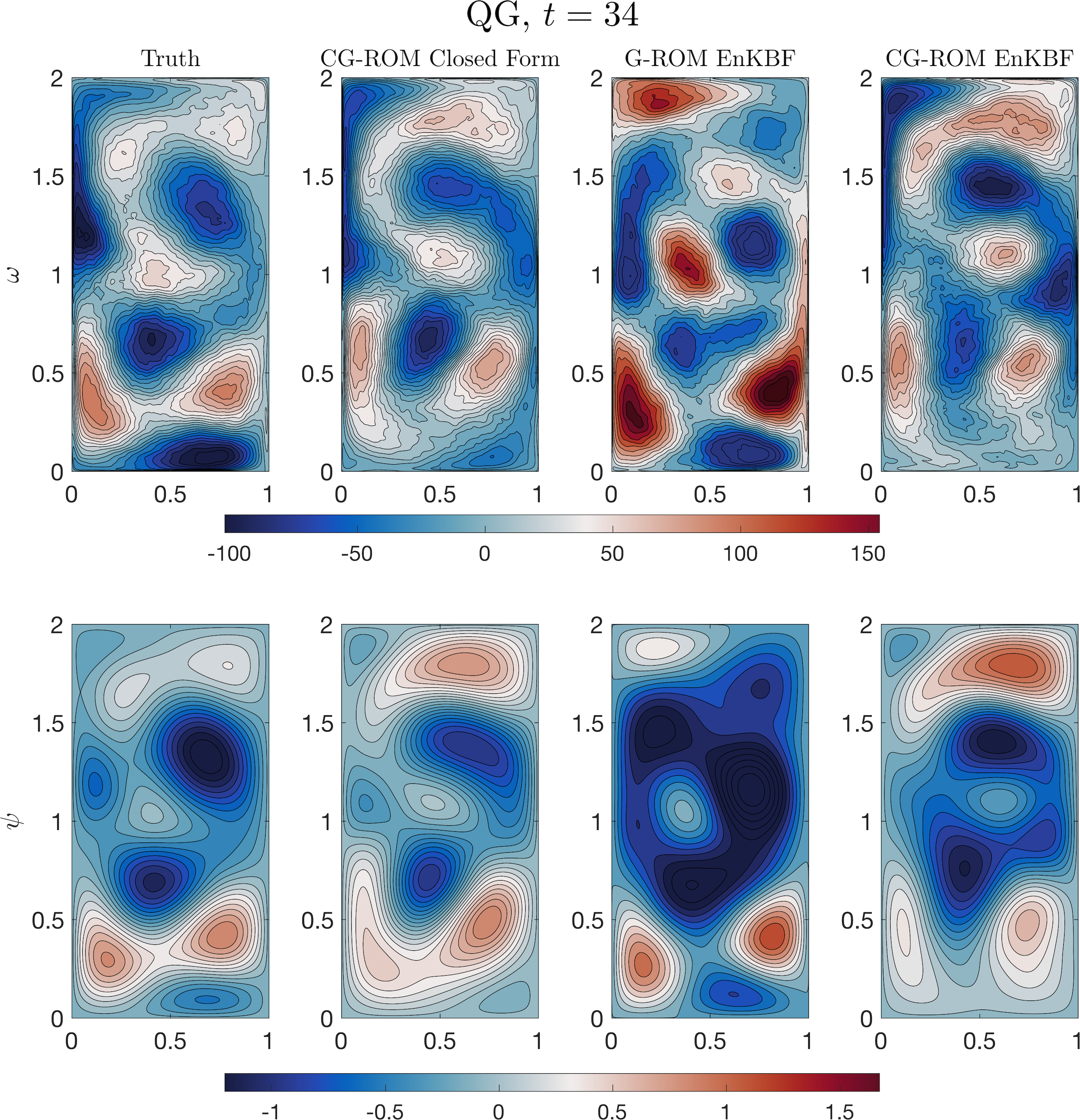}
        \caption{Time instance, $t = 34$}
    \end{subfigure}
    \begin{subfigure}[b]{0.48\textwidth}
        \includegraphics[width=\textwidth]{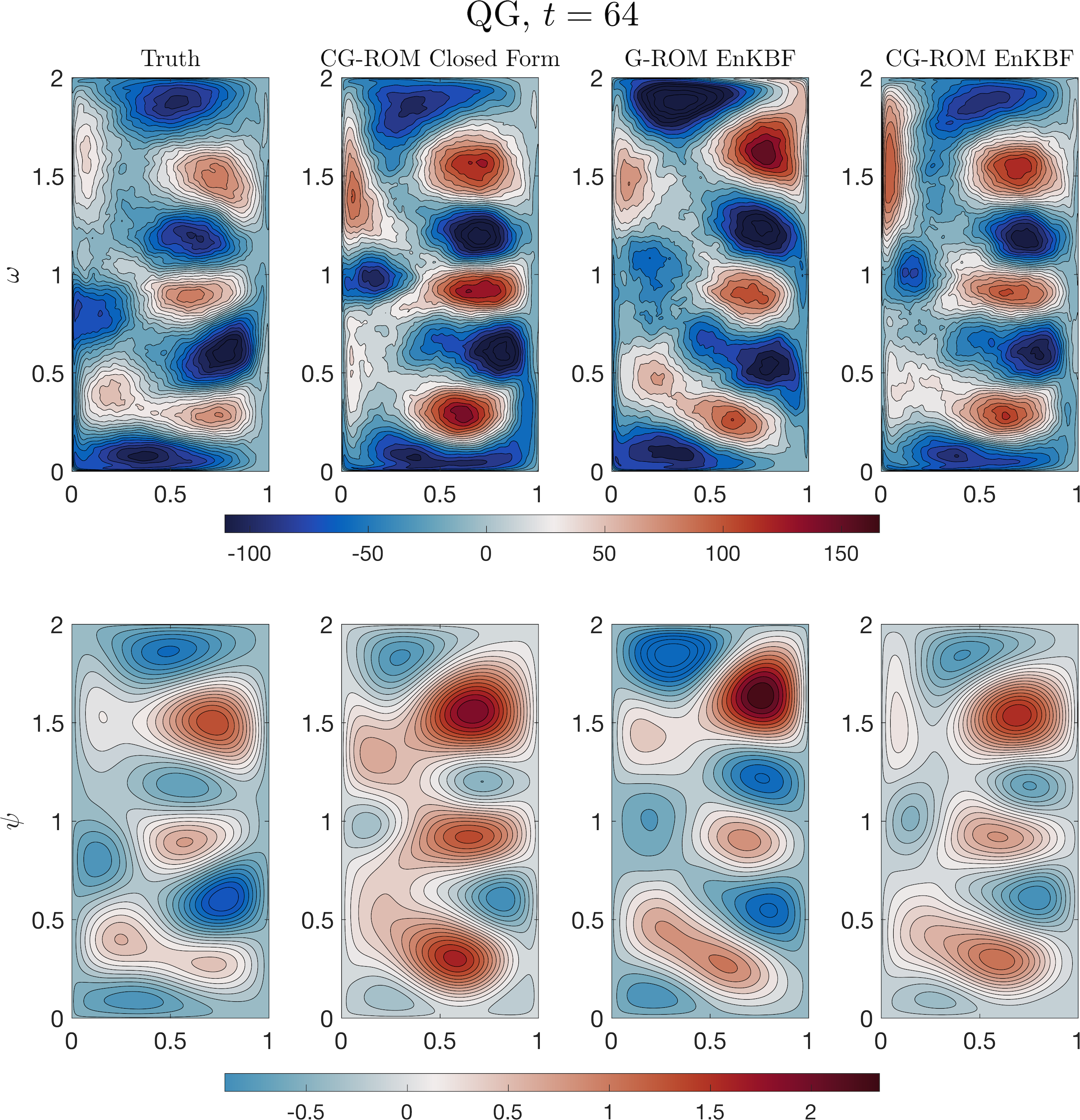}
         \caption{Time instance, $t = 64$}
    \end{subfigure}
        \caption{The reconstructed spatiotemporal patterns of $\omega$ (first row) and streamfunction $\psi$ (second row)
         from the posterior mean based on modes $a_{11},\ldots, a_{20}$, %comparing 
         compared with the truth. }
    \label{fig:qg-da-snapshots}
\end{figure}

Figure \ref{fig:qg-da-snapshots} shows the reconstructed spatiotemporal patterns from the posterior mean based on modes $a_{11},\ldots, a_{20}$. It indicates that the CG-ROM Closed Form can overall recover the streamlines (and thus velocity fields) and the vortices in an  accurate fashion. The CG-ROM EnKBF can capture the overall tendency but the sampling error leads to some biases in the amplitude. %On the other hand, 
In contrast, the G-ROM EnKBF leads to a much %bigger 
larger error in reproducing the spatiotemporal patterns of the truth.

%%%%%%%%%%%%%%%%%%%%%%%%%%%%%%%%%%%%%%%%%%%%%%%

%%%%%%%%%%%%%%%%%%%%%%%%%%%%%%%%%%%%%%%%%%%%%%%
% vorticity and streamfunction fields

\clearpage
\section{Discussion and Conclusions}\label{Sec:Conclusion}
In this paper, a new multiscale stochastic ROM framework, known as the CG-ROM, is developed. The new ROM focuses on recovering the mutliscale dynamical features as well as the statistical %feedbacks 
%\ti
{feedback} between different scales. The CG-ROM exploits cheap but effective conditional linear functions as the closure terms, %that 
which facilitates an efficient and accurate data assimilation scheme %, which 
that is crucial %for 
in recovering the unobserved states in turbulent systems. Physics constraints are incorporated into the CG-ROM to prevent the occurrence of pathological behavior of the model.

There are a few issues that %are remained 
remain as future work. First, both the trajectories of $\mathbf{v}$ and $\mathbf{w}$ are assumed%\ti
{known} in the training period for model calibration. Admittedly, the availability of the time series of the unobserved variable $\mathbf{w}$ can be explained as the result of the reanalysis. %Yet, 
However, it is more natural to assume that only the data of the observed variable $\mathbf{v}$ is accessible in practice. Then determining the model parameters is %not as straightforward as 
%\ti
{more challenging than} the case with full observations in \eqref{eqn:unconstrained-optimization-1}. Estimating the model parameters and recovering the unobserved states have to be carried out simultaneously. Fortunately, due to the analytic formulae of the state estimation, such a procedure can still be %solved 
%\ti{performed}
efficiently using an expectation-maximization algorithm \cite{chen2020learning}. It is then interesting to see how such an uncertainty affects the skill of the ROM. Second, the ROMs are often preferred to be parsimonious, which prevents the overfitting issue and %allows possibly the explainable physics. 
%\ti
{can yield more physical results.}
Such a constraint has not been included in this work, as the focus here is on reproducing the statistics and data assimilation %where 
for which the current setup is sufficient. Nevertheless, existing techniques, such as the LASSO (least absolute shrinkage and selection operator) regression \cite{brunton2016discovering} or information-theoretic based sparse identification methods \cite{sun2014causation, elinger2020information}, can be incorporated into the current framework. Lastly, as in other ROMs, there is often no rigorous way to determine the dimension of the ROMs. The data assimilation framework may have the potential to provide a different criterion to determine the optimal number of the state variables in the ROMs in addition to %use 
%\ti
{using} the explained variance as in most of the existing work.

%%%%%%%%%%%%%%%%%%%%%%%%%%%%%%%%%%%%%%%%%%%%%%%
%%%%%%%%%%%%%%%%%%%%%%%%%%%%%%%%%%%%%%%%%%%%%%%
\section*{Acknowledgments}
The research of N.C. is partially funded by the Office of VCRGE at UW-Madison and ONR N00014-21-1-2904.
The research of T.I. is partially funded by NSF through grant DMS-2012253 and CDS\&E-MSS-1953113.

%%%%%%%%%%%%%%%%%%%%%%%%%%%%%%%%%%%%%%%%%%%%%%%
%%%%%%%%%%%%%%%%%%%%%%%%%%%%%%%%%%%%%%%%%%%%%%%
%%%%%%%%%%%%%%%%%%%%%%%%%%%%%%%%%%%%%%%%%%%%%%%
%%%%%%%%%%%%%%%%%%%%%%%%%%%%%%%%%%%%%%%%%%%%%%%
\appendix

\section{Burgers equation}
\label{sec:eig}
\subsection{Eigenvalue Problems}
Consider the linear part of the viscous stochastic
Burgers equation~\eqref{eqn:sbe}, 
%the initial boundary value problem (IBVP) yields:
%\ti
{which yields the following initial boundary value problem (IBVP):}
\begin{align}
&u_t = \nu u_{xx}+\lambda u &(x,t)\in (0,1)\times\mathbb{R}^+,\label{eqn-sbe-lin-1}\\
&u(0,t) =u(L,t) =0, &t\ge 0,\label{eqn-sbe-lin-2}\\
&u(x,0) =u_0(x),&x\in(0,L).\label{eqn-sbe-lin-3}
\end{align}
% \ti{I don't think we need the next sentence and equation (A.4).  I would just eliminate them.  Or do we get rid of the time derivative in (A.4)?}
% The eigenvalue problem related to equations~\eqref{eqn-sbe-lin-1}-\eqref{eqn-sbe-lin-3} yields the following:
% \begin{align}
% u_t - \nu u_{xx}-\lambda u = 0.
% \end{align}
% With the eigenfunction expansion of $u(x,t)$, i.e., $u(x,t) = \sum_{j=1}^\infty s_j(t)w_j(t)$,
% % \begin{align}
% % u(x,t) = \sum_{j=1}^\infty s_j(t)w_j(t),
% % \end{align}
%\ti
{This IBVP yields the following eigenvalue problem:}
\begin{align}
-&\nu\frac{\partial^2 w_j}{\partial x^2} -\lambda w_j= \sigma_j w_j&x\in\Omega, \label{eqn-eigen-prob-1}\\
&w_j=0&x\in\partial\Omega,\label{eqn-eigen-prob-2}
\end{align}
where $\{w_j\}_{j=1}^\infty$ are eigenfunctions for the operator $\displaystyle\mathcal{L} = -\nu\frac{\partial^2}{\partial x^2}-\lambda$ on $\Omega= (0,l)$ with zero boundary conditions.
% , the eigenvalue problem yields:
% \ti{What happened with the $s_j(t)$ coefficients?}
Note that for \eqref{eqn-eigen-prob-1}-\eqref{eqn-eigen-prob-2}, $\displaystyle(\lambda+\sigma_j)/\nu>0$ must be true for all $j$. Letting $\displaystyle\alpha_j =\frac{\lambda+\sigma_j}{\nu}$, %then 
\eqref{eqn-eigen-prob-1}-\eqref{eqn-eigen-prob-2} become 
\begin{align}
&\frac{\partial^2w_j}{\partial^2 x}+\alpha_j w_j=0&x\in\Omega,\\
&w_j =0&x\in\partial\Omega.
\end{align}
Since $w_j(x)$ %yields 
is a nontrivial solution, $\alpha_j>0$ and the boundary condition imply
\begin{align}
&\alpha_j =\left(\frac{j\pi}{l}\right)^2,&w_j(x) = %\ti
{c_j} \, \sin\left(\frac{j\pi x}{l}\right), && j\in\mathbb{N}^+,
\label{eqn:eigenfunctions}
\end{align}
where $c_j$ are arbitrary constants.
We choose $\{w_j\}_{j=1}^\infty$ to be orthonormal in $L^2(\Omega)$, i.e.,
\begin{align}
\int_\Omega w_i(x) w_j(x) dx = \delta_{ij},\qquad\qquad{i,j=1,2,\ldots}
\end{align}
Consequently, the eigenfunctions in~\eqref{eqn:eigenfunctions} %yield the following:
%\ti
{can be written as follows:}
\begin{align}
&w_j(x) =\sqrt{\frac{2}{l}} \sin\left(\frac{j\pi x}{l}\right),&j=1,2,\ldots.
\end{align}

%%%%%%%%%%%%%%%%%%%%
\subsection{Unstable Eigen-modes}
To apply the Galerkin projection, we define the $L^2$ inner product, i.e., $(f,g)_{L^2} = \int_\Omega f(x) g(x) dx$.
% \begin{align}
% (f,g)_{L^2} = \int_\Omega fgdx
% \end{align}
Then, with a finite-dimensional approximation space spanned by $\{w_k\}_{k=1}^N$, the Galerkin projection of the viscous stochastic Burgers equation~\eqref{eqn:sbe} yields% ,
% \begin{align}
% (u_t, w_k)_{L^2} -\nu(u_{xx},w_k)_{L^2}-\lambda (u,w_k)_{L^2} +\gamma(uu_x,w_k)_{L^2} = 0,\\  k=1,2,\cdots,N.  \notag
% \end{align}
% Using the integration by parts, we have,
\begin{align}
(u_t, w_k)_{L^2} +\nu(\partial_xu,\partial_xw_k)_{L^2}-\lambda (u,w_k)_{L^2} +\gamma(u\partial_xu,w_k)_{L^2} = 0,\\  k=1,2,\cdots,N.  \notag
\end{align}
Writing $\displaystyle u(x,t) = \sum_{j=1}^N d_j(t)w_j(x)$, we have %,
\begin{align}
\sum_{j=1}^N \dot{d}_j(w_j, w_k)_{L^2} +\nu\sum_{j=1}^N d_j(\partial_xw_j,\partial_xw_k)_{L^2}-\lambda \sum_{j=1}^N d_j(w_j,w_k)_{L^2} +\label{appendix-Galerkin_Projection}\\\gamma\sum_{i=1}^N \sum_{j=1}^N d_id_j(w_i\partial_xw_j,w_k)_{L^2} = 0,\qquad k=1,2,\cdots,N.  \notag
\end{align}
With the inner products $(w_i\partial_xw_j,w_k)_{L^2}$,  $(w_j, w_k)_{L^2}$, and $(\partial_xw_j, \partial_xw_k)_{L^2}$ evaluated in equations~\eqref{eqn:rom-linear} and~\eqref{eqn:rom-nonlinear}, respectively, 
% The inner product $(w_i\partial_xw_j,w_k)_{L^2}$ yields the form:
% \begin{align}
% (w_i\partial_xw_j,w_k)_{L^2} &=  \begin{cases} 
%       \frac{j\pi}{4}\left(\frac{2}{l}\right)^{3/2} & \text{if  } i+j-k=0 \text{ or } i-j-k=0 \\
%         -\frac{j\pi}{4} \left(\frac{2}{l}\right)^{3/2} & \text{if  } i-j+k=0  \\
%       0 & \text{otherwise}
%   \end{cases} \label{tensor_basis}
% \end{align}
% Note the fact that $\{w_k\}_{k=1}^N$ are orthonormal basis,
% \begin{align}
% &(w_j, w_k)_{L^2} =\delta_{jk}, &(\partial_xw_j, \partial_xw_k)_{L^2} =\delta_{jk}\frac{k^2\pi^2}{l^2}
% \end{align}
%\subsection{$N=4$ case}
the Galerkin projection~\eqref{appendix-Galerkin_Projection} (i.e., the G-ROM~\eqref{eqn:G-ROM-weak-form} in Section 3.1.2) can be reformulated as a set of ODE: %equations:
\begin{align}
\dot{d_1} &= \left(\lambda-\nu \frac{\pi^2}{l^2}\right)d_1+ h.o.t.
%\gamma\tau\left( z_1z_2+ z_2z_3+ z_3z_4\right) 
\label{eqn-galerkin-eig1}\\
\dot{d_2} &= \left(\lambda-\nu \frac{4\pi^2}{l^2}\right)d_2
+ h.o.t.
%+\gamma\tau\left(- z_1^2+2 z_1z_3+2 z_2z_4\right)
\label{eqn-galerkin-eig2}\\
\dot{d_3} &= \left(\lambda-\nu \frac{9\pi^2}{l^2}\right)d_3
+ h.o.t.
%+\gamma\tau\left(-3z_1z_2+3z_1z_4\right) 
\label{eqn-galerkin-eig3}\\
\vdots  & \nonumber\\
\dot{d_n} &= \left(\lambda-\nu \frac{n^2\pi^2}{l^2}\right)d_n
+ h.o.t., 
%+\gamma\tau\left(-3z_1z_2+3z_1z_4\right) 
\label{eqn-galerkin-eig3}
% \dot{z_4} &= \left(\lambda-\nu \frac{16\pi^2}{l^2}\right)z_4+\gamma\tau\left(-3z_1z_3-2z^2_2\right) \label{Reduced_Eq4}
\end{align}
where $h.o.t.$ denotes the higher order terms and $1\le n\le N$.
As a result, the choice of the parameter pair, $\lambda$ and $\nu$, can determine the linear stability of the viscous stochastic Burgers equation. For example, the Regime I in Table~\ref{tab:regime-sbe} yields one unstable eigen-mode, while the Regime II yields three unstable eigen-modes. %hence 
Thus, the Regime II is more challenging than Regime I.

%%%%%%%%%%%%%%%%%%%%%%%%%%%%%%%
\section{Optimization Problem}
\subsection{Conditional Gaussian ROM with Physical Constraints}
In what follows, the details of the multiscale physics constraints are presented. In particular,
in Section~\ref{app-multi-physics-constraint}, the mathematical formulation of the multiscale physics constraints are derived and the matrix forms for implementation are presented. %and in 
In Section~\ref{app-multi-physics-constraint-eg}, a simple example of multiscale physics constraints is given.
 
% In particular, we assume that the quadratic form at each scale satisfy the followings:
% \begin{align}
%     &v^\ast \left(
%     v^\ast\left(
%     \mathbf{B}^{(v)}_{v\tau}
%     \oplus
%     \mathbf{C}^{(v)}_{w\tau}
%     \right) [v;w]
%     \right)=0,
%     \label{eqn:physical-constraint-1}
%     \\
%     &w^\ast \left(
%     v^\ast\left(
%     \mathbf{B}^{(w)}_{v\tau}
%     \oplus 
%     \mathbf{C}^{(w)}_{w\tau}
%     \right) 
%     [v;w]\right)=0.
%     \label{eqn:physical-constraint-2}
% \end{align}

%%%%%%%%%%%%%%%%%%%%
\subsection{Multiscale physical constraints\label{app-multi-physics-constraint}}
%Due to the difficulties in implementing constraints for the term $\left(\mathbf{B}^{(v)}_{v\tau}\right)$, 
The proposed conditional Gaussian ROMs %are intended to retain 
%\ti
{aim at preserving} the mutliscale dynamical features%\ti
{of the underlying system.}%; as a result, 
Therefore, %the physical constraints which can reserve a multiscale property between large and small scales, i.e., 
multiscale physics constraints, are considered.
In particular, the %imposed 
multiscale physics constraints%\ti
{that are imposed} for the quadratic terms are expected to satisfy the following:
\begin{align}
    &[v;w]^\ast 
    \begin{pmatrix}
       v^\ast\left(
    \mathbf{B}^{(v)}_{v\tau}
    \oplus 
    \mathbf{C}^{(v)}_{w\tau}
    \right)
    [v;w]    
    \\
    v^\ast\left(
    \mathbf{B}^{(w)}_{v\tau}
    \oplus 
    \mathbf{C}^{(w)}_{w\tau}
    \right) 
    [v;w]    
    \end{pmatrix}
=0.
    \label{eqn:physical-constraint-multi-1}
\end{align}
%where the energy is conserved when considering both large and small scales. 
%\ti
{This constraint enforces the conservation of energy with respect to both the large and the small scales.}
The discrete formulation of~\eqref{eqn:physical-constraint-multi-1} yields the following:
\begin{align}
\begin{aligned}
      &    \sum_{(i,j,k)=(1,1,1)}^{(r_1,r_1,r_1)}
    \left(\mathbf{B}^{(v)}_{v \tau}\right)_{(ijk)}v_iv_jv_{k}
    +
    {%\color{blue}
     \sum_{(i,j,k)=(1,1,1)}^{(r_1,r_2,r_1)}
    \left(\mathbf{C}^{(v)}_{w\tau}\right)_{(ijk)}v_iw_jv_{k}  
    }
    \\
& +{%\color{blue}
\sum_{(i,j,k)=(1,1,1)}^{(r_1,r_1,r_2)}
    \left(\mathbf{B}^{(w)}_{v\tau}\right)_{(ijk)}v_iv_jw_{k}
    }
    +
     \sum_{(i,j,k)=(1,1,1)}^{(r_1,r_2,r_2)}
    \left(\mathbf{C}^{(w)}_{w\tau}\right)_{(ijk)}v_iw_jw_{k}   
    =0.
    \label{eqn:physical-constraint--multi-2}
\end{aligned}
\end{align}
%Observing 
%\ti
{Taking into account} the symmetries of the nonlinear terms in equation~\eqref{eqn:physical-constraint--multi-2}, %the two terms in blue %color 
%\red
{the second and third terms on the LHS of~\eqref{eqn:physical-constraint--multi-2}}
should share their energy conservation:
%\ti{I don't know if we're allowed to use colored text.  Maybe we can just refer to those terms as ``the second and third term s on the LHS of (B.2)."}
\begin{align}
& \sum_{(i,j,k)=(1,1,1)}^{(r_1,r_2,r_1)}
    \left(\mathbf{C}^{(v)}_{w\tau}\right)_{(ijk)}v_iw_jv_{k} &+\sum_{(i,j,k)=(1,1,1)}^{(r_1,r_1,r_2)}
    \left(\mathbf{B}^{(w)}_{v\tau}\right)_{(ijk)}v_iv_jw_{k}
    =0.
    \label{eqn:physical-constraint-multi-5}
\end{align}
On the other hand, the other two terms should %have 
%\ti
{enforce} their own energy conservation:
\begin{align}
      &    \sum_{(i,j,k)=(1,1,1)}^{(r_1,r_1,r_1)}
    \left(\mathbf{B}^{(v)}_{v \tau}\right)_{(ijk)}v_iv_jv_{k}
    = 0
    &
       \sum_{(i,j,k)=(1,1,1)}^{(r_1,r_2,r_2)}
    \left(\mathbf{C}^{(w)}_{w\tau}\right)_{(ijk)}v_iw_jw_{k}   
    =0. \label{eqn:physical-constraint-multi-4}
\end{align}
In particular, these terms yield the following discrete relations:
%%%%%%%%%%%%%%%
\begin{align}
&{\mathbf{C}}^{(v)}_{w\tau}\text{ and } \mathbf{B}^{(w)}_{v\tau}:
\nonumber
\\
 &\left({\mathbf{C}}^{(v)}_{w\tau}\right)_{(iki)}
    +    \left(\mathbf{B}^{(w)}_{v\tau}\right)_{(iik)}=0,\quad i=1,\cdots, r_1, k=1,\cdots, r_2;
    \label{eqn:cg-constraint-vw-1}
\\
    &\left({\mathbf{C}}^{(v)}_{w\tau}\right)_{(ikj)}
    +    \left({\mathbf{C}}^{(v)}_{w\tau}\right)_{(jki)}
    +\left(\mathbf{B}^{(w)}_{v\tau}\right)_{(ijk)}+\left(\mathbf{B}^{(w)}_{v\tau}\right)_{(jik)}
    =0,
        \label{eqn:cg-constraint-vw-2}
\\
    &\hspace{6cm}
    i,j=1,\cdots, r_1,i\neq j; k=1,\cdots, r_2, \nonumber    
\end{align}
%%%%%%%%%%%%%%%%%
\begin{align}
&\mathbf{B}^{(v)}_{v\tau}:
\nonumber
\\
&\left(\mathbf{B}^{(v)}_{v\tau}\right)_{(iii)} =0, \qquad\qquad i=1,\cdots, r_1,
\label{eqn:cg-constraint-v-1}
\\
&\left(\mathbf{B}^{(v)}_{v\tau}\right)_{(ijj)}+\left(\mathbf{B}^{(v)}_{v\tau}\right)_{(jij)}
+\left(\mathbf{B}^{(v)}_{v\tau}\right)_{(jji)}=0,\qquad  i,j=1,\cdots, r_1,i\neq j,
\label{eqn:cg-constraint-v-2}
\\
&\left(\mathbf{B}^{(v)}_{v\tau}\right)_{(ijk)}+\left(\mathbf{B}^{(v)}_{v\tau}\right)_{(ikj)}
+\left(\mathbf{B}^{(v)}_{v\tau}\right)_{(jik)}+\left(\mathbf{B}^{(v)}_{v\tau}\right)_{(jki)}+\left(\mathbf{B}^{(v)}_{v\tau}\right)_{(kij)}
\label{eqn:cg-constraint-v-3}
\\
&\hspace{4cm}+\left(\mathbf{B}^{(v)}_{w\tau}\right)_{(kji)}=0,
\qquad i,j,k=1,\cdots, r_1,i\neq j \neq k.\nonumber
\end{align}
%%%%%%%%%%%%%%%%%
\begin{align} 
&\mathbf{C}^{(w)}_{w\tau}:
\nonumber
\\
&\left(\mathbf{C}^{(w)}_{w\tau}\right)_{(ijj)} =0, & i=1,\cdots, r_1; j=1,\cdots, r_2,
\label{eqn:cg-constraint-w-3}
\\
&\left(\mathbf{C}^{(w)}_{w\tau}\right)_{(ijk)}+\left(\mathbf{C}^{(w)}_{w\tau}\right)_{(ikj)} =0, & i=1,\cdots, r_1; j,k=1,\cdots, r_2, j\neq k.
\label{eqn:cg-constraint-w-4}
\end{align}
To further construct the constraint coefficient matrix $H$ in equation (13), let $\theta_L$ contain every element in $ \mathbf{B}^{(v)}_{v\tau}, \mathbf{C}^{(v)}_{w\tau},\mathbf{D}^{(v)}_{w\tau}, \mathbf{C}^{(v)}_{v\tau}$, and let $\theta_S$ contain every element in $ \mathbf{B}^{(w)}_{v\tau}, \mathbf{C}^{(w)}_{w\tau},\mathbf{D}^{(w)}_{w\tau}, \mathbf{C}^{(w)}_{v\tau}$. 
%\ti
{Then,} the physical constraints~\eqref{eqn:cg-constraint-vw-1}
 to~\eqref{eqn:cg-constraint-w-4} can be %reverted 
 %\ti
 {converted} into a matrix form:
 \begin{align}
     H \begin{pmatrix}
        \theta_L\\
        \theta_S
     \end{pmatrix}=0,
 \end{align}
 where $\theta = [\theta_L; \theta_S]$.
%  Similarly, if let $\theta_S$ contains every elements in $ \mathbf{B}^{(w)}_{v\tau}, \mathbf{C}^{(w)}_{w\tau},\mathbf{D}^{(w)}_{w\tau}, \mathbf{C}^{(w)}_{v\tau}$, 
% the physical constraints~\eqref{eqn:cg-constraint-vw-1}
%  to~\eqref{eqn:cg-constraint-w-4} can be reverted into a matrix form:
%  \begin{align}
%      H_S\theta_S=0.
%  \end{align}
In particular, we can formulate $\theta_L$%\ti
{and $\theta_S$ as follows}:
\begin{align}
\begin{aligned}
          &\theta_L =
    \biggl[
     \left(\mathbf{B}^{(v)}_{v\tau}\right)_{111},      \left(\mathbf{B}^{(v)}_{v\tau}\right)_{121},
    \cdots
    \left(\mathbf{B}^{(v)}_{v\tau}\right)_{r_1r_1r_1},
    \left(\mathbf{C}^{(v)}_{w\tau} \right)_{111},
\left(\mathbf{C}^{(v)}_{w\tau} \right)_{121},
    \cdots
    \left(\mathbf{C}^{(v)}_{w\tau} \right)_{r_1r_2r_1},\nonumber
        \\
    &
    \left( \mathbf{D}^{(v)}_{w\tau}\right)_{11},
    \cdots
     \left( \mathbf{D}^{(v)}_{w\tau}\right)_{r_1r_1},
     \left( \mathbf{C}^{(v)}_{v\tau}\right)_{11}, 
     \cdots
       \left( \mathbf{C}^{(v)}_{v\tau}\right)_{r_1r_2}
        \biggr]^\top
       \nonumber
       \\
 &\theta_S =
    \biggl[
            \left(\mathbf{B}^{(w)}_{v\tau}\right)_{111},      \left(\mathbf{B}^{(w)}_{v\tau}\right)_{121},
    \cdots
    \left(\mathbf{B}^{(w)}_{v\tau}\right)_{r_1r_1r_1},
    \left(\mathbf{C}^{(w)}_{w\tau} \right)_{111},
\left(\mathbf{C}^{(w)}_{w\tau} \right)_{121},
    \cdots
    \left(\mathbf{C}^{(w)}_{w\tau} \right)_{r_1r_2r_1},\nonumber
        \\
    &
    \left( \mathbf{D}^{(w)}_{w\tau}\right)_{11},
    \cdots
     \left( \mathbf{D}^{(w)}_{w\tau}\right)_{r_1r_1},
     \left( \mathbf{C}^{(w)}_{v\tau}\right)_{11}, 
     \cdots
       \left( \mathbf{C}^{(w)}_{v\tau}\right)_{r_1r_2}
     \biggr]^\top
\end{aligned}
\end{align}

\subsubsection{A Simple Example\label{app-multi-physics-constraint-eg}}
A simple example, e.g,% \ti
{with} $r_1=3, r=4$, can be used to illustrate the constraints from~\eqref{eqn:cg-constraint-v-1} to~\eqref{eqn:cg-constraint-w-4}.
\begin{align}
&\mathbf{B}^{(v)}_{v\tau}\in\mathbb{R}^{3\times3\times3}:
\nonumber
\\
&\left(\mathbf{B}^{(v)}_{v\tau}\right)_{(111)} =\left(\mathbf{B}^{(v)}_{v\tau}\right)_{(222)} =\left(\mathbf{B}^{(v)}_{v\tau}\right)_{(333)} = 0,
\label{eqn:cg-constraint-v-1-example}
\\
&\left(\mathbf{B}^{(v)}_{v\tau}\right)_{(122)}+\left(\mathbf{B}^{(v)}_{v\tau}\right)_{(212)}
+\left(\mathbf{B}^{(v)}_{v\tau}\right)_{(221)}=0,
\label{eqn:cg-constraint-v-2-example}
\\
&\left(\mathbf{B}^{(v)}_{v\tau}\right)_{(133)}+\left(\mathbf{B}^{(v)}_{v\tau}\right)_{(313)}
+\left(\mathbf{B}^{(v)}_{v\tau}\right)_{(331)}=0,
\\
&\left(\mathbf{B}^{(v)}_{v\tau}\right)_{(233)}+\left(\mathbf{B}^{(v)}_{v\tau}\right)_{(323)}
+\left(\mathbf{B}^{(v)}_{v\tau}\right)_{(332)}=0,
\\
&\left(\mathbf{B}^{(v)}_{v\tau}\right)_{(211)}+\left(\mathbf{B}^{(v)}_{v\tau}\right)_{(121)}
+\left(\mathbf{B}^{(v)}_{v\tau}\right)_{(112)}=0,
\\
&\left(\mathbf{B}^{(v)}_{v\tau}\right)_{(311)}+\left(\mathbf{B}^{(v)}_{v\tau}\right)_{(131)}
+\left(\mathbf{B}^{(v)}_{v\tau}\right)_{(113)}=0,
\\
&\left(\mathbf{B}^{(v)}_{v\tau}\right)_{(322)}+\left(\mathbf{B}^{(v)}_{v\tau}\right)_{(232)}
+\left(\mathbf{B}^{(v)}_{v\tau}\right)_{(223)}=0,
\\
&\left(\mathbf{B}^{(v)}_{v\tau}\right)_{(123)}+\left(\mathbf{B}^{(v)}_{v\tau}\right)_{(132)}
+\left(\mathbf{B}^{(v)}_{v\tau}\right)_{(213)}+\left(\mathbf{B}^{(v)}_{v\tau}\right)_{(231)}+\left(\mathbf{B}^{(v)}_{v\tau}\right)_{(312)}
\label{eqn:cg-constraint-v-3-example}
\\
&\hspace{4cm}+\left(\mathbf{B}^{(v)}_{w\tau}\right)_{(321)}=0.
\end{align}
%%%%%%%%%%%%%%%
\begin{align}
&\mathbf{B}^{(w)}_{v\tau}\in\mathbb{R}^{3\times3\times1},\, \mathbf{C}^{(v)}_{w\tau}\in\mathbb{R}^{3\times1\times3}:
\nonumber
\\
&\left(\mathbf{B}^{(w)}_{v\tau}\right)_{(111)} + \left(\mathbf{C}^{(v)}_{w\tau}\right)_{(111)} = 0,\\
&\left(\mathbf{B}^{(w)}_{v\tau}\right)_{(221)} + \left(\mathbf{C}^{(v)}_{w\tau}\right)_{(212)} = 0,\\
&\left(\mathbf{B}^{(w)}_{v\tau}\right)_{(331)} + \left(\mathbf{C}^{(v)}_{w\tau}\right)_{(313)} = 0,
\hspace{6cm}%&i=1,\cdots, r_1;k=1,\cdots,r_2
\label{eqn:cg-constraint-w-1-example}
\\
&\left(\mathbf{B}^{(w)}_{v\tau}\right)_{(121)}+\left(\mathbf{B}^{(w)}_{v\tau}\right)_{(211)}
+
\left(\mathbf{C}^{(v)}_{w\tau}\right)_{(112)}+\left(\mathbf{C}^{(v)}_{w\tau}\right)_{(211)}
=0
\label{eqn:cg-constraint-w-2-example}
\\
&\left(\mathbf{B}^{(w)}_{v\tau}\right)_{(131)}+\left(\mathbf{B}^{(w)}_{v\tau}\right)_{(311)}+
\left(\mathbf{C}^{(v)}_{w\tau}\right)_{(113)}+\left(\mathbf{C}^{(v)}_{w\tau}\right)_{(311)}= 0
\\
&\left(\mathbf{B}^{(w)}_{v\tau}\right)_{(231)}+\left(\mathbf{B}^{(w)}_{v\tau}\right)_{(321)}
+\left(\mathbf{C}^{(v)}_{w\tau}\right)_{(213)}+\left(\mathbf{C}^{(v)}_{w\tau}\right)_{(312)}
=0
\end{align}
%%%%%%%%%%%%%%%
\begin{align} 
&\mathbf{C}^{(w)}_{w\tau}\in\mathbb{R}^{3\times1\times1}:
\nonumber
\\
&\left(\mathbf{C}^{(w)}_{w\tau}\right)_{(111)} =\left(\mathbf{C}^{(w)}_{w\tau}\right)_{(211)}=\left(\mathbf{C}^{(w)}_{w\tau}\right)_{(311)}  = 0,\hspace{3.5cm} %& i=1,\cdots, r_1; j=1,\cdots, r_2,
\label{eqn:cg-constraint-w-3-example}
\end{align}
%%%%%%%%%%%%%%%%%%%%%%%%%%%%%%%%

%%%%%%%%%%%%%%%%%%%%%%%%%%%%%%%%%%%%%%%%%

%%%%%%%%%%%%%%%%%%%%%%%%%%%%%%%%%%%%%%%%%

\subsection{Optimization Problem Formulation}
In this section, the details of constructing the optimization problem in %\red
{Section~\ref{section:ss-data-driven model calibration}} and %\red
{Section~\ref{sectoin:ss-physics-constraints}} are presented. 
%\ti{Please use the Section~\ref{} format.}
In Section~\ref{app-optimization-general}, the matrix formulations of $\mathbf{M}^j_L$ in the constrained optimization problem (14)-(15) are sketched.
In Section~\ref{app-optimization-general-rank-problem}, the ill conditioning in the general formulation of the optimization problem is identified and the mathematical explanation is provided.
In Section~\ref{app-optimization-special}, a special regularization for the constrained optimization problem is proposed, which can overcome the ill conditioning described in Section~\ref{app-optimization-general-rank-problem}.

\subsubsection{General Formulation\label{app-optimization-general}}
Following the notations in 
%\red
{Section~\ref{section:ss-data-driven model calibration}} and %\red
{Section~\ref{sectoin:ss-physics-constraints}},
%Section 2.3 and Section 2.4,
the $\mathbf{M}^j_L$, the coefficient matrix for $\theta_L$, can be expressed as: 
%\ti{Please use the Section~\ref{} format.}
\begin{align}
\scriptsize
    &\mathbf{M}^j_L=
    \Delta T\cdot
    \\
    &
    \begin{bmatrix}
\left(v^{(j)}\right)^\top\bigotimes \left(u^{(j)}\right)^\top & \mathbf{0}&\cdots&\mathbf{0}& \left(u^{(j)}\right)^\top & \mathbf{0}&\cdots&\mathbf{0} \\
 \mathbf{0} &\left(v^{(j)}\right)^\top\bigotimes \left(u^{(j)}\right)^\top &\cdots&\mathbf{0}&
 \mathbf{0}&\left(u^{(j)}\right)^\top &\cdots&\mathbf{0} \\
 \vdots&\ddots&\ldots& \vdots& \vdots&\vdots&\ddots&\vdots\\
  \mathbf{0}& \mathbf{0}&\dots &\left(v^{(j)}\right)^\top\bigotimes \left(u^{(j)}\right)^\top &\mathbf{0}&\mathbf{0}&\ldots&\left(u^{(j)}\right)^\top
    \end{bmatrix}
\end{align}
where $v^{(j)}$ is the snapshot of $\boldsymbol v$ at $t_j$, $u^{(j)}$ is the snapshot of $\boldsymbol u = [\boldsymbol v;\boldsymbol w]$ at $t_j$, and $\bigotimes$ is the Kronecker tensor product. 
Using the notations from tensor products, $\mathbf{M}^j_L$ can also be expressed as $\mathbf{M}^j_L = \Delta T\cdot\left(\mathbf{I}_{r_1\times r_1}\bigotimes \left(\left(v^{(j)}\right)^\top\bigotimes \left(u^{(j)}\right)^\top\right)\right) \bigoplus \left(\mathbf{I}_{r_1\times r_1}\bigotimes\left(u^{(j)}\right)^\top \right)$, where $\mathbf{I}_{r_1\times r_1}$ is the $r_1\times r_1$ identity matrix.

\subsubsection{Ill Conditioning Problem \label{app-optimization-general-rank-problem}}
When solving the constrained optimization problem, $\sum_j(\mathbf{M}^{j}_L)^\ast \mathbf{\Sigma}^{-1} \mathbf{M}^{j}_L$ is expected to be full rank. However, due to the general formulation presented in  Section~\ref{app-optimization-general}, $\sum_j(\mathbf{M}^{j}_L)^\ast \mathbf{\Sigma}^{-1} \mathbf{M}^{j}_L$ will inevitably be rank deficient, which leads to the ill conditioning.
% Since $\mathbf{M}^{j}_L\in\mathbb{R}^{r_1\times(r_1^2r+r_1r)},\mathbf{\Sigma}\in\mathbb{R}^{r_1\times r_1}$,
% the rank of $(\mathbf{M}^{j}_L)^\ast \mathbf{\Sigma}^{-1} \mathbf{M}^{j}_L$ will be $r_1$ while its size is $(r_1^2r+r_1r)\times (r_1^2r+r_1r)$. 
% Hence, $(\mathbf{M}^{j}_L)^\ast \mathbf{\Sigma}^{-1} \mathbf{M}^{j}_L$ is rank deficient. So in the iteration, $\mathcal{K}^{(k)}$ in equation (20) can be rank deficient.
To track the source of the ill conditioning, it is essential to check the matrix form of $(\mathbf{M}^j)^\ast (\mathbf{\Sigma}^{(k)})^{-1}\mathbf{M}^j$. 
%is essential to check.  
% To see it we can directly write the $(\mathbf{M}^j)^\ast (\mathbf{\Sigma}^{(k)})^{-1}\mathbf{M}^j$ as follows:
In particular, $(\mathbf{M}^j)^\ast (\mathbf{\Sigma}^{(k)})^{-1}\mathbf{M}^j$ yields the following:
\begin{align}
\begin{aligned}
&&&(\mathbf{M}^j)^\ast (\mathbf{\Sigma}^{(k)})^{-1}\mathbf{M}^j
    \\
&&&
=
\begin{bmatrix}
\alpha_j^\ast &0 &\cdots &0\\
0 &\alpha_j^\ast &\cdots &0\\
\vdots &\vdots &\ddots&\vdots\\
0 &0 &\cdots &\alpha_j^\ast\\
\beta_j^\ast &0 &\cdots &0\\
0 &\beta_j^\ast &\cdots &0\\
\vdots &\vdots &\ddots&\vdots\\
0 &0 &\cdots &\beta_j^\ast\\
\end{bmatrix}
\begin{bmatrix}
\mathbf{\Sigma}_{11}&\cdots&0\\
\vdots&\ddots&\vdots\\
0&\cdots&\mathbf{\Sigma}_{r_1,r_1}
\end{bmatrix}
\begin{bmatrix}
\alpha_j &0 &\cdots &0&\beta_j &0 &\cdots &0\\
0 &\alpha_j&\cdots &0 &0 &\beta_j &\cdots &0\\
\vdots &\vdots &\ddots&\vdots&\vdots &\vdots &\ddots&\vdots\\
0 &0 &\cdots &\alpha_j&0 &0 &\cdots &\beta_j\\
\end{bmatrix}
\\
&&&=
\begin{bmatrix}
\mathbf{\Sigma}_{11}\alpha_j^\ast &0 &\cdots &0\\
0 &\mathbf{\Sigma}_{22}\alpha_j^\ast &\cdots &0\\
\vdots &\vdots &\ddots&\vdots\\
0 &0 &\cdots &\mathbf{\Sigma}_{r_1,r_1}\alpha_j^\ast\\
\mathbf{\Sigma}_{11}\beta_j^\ast &0 &\cdots &0\\
0 &\mathbf{\Sigma}_{22}\beta_j^\ast &\cdots &0\\
\vdots &\vdots &\ddots&\vdots\\
0 &0 &\cdots &\mathbf{\Sigma}_{r_1,r_1}\beta_j^\ast\\
\end{bmatrix}
\begin{bmatrix}
\alpha_j &0 &\cdots &0&\beta_j &0 &\cdots &0\\
0 &\alpha_j&\cdots &0 &0 &\beta_j &\cdots &0\\
\vdots &\vdots &\ddots&\vdots&\vdots &\vdots &\ddots&\vdots\\
0 &0 &\cdots &\alpha_j&0 &0 &\cdots &\beta_j\\
\end{bmatrix}
\\
&&&=
\begin{bmatrix}
\mathbf{\Sigma}_{11}\alpha_j^\ast\alpha_j &0 &\cdots &0&\mathbf{\Sigma}_{11}\alpha_j^\ast\beta_j &0 &\cdots &0\\
0 &\mathbf{\Sigma}_{22}\alpha_j^\ast\alpha_j &\cdots &0&0 &\mathbf{\Sigma}_{22}\alpha_j^\ast\beta_j &\cdots &0\\
\vdots &\vdots &\ddots&\vdots&\vdots &\vdots &\ddots&\vdots\\
0 &0 &\cdots &\mathbf{\Sigma}_{r_1,r_1}\alpha_j^\ast\alpha_j&0 &0 &\cdots &\mathbf{\Sigma}_{r_1,r_1}\alpha_j^\ast\beta_j\\
\mathbf{\Sigma}_{11}\beta_j^\ast\alpha_j &0 &\cdots &0&\mathbf{\Sigma}_{11}\beta_j^\ast\beta_j &0 &\cdots &0\\
0 &\mathbf{\Sigma}_{22}\beta_j^\ast\alpha_j &\cdots &0&0 &\mathbf{\Sigma}_{22}\beta_j^\ast\beta_j &\cdots &0\\
\vdots &\vdots &\ddots&\vdots&\vdots &\vdots &\ddots&\vdots\\
0 &0 &\cdots &\mathbf{\Sigma}_{r_1,r_1}\beta_j^\ast \alpha_j&0 &0 &\cdots &\mathbf{\Sigma}_{r_1,r_1}\beta_j^\ast \beta_j\\
\end{bmatrix}
\,,
\end{aligned}
\end{align}
where
\begin{align}
    \alpha_j &= \left(v^{(j)}\right)^\top\bigotimes \left(u^{(j)}\right)^\top
    &&\in\mathbb{R}^{1\times r_1r}
    \\
    \beta_j &=    \left(u^{(j)}\right)^\top
    &&\in\mathbb{R}^{1\times r} .
\end{align}
%and, moreover, 
%\ti
{Moreover,} the products between $\alpha_j$ and $\beta_j$ can be written as 
\begin{align}
\begin{aligned}
    \alpha_j^\ast \alpha_j &= \left[ v^{(j)}\bigotimes u^{(j)}\right]
    \bigotimes
    \left[
    \left(v^{(j)}\right)^\top\bigotimes \left(u^{(j)}\right)^\top
    \right]
    \\
    &=
    \begin{bmatrix}
    v_1^{(j)}u_1^{(j)} v_1^{(j)}u_1^{(j)} &v_1^{(j)}u_1^{(j)} v_1^{(j)}u_2^{(j)} &v_1^{(j)}u_1^{(j)} v_1^{(j)}u_3^{(j)}&\cdots &v_1^{(j)}u_1^{(j)} v_{r_1}^{(j)}u_r^{(j)}\\
    v_1^{(j)}u_2^{(j)} v_1^{(j)}u_1^{(j)} &v_1^{(j)}u_2^{(j)} v_1^{(j)}u_2^{(j)} &v_1^{(j)}u_2^{(j)} v_1^{(j)}u_3^{(j)}&\cdots &v_1^{(j)}u_2^{(j)} v_{r_1}^{(j)}u_r^{(j)}\\
    v_1^{(j)}u_3^{(j)} v_1^{(j)}u_1^{(j)} &v_1^{(j)}u_3^{(j)} v_1^{(j)}u_2^{(j)} &v_1^{(j)}u_3^{(j)} v_1^{(j)}u_3^{(j)}&\cdots &v_1^{(j)}u_3^{(j)} v_{r_1}^{(j)}u_r^{(j)}\\
    \vdots & \vdots & \vdots & \ddots &\vdots \\
    v_2^{(j)}u_1^{(j)} v_1^{(j)}u_1^{(j)} &v_2^{(j)}u_1^{(j)} v_1^{(j)}u_2^{(j)} &v_2^{(j)}u_1^{(j)} v_1^{(j)}u_3^{(j)}&\cdots &v_2^{(j)}u_1^{(j)} v_{r_1}^{(j)}u_r^{(j)}\\
    v_2^{(j)}u_2^{(j)} v_1^{(j)}u_1^{(j)} &v_2^{(j)}u_2^{(j)} v_1^{(j)}u_2^{(j)} &v_2^{(j)}u_2^{(j)} v_1^{(j)}u_3^{(j)}&\cdots &v_2^{(j)}u_2^{(j)} v_{r_1}^{(j)}u_r^{(j)}\\   
    \vdots & \vdots & \vdots & \ddots &\vdots \\
    \vdots & \vdots & \vdots & \ddots &\vdots \\
    v_{r_1}^{(j)}u_r^{(j)} v_1^{(j)}u_1^{(j)} &v_{r_1}^{(j)}u_r^{(j)} v_1^{(j)}u_2^{(j)} &v_{r_1}^{(j)}u_r^{(j)} v_1^{(j)}u_3^{(j)}&\cdots &v_{r_1}^{(j)}u_r^{(j)} v_{r_1}^{(j)}u_r^{(j)}\\
    \end{bmatrix}
    &&\in\mathbb{R}^{r_1r\times r_1r},
\end{aligned}
\end{align}
\begin{align}
\begin{aligned}
    \beta_j^\ast \alpha_j &= \left[ v^{(j)}\bigotimes u^{(j)}\right]
    \bigotimes
    \left[
     \left(u^{(j)}\right)^\top
    \right]
    \\
    &=
    \begin{bmatrix}
    v_1^{(j)}u_1^{(j)} u_1^{(j)} &v_1^{(j)}u_1^{(j)} u_2^{(j)} &v_1^{(j)}u_1^{(j)} u_3^{(j)}&\cdots &v_1^{(j)}u_1^{(j)} u_r^{(j)}\\
    v_1^{(j)}u_2^{(j)} u_1^{(j)} &v_1^{(j)}u_2^{(j)} u_2^{(j)} &v_1^{(j)}u_2^{(j)} u_3^{(j)}&\cdots &v_1^{(j)}u_2^{(j)} u_r^{(j)}\\
    v_1^{(j)}u_3^{(j)} u_1^{(j)} &v_1^{(j)}u_3^{(j)} u_2^{(j)} &v_1^{(j)}u_3^{(j)} u_3^{(j)}&\cdots &v_1^{(j)}u_3^{(j)} u_r^{(j)}\\
    \vdots & \vdots & \vdots & \ddots &\vdots \\
    v_2^{(j)}u_1^{(j)} u_1^{(j)} &v_2^{(j)}u_1^{(j)} u_2^{(j)} &v_2^{(j)}u_1^{(j)} u_3^{(j)}&\cdots &v_2^{(j)}u_1^{(j)} u_r^{(j)}\\
    v_2^{(j)}u_2^{(j)} u_1^{(j)} &v_2^{(j)}u_2^{(j)} u_2^{(j)} &v_2^{(j)}u_2^{(j)} u_3^{(j)}&\cdots &v_2^{(j)}u_2^{(j)} u_r^{(j)}\\   
    \vdots & \vdots & \vdots & \ddots &\vdots \\
    \vdots & \vdots & \vdots & \ddots &\vdots \\
    v_{r_1}^{(j)}u_r^{(j)} u_1^{(j)} &v_{r_1}^{(j)}u_r^{(j)} u_2^{(j)} &v_{r_1}^{(j)}u_r^{(j)} u_3^{(j)}&\cdots &v_{r_1}^{(j)}u_r^{(j)} u_r^{(j)}\\
    \end{bmatrix}
    &&\in\mathbb{R}^{r_1r\times r}
\end{aligned}
\end{align}
\begin{align}
\begin{aligned}
    \alpha_j^\ast \beta_j &= \left[ u^{(j)}\right]
    \bigotimes
    \left[
    \left(v^{(j)}\right)^\top\bigotimes \left(u^{(j)}\right)^\top
    \right]
    \\
    &=
    \begin{bmatrix}
    u_1^{(j)} v_1^{(j)}u_1^{(j)} &u_1^{(j)} v_1^{(j)}u_2^{(j)} &u_1^{(j)} v_1^{(j)}u_3^{(j)}&\cdots &u_1^{(j)} v_{r_1}^{(j)}u_r^{(j)}\\
    u_2^{(j)} v_1^{(j)}u_1^{(j)} &u_2^{(j)} v_1^{(j)}u_2^{(j)} &u_2^{(j)} v_1^{(j)}u_3^{(j)}&\cdots &u_2^{(j)} v_{r_1}^{(j)}u_r^{(j)}\\
    \vdots & \vdots & \vdots & \ddots &\vdots \\
    u_r^{(j)} v_1^{(j)}u_1^{(j)} &u_r^{(j)} v_1^{(j)}u_2^{(j)} &u_r^{(j)} v_1^{(j)}u_3^{(j)}&\cdots &u_r^{(j)} v_{r_1}^{(j)}u_r^{(j)}\\
    \end{bmatrix}
    &&\in\mathbb{R}^{r_1r\times r_1r},
\end{aligned}
\end{align}
\begin{align}
\begin{aligned}
    \beta_j^\ast \beta_j &=  u^{(j)}
    \bigotimes
    \left(u^{(j)}\right)^\top
    \\
    &=
    \begin{bmatrix}
    u_1^{(j)} u_1^{(j)} &u_1^{(j)} u_2^{(j)} &u_1^{(j)} u_3^{(j)}&\cdots &u_1^{(j)} u_r^{(j)}\\
    u_2^{(j)} u_1^{(j)} &u_2^{(j)} u_2^{(j)} &u_2^{(j)} u_3^{(j)}&\cdots &u_2^{(j)} u_r^{(j)}\\
    \vdots & \vdots & \vdots & \ddots &\vdots \\
    u_r^{(j)} u_1^{(j)} &u_r^{(j)} u_2^{(j)} &u_r^{(j)} u_3^{(j)}&\cdots &u_r^{(j)} u_r^{(j)}\\
    \end{bmatrix}
    &&\in\mathbb{R}^{r\times r}.
\end{aligned}
\end{align}
The rank deficiency occurs from $\alpha_j^\ast \alpha_j$ where the repeated entries appear in different rows. The entry-wise expression of $\alpha_j^\ast \alpha_j$ yields%:
\begin{align}
\begin{aligned}
    &\biggl(\alpha_j^\ast \alpha_j\biggr)_{p,q} = \left(v_{l_1}u_{l_2}\right)\left(v_{m_1}u_{m_2}\right),&& p= (r(l_1-1)+l_2)
   \\
  &&&    q= (r(m_1-1)+m_2),
\end{aligned}
\end{align}
where $1\le l_1,m_1\le r_1$, %and 
$1\le l_2,m_2\le r$, %and 
$p$ is the row index, and $q$ is the column index of $\alpha_j^\ast \alpha_j$, which is also the block matrix in $(\mathbf{M}^j)^\ast (\mathbf{\Sigma}^{(k)})^{-1}\mathbf{M}^j$. 
%\ti
{Note that, since $v_{l_1} = u_{l_1}, 1\le l_1\le r_1$, the following equality holds:} 
\begin{align}
    v_{l_1}u_{l_2} = v_{l_2}u_{l_1}, \qquad\text{provided that }1\le l_1,l_2\le r_1.
\end{align}
Therefore, for $l_1\neq l_2$ and $1\le l_1,l_2\le r_1$ ($r_1 \ge 2$), there must exist $p_1 =(r(l_1-1)+l_2) $ and $p_2 =(r(l_2-1)+l_1)$, such that 
\begin{align}
    \biggl(\alpha_j^\ast \alpha_j\biggr)_{p_1,q} =     \biggl(\alpha_j^\ast \alpha_j\biggr)_{p_2,q},\qquad p_1\neq p_2.
\end{align}
Then $\frac{r_1(r_1-1)}{2}$ pairs of rows are the same in each sub-block ($r_1$ sub-blocks in total) of $(\mathbf{M}_L^j)^\ast (\mathbf{\Sigma}_L^{(k)})^{-1}\mathbf{M}_L^j$, and hence the deficient rank of matrix $\mathcal{K}_L$ will be $\frac{r_1^2(r_1-1)}{2}$.

% For example, if we choose $r_1=2,r=5$, then $\mathcal{K}_L$ is of size $30\times 30$ which will share  $\frac{r_1^2(r_1-1)}{2}=2$ pairs of same rows; hence the rank of $\mathcal{K}_L$ will be no larger than $28$. This is exactly the situation in the numerical test.

%-------

\begin{remark}
To illustrate the ill conditioning, we use $r_1=2$ and $r=4$ as a simple case. 
Let 
\begin{align}
    &&u = 
\begin{pmatrix}
    v_1\\
    v_2\\
    w_1\\
    w_2
\end{pmatrix},
&&
v = 
\begin{pmatrix}
    v_1\\
    v_2
\end{pmatrix},
&&v = 
\begin{pmatrix}
    w_1\\
    w_2
\end{pmatrix}.
\end{align}
Then

\begin{align}
&\mathbf{M}^j_L=
\begin{pmatrix}
    \alpha^{(j)}& \mathbf{0}_{1\times 8} 
    &\beta^{(j)}&\mathbf{0}_{1\times 4} 
    \\
   \mathbf{0}_{1\times 8} &  \alpha^{(j)}
    & \mathbf{0}_{1\times 4}  &\beta^{(j)}
\end{pmatrix},
\end{align}
with
\begin{align}
    \alpha^{(j)}
    &=
\begin{pmatrix}
    v_1^{(j)}v_1^{(j)} & v_1^{(j)}v_2^{(j)} & v_1^{(j)}w_1^{(j)} & v_1^{(j)}w_2^{(j)}
    &v_2^{(j)}v_1^{(j)} & v_2^{(j)}v_2^{(j)} & v_2^{(j)}w_1^{(j)} & v_2^{(j)}w_2^{(j)}
\end{pmatrix},
\\
    \beta^{(j)}
    &=
    \begin{pmatrix}
       v_1&v_2&w_1&w_2
    \end{pmatrix},
\\
\mathbf{0}_{1\times 8}
&=
\begin{pmatrix}
   0&0&0&0&0&0&0&0
\end{pmatrix},
    \\
\mathbf{0}_{1\times 4}
&=
\begin{pmatrix}
   0&0&0&0
\end{pmatrix}.
\end{align}
%And the 
The unknown vector $\theta_L$ yields the following:
%\ti
{I don't think we're allowed colored text.}
\begin{align}
\begin{aligned}
          \theta_L
    =
%\left[      
\left[
%\color{red}
\left(
       \mathbf{B}^{(v)}_{v\tau}\right)_{111},
       \quad
       \left(\mathbf{B}^{(v)}_{v\tau}\right)_{121},
       \quad
       \left(\mathbf{C}^{(v)}_{w\tau}\right)_{(111)},
       \quad
       \left(\mathbf{C}^{(v)}_{w\tau}\right)_{(121)},
       \quad \right.
       \\
       %\color{red}
       \left.
       \left(\mathbf{B}^{(v)}_{v\tau}\right)_{211},
      \quad
       \left(\mathbf{B}^{(v)}_{v\tau}\right)_{221},
       \quad
       \left(\mathbf{C}^{(v)}_{w\tau}\right)_{(211)},
        \quad      
       \left(\mathbf{C}^{(v)}_{w\tau}\right)_{(221)},\right.
       \\
       %\color{red}
       \left.
       \left(\mathbf{B}^{(v)}_{v\tau}\right)_{112},
       \quad
       \left(\mathbf{B}^{(v)}_{v\tau}\right)_{122},
       \quad
       \left(\mathbf{C}^{(v)}_{w\tau}\right)_{(112)},
       \quad
       \left(\mathbf{C}^{(v)}_{w\tau}\right)_{(122)},
       \quad \right.
       \\
       %\color{red}
       \left.
       \left(\mathbf{B}^{(v)}_{v\tau}\right)_{212},
      \quad
       \left(\mathbf{B}^{(v)}_{v\tau}\right)_{222},
       \quad
       \left(\mathbf{C}^{(v)}_{w\tau}\right)_{(212)},
        \quad      
       \left(\mathbf{C}^{(v)}_{w\tau}\right)_{(222)},
       \quad \right.
              \\
       %\color{blue}
       \left.
        \left(\mathbf{C}^{(v)}_{v\tau}\right)_{(11)},\quad
        \left(\mathbf{C}^{(v)}_{v\tau}\right)_{(12)},\quad
        \left(\mathbf{D}^{(v)}_{w\tau}\right)_{(11)},\quad
        \left(\mathbf{D}^{(v)}_{w\tau}\right)_{(12)},\right.
       \\
              %\color{blue}
        \left.
      \left(\mathbf{C}^{(v)}_{v\tau}\right)_{(21)},\quad
        \left(\mathbf{C}^{(v)}_{v\tau}\right)_{(22)},\quad
        \left(\mathbf{D}^{(v)}_{w\tau}\right)_{(21)},\quad
        \left(\mathbf{D}^{(v)}_{w\tau}\right)_{(22)}
        \right],
       %\right]
\end{aligned}
\label{eqn:theta-l}
\end{align}
where the %{\textcolor{red}{red}} %parts means 
%\red
{first four rows in~\eqref{eqn:theta-l} }
terms denote the nonlinear parameters in $\theta_L$, and %color 
the %{\textcolor{blue}{blue}} %parts means 
%\red
{last two rows in~\eqref{eqn:theta-l}} 
terms denote the linear parameters in $\theta_L$.
The $j$th row of product $(\mathbf{M}_L^j)^\top (\mathbf{\Sigma}^{(k)})^{-1}\mathbf{M}_L^j$ is related to the $j$th row of the residual.
In particular, the first row of the product $(\mathbf{M}_L^j)^\top (\mathbf{\Sigma}^{(k)})^{-1}\mathbf{M}_L^j$ 
yields the following:
\begin{align}
    \begin{aligned}
    &\left({(\mathbf{M}_L^j)^\ast} {(\mathbf{\Sigma}^{(k)})^{-1}}{\mathbf{M}_L^j}\right)_{1\cdot}=
          \\
    &\left[
    {\mathbf{\Sigma}_{11}}{v_1v_1}{v_1v_1}
    ,\quad
    {\mathbf{\Sigma}_{11}}{v_1v_1}{v_1v_2}
    ,\quad
    {\mathbf{\Sigma}_{11}}{v_1v_1}{v_1w_1}
    ,\quad
    {\mathbf{\Sigma}_{11}}{v_1v_1}{v_1w_2}
    ,\right.
    \\
    &\left.
    {\mathbf{\Sigma}_{11}}{v_1v_1}{v_2v_1}
    ,\quad
    {\mathbf{\Sigma}_{11}}{v_1v_1}{v_2v_2}
    ,\quad
    {\mathbf{\Sigma}_{11}}{v_1v_1}{v_2w_1}
    ,\quad
    {\mathbf{\Sigma}_{11}}{v_1v_1}{v_2w_2},
    \right.
    \\
    &\left.
    0,\quad 0,\quad 0,\quad 0,\quad 0,\quad 0,\quad 0,\quad 0,
    \right.
    \\
    &\left.
    {\mathbf{\Sigma}_{11}}{v_1v_1}{v_1}
    ,\quad
    {\mathbf{\Sigma}_{11}}{v_1v_1}{v_2}
    ,\quad
    {\mathbf{\Sigma}_{11}}{v_1v_1}{w_1}
    ,\quad
    {\mathbf{\Sigma}_{11}}{v_1v_1}{w_2},
    \right.
    \\
    &\left.
    0,\quad 0,\quad 0,\quad 0
    \right].
\end{aligned}
\end{align}
%%%%%
%\clearpage
\noindent
The problem occurs %at 
in the second %row 
and fifth rows.
The second row yields:
\begin{align}
    \begin{aligned}
    &\left({
    (\mathbf{M}_L^j)^\ast} {(\mathbf{\Sigma}^{(k)})^{-1}}{\mathbf{M}_L^j}\right)_{2\cdot}=
          \\
    &\left[
    {\mathbf{\Sigma}_{11}}{v_1v_2}{v_1v_1}
    ,\quad
    {\mathbf{\Sigma}_{11}}{v_1v_2}{v_1v_2}
    ,\quad
    {\mathbf{\Sigma}_{11}}{v_1v_2}{v_1w_1}
    ,\quad
    {\mathbf{\Sigma}_{11}}{v_1v_2}{v_1w_2}
    ,\right.
    \\
    &\left.
    {\mathbf{\Sigma}_{11}}{v_1v_2}{v_2v_1}
    ,\quad
    {\mathbf{\Sigma}_{11}}{v_1v_2}{v_2v_2}
    ,\quad
    {\mathbf{\Sigma}_{11}}{v_1v_2}{v_2w_1}
    ,\quad
    {\mathbf{\Sigma}_{11}}{v_1v_2}{v_2w_2},
    \right.
    \\
    &\left.
    0,\quad 0,\quad 0,\quad 0,\quad 0,\quad 0,\quad 0,\quad 0,
    \right.
    \\
    &\left.
    {\mathbf{\Sigma}_{11}}{v_1v_2}{v_1}
    ,\quad
    {\mathbf{\Sigma}_{11}}{v_1v_2}{v_2}
    ,\quad
    {\mathbf{\Sigma}_{11}}{v_1v_2}{w_1}
    ,\quad
    {\mathbf{\Sigma}_{11}}{v_1v_2}{w_2},
    \right.
    \\
    &\left.
    0,\quad 0,\quad 0,\quad 0
    \right].
\end{aligned}
\end{align}
%%%%%
The fifth row yields:
\begin{align}
    \begin{aligned}
    &\left({(\mathbf{M}_L^j)^\ast} {(\mathbf{\Sigma}^{(k)})^{-1}}{\mathbf{M}_L^j}\right)_{5\cdot}=
          \\
    &\left[
    {\mathbf{\Sigma}_{11}}{v_2v_1}{v_1v_1}
    ,\quad
    {\mathbf{\Sigma}_{11}}{v_2v_1}{v_1v_2}
    ,\quad
    {\mathbf{\Sigma}_{11}}{v_2v_1}{v_1w_1}
    ,\quad
    {\mathbf{\Sigma}_{11}}{v_2v_1}{v_1w_2}
    ,\right.
    \\
    &\left.
    {\mathbf{\Sigma}_{11}}{v_2v_1}{v_2v_1}
    ,\quad
    {\mathbf{\Sigma}_{11}}{v_2v_1}{v_2v_2}
    ,\quad
    {\mathbf{\Sigma}_{11}}{v_2v_1}{v_2w_1}
    ,\quad
    {\mathbf{\Sigma}_{11}}{v_2v_1}{v_2w_2},
    \right.
    \\
    &\left.
    0,\quad 0,\quad 0,\quad 0,\quad 0,\quad 0,\quad 0,\quad 0,
    \right.
    \\
    &\left.
    {\mathbf{\Sigma}_{11}}{v_2v_1}{v_1}
    ,\quad
    {\mathbf{\Sigma}_{11}}{v_2v_1}{v_2}
    ,\quad
    {\mathbf{\Sigma}_{11}}{v_2v_1}{w_1}
    ,\quad
    {\mathbf{\Sigma}_{11}}{v_2v_1}{w_2},
    \right.
    \\
    &\left.
    0,\quad 0,\quad 0,\quad 0
    \right].
\end{aligned}
\end{align}
The terms $v_2v_1$ and $v_1v_2$ are equal. %equivalent by commuting. 
These two terms  are corresponding to the second and %the 
fifth elements of $\theta_L$, i.e., $\left(\mathbf{B}^{(v)}_{v\tau}\right)_{121}$ and $\left(\mathbf{B}^{(v)}_{v\tau}\right)_{211}$.
In fact, this issue may occur in solving the least square problem if all the nonlinear unknowns are involved, i.e., all entries of the unknown tensors are assumed non-zeros.

\end{remark}
%%%%%%%%%%%%%%%%%%%%%%%%%%%%%%%%%%%%%%%%
% \begin{remark}
% Given any least square problem
% \begin{align}
%     \min_{x} f(x) = 
%     \frac{1}{2}
%     \biggl\|
%     \mathbf{M} \theta-\bb
%     \biggr\|^2
% \end{align}
% where $\theta$ is the unknown parameter; 
% If the minimizer $\theta^\dag$ exists, then the following normal equation must be true:
% \begin{align}
%     \mathbf{M}^\top \mathbf{M}  \theta^\dag
%     = \mathbf{M}^\top\bb.
% \end{align}
% However, $\mathbf{M}^\top \mathbf{M}$ may have more than one pair of rows that are the same.
% \end{remark}
%%%%%%%%%%%%%%%%%%%%%%%%%%%%%%%%%%%%%%%%
%%%%%%%%%%%%%%%%%%%%%%%%%%%%%%%%%%%%%%%%
\subsubsection{Special Regularization in Optimization Problem\label{app-optimization-special} }
To overcome the rank-deficiency from the formulation of $M_L$, we consider that $\mathbf{B}^{(v)}_{v\tau}$ and $\mathbf{B}^{(w)}_{v\tau}$ are upper triangular matrices %at 
%\ti
{in} each dimension, i.e., $\left(\mathbf{B}^{(v)}_{v\tau}\right)_{ijk}$ and $\left(\mathbf{B}^{(w)}_{v\tau}\right)_{ijk}$ %is an 
are upper triangular matrices for a fixed index $k$. 
In particular, these terms share the following:
\begin{align}
&\begin{aligned}
      &\left(\mathbf{B}^{(v)}_{v\tau}\right)_{ijk}
    =
    0,
    &\text{provided that } i>j;   
    & \, i,j,k=1,\ldots,r_1
\end{aligned}
\\
&\begin{aligned}
      &\left(\mathbf{B}^{(w)}_{v\tau}\right)_{ijk}
    =
    0,
    &\text{provided that } i>j;  
    &\, i,j=1,\cdots,r_1; k=1,\ldots,r_2.
\end{aligned}
\end{align}
Note that, since the constraints for $\left(\mathbf{B}^{(w)}_{v\tau}\right)_{ijk}$ require that $\left(\mathbf{B}^{(w)}_{v\tau}\right)_{iii}=0$ and $\left(\mathbf{B}^{(w)}_{v\tau}\right)_{ijk}+\left(\mathbf{B}^{(w)}_{v\tau}\right)_{jik}=0,i\neq j$,
%then 
$\mathbf{B}^{(w)}_{v\tau}$ vanishes.
%\ti{Check.}
Hence, the discrete formulation of the new multiscale physics constraints yield the following:
%% large scale
\begin{align}
&\mathbf{B}^{(v)}_{v\tau}:
\nonumber
\\
&\left(\mathbf{B}^{(v)}_{v\tau}\right)_{(iii)} =0, \qquad\qquad i=1,\cdots, r_1,
\label{eqn:cg-constraint2-v-1}
\\
&\left(\mathbf{B}^{(v)}_{v\tau}\right)_{(ijj)}+
+\left(\mathbf{B}^{(v)}_{v\tau}\right)_{(jji)}=0,\qquad  i,j=1,\cdots, r_1,i<j,
\label{eqn:cg-constraint2-v-2}
\\
&\left(\mathbf{B}^{(v)}_{v\tau}\right)_{(ijk)}+\left(\mathbf{B}^{(v)}_{v\tau}\right)_{(ikj)}
+\left(\mathbf{B}^{(v)}_{v\tau}\right)_{(jki)}
=0,
\qquad i,j,k=1,\cdots, r_1,i< j < k. \label{eqn:cg-constraint2-v-3}
\end{align}
\begin{align}
&\mathbf{C}^{(v)}_{w\tau}\, \text{ and } \,\mathbf{B}^{(w)}_{v\tau}:
\nonumber
\\
 &\left({\mathbf{C}}^{(v)}_{w\tau}\right)_{(iki)}
    +    \left(\mathbf{B}^{(w)}_{v\tau}\right)_{(iik)}=0,\quad i=1,\cdots, r_1, k=1,\cdots, r_2;
    \\
    &\left({\mathbf{C}}^{(v)}_{w\tau}\right)_{(ikj)}
    +    \left({\mathbf{C}}^{(v)}_{w\tau}\right)_{(jki)}
    +\left(\mathbf{B}^{(w)}_{v\tau}\right)_{(ijk)}
    =0,\\
    &\hspace{6cm}
    i<j, 
    i,j=1,\cdots, r_1, k=1,\cdots, r_2.\nonumber    
\end{align}
\begin{align} 
&\mathbf{C}^{(w)}_{w\tau}:
\nonumber
\\
&\left(\mathbf{C}^{(w)}_{w\tau}\right)_{(ijj)} =0, & i=1,\cdots, r_1; j=1,\cdots, r_2,
\label{eqn:cg-constraint2-w-3}
\\
&\left(\mathbf{C}^{(w)}_{w\tau}\right)_{(ijk)}+\left(\mathbf{C}^{(w)}_{w\tau}\right)_{(ikj)} =0, & i=1,\cdots, r_1; j,k=1,\cdots, r_2, j\neq k.
\label{eqn:cg-constraint2-w-4}
\end{align}

%%%%%%%%%%%%%%%%%%%%%%%%%%%%%%%%%%%%%%%%%%%%%%%
%%%%%%%%%%%%%%%%%%%%%%%%%%%%%%%%%%%%%%%%%%%%%%%

%%%%%%%%%%%%%%%%%%%%%%%%%%%%%%%%%%%%%%%%%%%%%%%
%%%%%%%%%%%%%%%%%%%%%%%%%%%%%%%%%%%%%%%%%%%%%%%

\bibliography{ref}

\begin{thebibliography}{10}
\expandafter\ifx\csname url\endcsname\relax
  \def\url#1{\texttt{#1}}\fi
\expandafter\ifx\csname urlprefix\endcsname\relax\def\urlprefix{URL }\fi
\expandafter\ifx\csname href\endcsname\relax
  \def\href#1#2{#2} \def\path#1{#1}\fi

\bibitem{vallis2017atmospheric}
G.~K. Vallis, Atmospheric and oceanic fluid dynamics, Cambridge University
  Press, 2017.

\bibitem{strogatz2018nonlinear}
S.~H. Strogatz, Nonlinear dynamics and chaos with student solutions manual:
  With applications to physics, biology, chemistry, and engineering, CRC press,
  2018.

\bibitem{wilcox1988multiscale}
D.~C. Wilcox, Multiscale model for turbulent flows, AIAA journal 26~(11) (1988)
  1311--1320.

\bibitem{sheard2009principles}
S.~A. Sheard, A.~Mostashari, Principles of complex systems for systems
  engineering, Systems Engineering 12~(4) (2009) 295--311.

\bibitem{ghil2012topics}
M.~Ghil, S.~Childress, Topics in geophysical fluid dynamics: atmospheric
  dynamics, dynamo theory, and climate dynamics, Springer Science \& Business
  Media, 2012.

\bibitem{majda2016introduction}
A.~J. Majda, Introduction to turbulent dynamical systems in complex systems,
  Springer, 2016.

\bibitem{tao2009multiscale}
W.-K. Tao, J.-D. Chern, R.~Atlas, D.~Randall, M.~Khairoutdinov, J.-L. Li, D.~E.
  Waliser, A.~Hou, X.~Lin, C.~Peters-Lidard, et~al., A multiscale modeling
  system: Developments, applications, and critical issues, Bulletin of the
  American Meteorological Society 90~(4) (2009) 515--534.

\bibitem{majda2014new}
A.~J. Majda, I.~Grooms, New perspectives on superparameterization for
  geophysical turbulence, Journal of Computational Physics 271 (2014) 60--77.

\bibitem{salmon1998lectures}
R.~Salmon, Lectures on geophysical fluid dynamics, Oxford University Press,
  1998.

\bibitem{dijkstra2013nonlinear}
H.~A. Dijkstra, Nonlinear climate dynamics, Cambridge University Press, 2013.

\bibitem{palmer1993nonlinear}
T.~Palmer, A nonlinear dynamical perspective on climate change, Weather 48~(10)
  (1993) 314--326.

\bibitem{farazmand2019extreme}
M.~Farazmand, T.~P. Sapsis, Extreme events: Mechanisms and prediction, Applied
  Mechanics Reviews 71~(5).

\bibitem{trenberth2015attribution}
K.~E. Trenberth, J.~T. Fasullo, T.~G. Shepherd, Attribution of climate extreme
  events, Nature Climate Change 5~(8) (2015) 725--730.

\bibitem{moffatt2021extreme}
H.~Moffatt, Extreme events in turbulent flow, Journal of Fluid Mechanics 914.

\bibitem{majda2003introduction}
A.~Majda, Introduction to PDEs and Waves for the Atmosphere and Ocean, Vol.~9,
  American Mathematical Soc., 2003.

\bibitem{manneville1979intermittency}
P.~Manneville, Y.~Pomeau, Intermittency and the {L}orenz model, Physics Letters
  A 75~(1-2) (1979) 1--2.

\bibitem{majda2012challenges}
A.~J. Majda, Challenges in climate science and contemporary applied
  mathematics, Communications on Pure and Applied Mathematics 65~(7) (2012)
  920--948.

\bibitem{majda2018model}
A.~J. Majda, N.~Chen, Model error, information barriers, state estimation and
  prediction in complex multiscale systems, Entropy 20~(9) (2018) 644.

\bibitem{hesthaven2015certified}
J.~S. Hesthaven, G.~Rozza, B.~Stamm, Certified Reduced Basis Methods for
  Parametrized Partial Differential Equations, Springer, 2015.

\bibitem{HLB96}
P.~Holmes, J.~L. Lumley, G.~Berkooz, Turbulence, Coherent Structures, Dynamical
  Systems and Symmetry, Cambridge, 1996.

\bibitem{noack2011reduced}
B.~R. Noack, M.~Morzynski, G.~Tadmor, Reduced-order modelling for flow control,
  Vol. 528, Springer Science \& Business Media, 2011.

\bibitem{quarteroni2015reduced}
A.~Quarteroni, A.~Manzoni, F.~Negri, Reduced Basis Methods for Partial
  Differential Equations: An Introduction, Vol.~92, Springer, 2015.

\bibitem{taira2020modal}
K.~Taira, M.~S. Hemati, S.~L. Brunton, Y.~Sun, K.~Duraisamy, S.~Bagheri, S.~T.
  Dawson, C.-A. Yeh, Modal analysis of fluid flows: Applications and outlook,
  AIAA journal 58~(3) (2020) 998--1022.

\bibitem{yan2009linear}
X.~Yan, X.~Su, Linear regression analysis: theory and computing, World
  Scientific, 2009.

\bibitem{freedman2009statistical}
D.~A. Freedman, Statistical models: theory and practice, Cambridge University
  Press, 2009.

\bibitem{hasselmann1988pips}
K.~Hasselmann, {PIPs and POPs: The reduction of complex dynamical systems using
  principal interaction and oscillation patterns}, Journal of Geophysical
  Research: Atmospheres 93~(D9) (1988) 11015--11021.

\bibitem{kwasniok1996reduction}
F.~Kwasniok, The reduction of complex dynamical systems using principal
  interaction patterns, Physica D: Nonlinear Phenomena 92~(1-2) (1996) 28--60.

\bibitem{rowley2009spectral}
C.~W. Rowley, I.~Mezi{\'c}, S.~Bagheri, P.~Schlatter, D.~S. Henningson,
  Spectral analysis of nonlinear flows, Journal of Fluid Mechanics 641 (2009)
  115--127.

\bibitem{schmid2010dynamic}
P.~J. Schmid, Dynamic mode decomposition of numerical and experimental data,
  Journal of Fluid Mechanics 656 (2010) 5--28.

\bibitem{ahmed2021closures}
S.~E. Ahmed, S.~Pawar, O.~San, A.~Rasheed, T.~Iliescu, B.~R. Noack, {On
  closures for reduced order models--A spectrum of first-principle to
  machine-learned avenues}, Physics of Fluids 33~(9) (2021) 091301.

\bibitem{carlberg2013gnat}
K.~Carlberg, C.~Farhat, J.~Cortial, D.~Amsallem, {The GNAT method for nonlinear
  model reduction: effective implementation and application to computational
  fluid dynamics and turbulent flows}, Journal of Computational Physics 242
  (2013) 623--647.

\bibitem{mou2021data}
C.~Mou, B.~Koc, O.~San, L.~G. Rebholz, T.~Iliescu, Data-driven variational
  multiscale reduced order models, Computer Methods in Applied Mechanics and
  Engineering 373 (2021) 113470.

\bibitem{xie2018data}
X.~Xie, M.~Mohebujjaman, L.~G. Rebholz, T.~Iliescu, Data-driven filtered
  reduced order modeling of fluid flows, SIAM Journal on Scientific Computing
  40~(3) (2018) B834--B857.

\bibitem{majda2001mathematical}
A.~J. Majda, I.~Timofeyev, E.~Vanden~Eijnden, A mathematical framework for
  stochastic climate models, Communications on Pure and Applied Mathematics
  54~(8) (2001) 891--974.

\bibitem{majda1999models}
A.~J. Majda, I.~Timofeyev, E.~V. Eijnden, Models for stochastic climate
  prediction, Proceedings of the National Academy of Sciences 96~(26) (1999)
  14687--14691.

\bibitem{lu2020data}
F.~Lu, Data-driven model reduction for stochastic burgers equations, Entropy
  22~(12) (2020) 1360.

\bibitem{lin2021data}
K.~K. Lin, F.~Lu, Data-driven model reduction, wiener projections, and the
  {K}oopman-{M}ori-{Z}wanzig formalism, Journal of Computational Physics 424
  (2021) 109864.

\bibitem{majda2012physics}
A.~J. Majda, J.~Harlim, Physics constrained nonlinear regression models for
  time series, Nonlinearity 26~(1) (2012) 201.

\bibitem{harlim2014ensemble}
J.~Harlim, A.~Mahdi, A.~J. Majda, An ensemble {K}alman filter for statistical
  estimation of physics constrained nonlinear regression models, Journal of
  Computational Physics 257 (2014) 782--812.

\bibitem{kondrashov2015data}
D.~Kondrashov, M.~D. Chekroun, M.~Ghil, {Data-driven non-Markovian closure
  models}, Physica D: Nonlinear Phenomena 297 (2015) 33--55.

\bibitem{palmer2001nonlinear}
T.~N. Palmer, A nonlinear dynamical perspective on model error: A proposal for
  non-local stochastic-dynamic parametrization in weather and climate
  prediction models, Quarterly Journal of the Royal Meteorological Society
  127~(572) (2001) 279--304.

\bibitem{majda2011improving}
A.~J. Majda, B.~Gershgorin, Improving model fidelity and sensitivity for
  complex systems through empirical information theory, Proceedings of the
  National Academy of Sciences 108~(25) (2011) 10044--10049.

\bibitem{crommelin2008subgrid}
D.~Crommelin, E.~Vanden-Eijnden, {Subgrid-scale parameterization with
  conditional Markov chains}, Journal of the Atmospheric Sciences 65~(8) (2008)
  2661--2675.

\bibitem{phillips2004evaluating}
T.~J. Phillips, G.~L. Potter, D.~L. Williamson, R.~T. Cederwall, J.~S. Boyle,
  M.~Fiorino, J.~J. Hnilo, J.~G. Olson, S.~Xie, J.~J. Yio, Evaluating
  parameterizations in general circulation models: Climate simulation meets
  weather prediction, Bulletin of the American Meteorological Society 85~(12)
  (2004) 1903--1916.

\bibitem{branicki2013non}
M.~Branicki, N.~Chen, A.~J. Majda, Non-{G}aussian test models for prediction
  and state estimation with model errors, Chinese Annals of Mathematics, Series
  B 34~(1) (2013) 29--64.

\bibitem{evensen2009data}
G.~Evensen, Data assimilation: the ensemble {K}alman filter, Springer Science
  \& Business Media, 2009.

\bibitem{kalnay2003atmospheric}
E.~Kalnay, Atmospheric modeling, data assimilation and predictability,
  Cambridge University Press, 2003.

\bibitem{law2015data}
K.~Law, A.~Stuart, K.~Zygalakis, Data assimilation, Cham, Switzerland: Springer
  214.

\bibitem{majda2012filtering}
A.~J. Majda, J.~Harlim, Filtering complex turbulent systems, Cambridge
  University Press, 2012.

\bibitem{majda2012fundamental}
A.~J. Majda, Y.~Yuan, Fundamental limitations of ad hoc linear and quadratic
  multi-level regression models for physical systems, Discrete and Continuous
  Dynamical Systems B 17~(4) (2012) 1333--1363.

\bibitem{chen2018conditional}
N.~Chen, A.~Majda, Conditional {G}aussian systems for multiscale nonlinear
  stochastic systems: Prediction, state estimation and uncertainty
  quantification, Entropy 20~(7) (2018) 509.

\bibitem{weinan2000invariant}
W.~E, K.~Khanin, A.~Mazel, Y.~Sinai, Invariant measures for {B}urgers equation
  with stochastic forcing, Annals of Mathematics (2000) 877--960.

\bibitem{weinan1997probability}
W.~E, K.~Khanin, A.~Mazel, Y.~Sinai, Probability distribution functions for the
  random forced {B}urgers equation, Physical Review Letters 78~(10) (1997)
  1904.

\bibitem{chen2016filtering}
N.~Chen, A.~J. Majda, Filtering nonlinear turbulent dynamical systems through
  conditional {G}aussian statistics, Monthly Weather Review 144~(12) (2016)
  4885--4917.

\bibitem{chen2014predicting}
N.~Chen, A.~J. Majda, D.~Giannakis, Predicting the cloud patterns of the
  {M}adden-{J}ulian oscillation through a low-order nonlinear stochastic model,
  Geophysical Research Letters 41~(15) (2014) 5612--5619.

\bibitem{chen2015predicting}
N.~Chen, A.~J. Majda, Predicting the real-time multivariate {M}adden--{J}ulian
  oscillation index through a low-order nonlinear stochastic model, Monthly
  Weather Review 143~(6) (2015) 2148--2169.

\bibitem{chen2014information}
N.~Chen, A.~J. Majda, X.~T. Tong, Information barriers for noisy {L}agrangian
  tracers in filtering random incompressible flows, Nonlinearity 27~(9) (2014)
  2133.

\bibitem{chen2015noisy}
N.~Chen, A.~J. Majda, X.~T. Tong, Noisy {L}agrangian tracers for filtering
  random rotating compressible flows, Journal of Nonlinear Science 25~(3)
  (2015) 451--488.

\bibitem{branicki2013dynamic}
M.~Branicki, A.~J. Majda, Dynamic stochastic superresolution of sparsely
  observed turbulent systems, Journal of Computational Physics 241 (2013)
  333--363.

\bibitem{keating2012new}
S.~R. Keating, A.~J. Majda, K.~S. Smith, New methods for estimating ocean eddy
  heat transport using satellite altimetry, Monthly Weather Review 140~(5)
  (2012) 1703--1722.

\bibitem{kaercher2018reduced}
M.~Kaercher, S.~Boyaval, M.~A. Grepl, K.~Veroy, {Reduced basis approximation
  and a posteriori error bounds for 4D-Var data assimilation}, Optim. Eng.
  (2018) 1--33.

\bibitem{maday2015parameterized}
Y.~Maday, A.~T. Patera, J.~D. Penn, M.~Yano, A parameterized-background
  data-weak approach to variational data assimilation: formulation, analysis,
  and application to acoustics, Int. J. Num. Meth. Engng. 102~(5) (2015)
  933--965.

\bibitem{pagani2017efficient}
S.~Pagani, A.~Manzoni, A.~Quarteroni, {Efficient state/parameter estimation in
  nonlinear unsteady PDEs by a reduced basis ensemble Kalman filter}, SIAM-ASA
  J. Uncertain. 5~(1) (2017) 890--921.

\bibitem{popov2021multifidelity}
A.~A. Popov, C.~Mou, T.~Iliescu, A.~Sandu, {A multifidelity ensemble Kalman
  filter with reduced order control variates}, SIAM J. Sci. Comput. 43~(2)
  (2021) A1134--A1162.

\bibitem{stefanescu2015pod}
R.~{\c{S}}tef{\u{a}}nescu, A.~Sandu, I.~M. Navon, {POD/DEIM reduced-order
  strategies for efficient four dimensional variational data assimilation}, J.
  Comput. Phys. 295 (2015) 569--595.

\bibitem{xiao2018parameterised}
D.~Xiao, J.~Du, F.~Fang, C.~C. Pain, J.~Li, {Parameterised non-intrusive
  reduced order methods for ensemble Kalman filter data assimilation}, Comput.
  \& Fluids 177 (2018) 69--77.

\bibitem{zerfas2019continuous}
C.~Zerfas, L.~G. Rebholz, M.~Schneier, T.~Iliescu, Continuous data assimilation
  reduced order models of fluid flow, Comput. Meth. Appl. Mech. Eng. 357 (2019)
  112596.

\bibitem{lahoz2010data}
W.~Lahoz, B.~Khattatov, R.~M{\'e}nard, Data assimilation and information, in:
  Data Assimilation, Springer, 2010, pp. 3--12.

\bibitem{whitaker2002ensemble}
J.~S. Whitaker, T.~M. Hamill, Ensemble data assimilation without perturbed
  observations, Monthly weather review 130~(7) (2002) 1913--1924.

\bibitem{hunt2007efficient}
B.~R. Hunt, E.~J. Kostelich, I.~Szunyogh, Efficient data assimilation for
  spatiotemporal chaos: A local ensemble transform kalman filter, Physica D:
  Nonlinear Phenomena 230~(1-2) (2007) 112--126.

\bibitem{houtekamer2016review}
P.~L. Houtekamer, F.~Zhang, Review of the ensemble kalman filter for
  atmospheric data assimilation, Monthly Weather Review 144~(12) (2016)
  4489--4532.

\bibitem{liptser2013statistics}
R.~S. Liptser, A.~N. Shiryaev, Statistics of random processes II: Applications,
  Vol.~6, Springer Science \& Business Media, 2013.

\bibitem{kalman1961new}
R.~E. Kalman, R.~S. Bucy, New results in linear filtering and prediction
  theory, Journal of Basic Engineering 83~(1) (1961) 95--108.

\bibitem{saha2010ncep}
S.~Saha, S.~Moorthi, H.-L. Pan, X.~Wu, J.~Wang, S.~Nadiga, P.~Tripp,
  R.~Kistler, J.~Woollen, D.~Behringer, et~al., The ncep climate forecast
  system reanalysis, Bulletin of the American Meteorological Society 91~(8)
  (2010) 1015--1058.

\bibitem{chen2020learning}
N.~Chen, Learning nonlinear turbulent dynamics from partial observations via
  analytically solvable conditional statistics, Journal of Computational
  Physics 418 (2020) 109635.

\bibitem{boyd2004convex}
S.~Boyd, L.~Vandenberghe, Convex optimization, Cambridge University Press,
  2004.

\bibitem{bergemann2012ensemble}
K.~Bergemann, S.~Reich, {An ensemble {K}alman-Bucy filter for continuous data
  assimilation}, Meteorologische Zeitschrift 21 (2012) 213--219.

\bibitem{greatbatch2000four}
R.~J. Greatbatch, B.~T. Nadiga, Four-gyre circulation in a barotropic model
  with double-gyre wind forcing, Journal of Physical Oceanography 30~(6) (2000)
  1461--1471.

\bibitem{mou2020data}
C.~Mou, H.~Liu, D.~R. Wells, T.~Iliescu, Data-driven correction reduced order
  models for the quasi-geostrophic equations: A numerical investigation,
  International Journal of Computational Fluid Dynamics 34~(2) (2020) 147--159.

\bibitem{san2015stabilized}
O.~San, T.~Iliescu, A stabilized proper orthogonal decomposition reduced-order
  model for large scale quasigeostrophic ocean circulation, Advances in
  Computational Mathematics (2015) 1289--1319.

\bibitem{san2011approximate}
O.~San, A.~E. Staples, Z.~Wang, T.~Iliescu, Approximate deconvolution large
  eddy simulation of a barotropic ocean circulation model, Ocean Modelling 40
  (2011) 120--132.

\bibitem{kullback1951information}
S.~Kullback, R.~A. Leibler, On information and sufficiency, The Annals of
  Mathematical Statistics 22~(1) (1951) 79--86.

\bibitem{kullback1987letter}
S.~Kullback, Letter to the editor: The kullback-leibler distance, American
  Statistician.

\bibitem{kleeman2011information}
R.~Kleeman, Information theory and dynamical system predictability, Entropy
  13~(3) (2011) 612--649.

\bibitem{majda2005information}
A.~Majda, R.~V. Abramov, M.~J. Grote, Information theory and stochastics for
  multiscale nonlinear systems, Vol.~25, American Mathematical Soc., 2005.

\bibitem{brunton2016discovering}
S.~L. Brunton, J.~L. Proctor, J.~N. Kutz, Discovering governing equations from
  data by sparse identification of nonlinear dynamical systems, Proceedings of
  the National Academy of Sciences 113~(15) (2016) 3932--3937.

\bibitem{sun2014causation}
J.~Sun, E.~M. Bollt, Causation entropy identifies indirect influences,
  dominance of neighbors and anticipatory couplings, Physica D: Nonlinear
  Phenomena 267 (2014) 49--57.

\bibitem{elinger2020information}
J.~Elinger, Information theoretic causality measures for parameter estimation
  and system identification, Ph.D. thesis, Georgia Institute of Technology
  (2020).

\end{thebibliography}

\end{document}